\documentclass[twocolumn,tighten,twocolappendix]{aastex63}

\usepackage{amsmath}
\usepackage{bbm,bm}
\usepackage{enumitem}
\usepackage{bbold}

\graphicspath{{./}{figures/}}
\newcommand{\p}{\partial}
\newcommand{\di}{\mathrm{d}}

\newcommand{\pc}{\,{\rm pc}}

\newcommand{\second}{\,{\rm s}}
\newcommand{\yr}{\,{\rm yr}}
\newcommand{\Myr}{\,{\rm Myr}}

\newcommand{\nm}{\,{\rm nm}}
\newcommand{\erg}{\,{\rm erg}}
\newcommand{\cm}{\,{\rm cm}}
\newcommand{\gram}{\,{\rm g}}
\newcommand{\kms}{\,{\rm km}\,{\rm s}^{-1}}
\newcommand{\Msun}{\,M_{\odot}}
\newcommand{\Lsun}{\,L_{\odot}}
\newcommand{\Kel}{\,{\rm K}}
\newcommand{\kB}{{\,k_{\rm B}}}
\newcommand{\eV}{{\,{\rm eV}}}

\newcommand{\pcc}{\,{\rm cm}^{-3}}
\newcommand{\Sunit}{\,M_{\odot}\,{\rm pc^{-2}}}
\newcommand{\Bunit}{\,\mu {\rm G}}

\newcommand{\HII}{\ion{H}{2}\ignorespaces}

\newcommand{\Mcl}{M_{\rm 0}}
\newcommand{\Rcl}{R_{\rm 0}}
\newcommand{\Sigmacl}{\Sigma_{\rm 0}}
\newcommand{\sigmacl}{\sigma_{\rm 1d,0}}
\newcommand{\nHcl}{n_{\rm H,0}}
\newcommand{\tffcl}{t_{\rm ff,0}}
\newcommand{\muBcl}{\mu_{\Phi,0}}

\newcommand{\avir}{\alpha_{\rm vir}}
\newcommand{\avircl}{\alpha_{\rm vir,0}}
\newcommand{\avirt}{\tilde{\alpha}_{\rm vir}}
\newcommand{\avirttot}{\tilde{\alpha}_{\rm vir, tot}}

\newcommand{\eff}{\varepsilon_{\rm ff,0}}
\newcommand{\muB}{\mu_{\Phi}}
\newcommand{\tprime}{t^{\prime}}

\newcommand{\muH}{\mu_{\rm H}}
\newcommand{\nH}{n_{\rm H}}
\newcommand{\nHH}{n_{\rm H_2}}
\newcommand{\nHI}{n_{\rm H^0}}

\newcommand{\nEL}{n_{\rm e}}
\newcommand{\xHH}{x_{\rm H_2}}
\newcommand{\xHI}{x_{\rm H^0}}
\newcommand{\xHII}{x_{\rm H^+}}
\newcommand{\xEL}{x_{\rm e}}
\newcommand{\xCII}{x_{\rm C^+}}

\begin{document}

\title{Star Formation Efficiency and Dispersal of Giant Molecular Clouds with UV
  Radiation Feedback:\\
  Dependence on Gravitational Boundedness and Magnetic Fields}

\author[0000-0001-6228-8634]{Jeong-Gyu Kim} %
\author[0000-0002-0509-9113]{Eve C.~Ostriker} %
\author{Nina Filippova} %

\affiliation{Department of Astrophysical Sciences, Princeton University,
Princeton, NJ 08544, USA}

\begin{abstract}
  Molecular clouds are supported by turbulence and magnetic fields, but
  quantifying their influence on cloud lifecycle and star formation efficiency
  (SFE) remains an open question. We perform radiation magnetohydrodynamic
  simulations of star-forming giant molecular clouds (GMCs) with UV radiation
  feedback, in which the propagation of UV radiation via ray-tracing is coupled
  to hydrogen photochemistry. We consider 10 GMC models that vary in either
  initial virial parameter ($1 \le \alpha_{\rm vir,0} \le 5$) or dimensionless
  mass-to-magnetic flux ratio ($0.5 \le \mu_{\Phi,0} \le 8$ and $\infty$);
  the initial mass $10^5 M_{\odot}$ and radius $20\,{\rm pc}$ are fixed. Each
  model is run with five different initial turbulence realizations. In most
  models, the duration of star formation and the timescale for molecular gas
  removal (primarily by photoevaporation) are $4$-$8\,{\rm Myr}$. Both the final
  SFE ($\varepsilon_*$) and time-averaged SFE per freefall time
  ($\varepsilon_{\rm ff}$) are reduced by strong turbulence and magnetic fields.
  The median $\varepsilon_*$ ranges between $2.1\%$ and $9.5\%$. The median
  $\varepsilon_{\rm ff}$ ranges between $1.0\%$ and $8.0\%$ and anticorrelates
  with $\alpha_{\rm vir,0}$, in qualitative agreement with previous analytic
  theory and simulations. However, the time-dependent $\alpha_{\rm vir}(t)$ and
  $\varepsilon_{\rm ff,obs}(t)$ based on instantaneous gas properties and
  cluster luminosity are positively correlated due to rapid evolution, making
  observational validation of star formation theory difficult. Our median
  $\varepsilon_{\rm ff,obs}(t)\approx 2\%$ is similar to observed values. We
  show that the traditional virial parameter estimates the true gravitational
  boundedness within a factor of 2 on average, but neglect of magnetic support
  and velocity anisotropy can sometimes produce large departures. Magnetically
  subcritical GMCs
  are unlikely to represent sites of massive star formation
  given their unrealistic columnar outflows, prolonged lifetime, and low escape
  fraction of radiation.
\end{abstract}


\section{Introduction}\label{s:intro}

Giant molecular clouds (GMCs) are the primary reservoir of the cold, molecular
interstellar medium (ISM) and sites of ongoing star formation. The gas
distribution and velocity structure of GMCs are highly complex and hierarchical,
arising from the interplay of supersonic turbulence, magnetic fields, stellar
feeedback, and gravity. Both turbulent flows and magnetic fields are thought to
play crucial roles in structure formation as well as in preventing excessive
star formation by providing support against gravity
\citep{maclow04,mckee07,hennebelle19,krumholz19b}.
For example, turbulence can both create
and disperse local density enhancements such as sheets, filaments, and cores,
some of which become sufficiently dense to be susceptible to gravitational
contraction. Magnetic fields can guide gas flows and help organize mass into
filaments, but also prevent collapse if sufficiently strong.

Massive stars formed in GMCs produce intense ultraviolet (UV) radiation and
stellar winds during their evolution, and explode as supernovae at the end of
their lives. The energy and momentum injected by these processes can be
destructive to the natal clouds and quench star formation activity. Therefore,
the efficiency with which each feedback mechanism (individually and in
combination) couples to the ISM within turbulent and magnetized molecular clouds
is key to understanding several important issues, such as the lifecycle of
molecular clouds and formation of star clusters
\citep{krumholz19,adamo20b,chevance20b}.

Extensive surveys of molecular lines in the Milky Way and nearby galaxies have shown that GMCs exhibit a range of physical conditions.
For example, GMCs (and cluster-forming clumps within GMCs) in the main disk of the Milky Way and nearby disk galaxies have masses
$M \sim 10^3$--$10^6\Msun$, size $R \sim 1$--$10^2\pc$, 
surface density
$\Sigma \sim 10$--$10^3\Sunit$, and one dimensional (1D) velocity dispersion
$\sigma_{\rm 1d} \sim 1$--$8 \kms$
\citep[e.g.,][]{bolatto08,heyer09,fukui10,roman-duval10,heyer15,mivilledeschenes17}. From
extragalactic surveys, certain cloud properties such as surface density,
velocity dispersion, and internal turbulent pressure widely vary as a function
of galactic environment \citep{sun18,sun20a,sun20}. The virial parameter
$\avir$, which measures the relative importance of the turbulent kinetic energy
to the self-gravitational energy (albeit with some simplifying assumptions), has
a relatively narrow distribution with a typical value $\sim 0.5$--$5$
\citep{roman-duval10,heyer15,sun18,sun20}. This suggests that molecular clouds
are self-gravitating yet internal turbulent support is sufficient to prevent
global collapse. While self-gravity is important to GMCs' dynamical state
\citep{sun20a}, they are not necessarily virialized or even gravitationally
bound as isolated systems. Nevertheless, overdense clumps in filaments within
GMCs are more strongly bound \citep{kauffmann13}, and these are susceptible to
collapse and star formation.

Although observational constraints are indirect and weak, GMCs are permeated by
dynamically important magnetic fields \citep[see the review by][]{crutcher12}.
Linear polarization of dust thermal emission and spectral lines shows that
molecular clouds have a well-defined mean field direction that is correlated
with the direction of larger scale galactic magnetic fields
\citep{li06,li11,li13}. The orientation of the local magnetic field is
perpendicular to dense filaments, whereas it tends to be aligned with
low-density filaments connected to dense filaments
\citep{palmerim13,planckXXXV16,wardthompson17,soler19}.
Observing Zeeman splitting of
spectral lines is the only direct method to obtain the line-of-sight component
of magnetic field strength. The geometry-corrected total magnetic field strength
$B_{\rm tot}$ is relatively constant $\sim 6 \Bunit$ in diffuse \ion{H}{1}
clouds \citep{heiles05}, but increases roughly as
$B_{\rm tot} \propto \nH^{2/3}$ in denser molecular gas \citep{crutcher10}. The
average dimensionless mass-to-magnetic flux ratio
$\mu_{\Phi} \propto M_{\rm gas}/\Phi \propto \Sigma/B_{\rm tot}$, which measures
the relative importance of magnetic fields to gravity, exceeds unity
(supercritical) by a factor $\sim 2$--$3$ in molecular cores
\citep{crutcher99,crutcher10}. Recently, \cite{thompson19} measured the magnetic
strength of low-density (inter-core) regions of nearby molecular clouds obtained
from the OH Zeeman effect, finding the total magnetic strength of a few $\Bunit$ to
$30\Bunit$ with mean $15\Bunit$.

The efficiency and rate of star formation (SFE and SFR) on the scale of
molecular clouds are both low
\citep[e.g.,][]{mooney88,evans91,murray11,kennicutt12,utomo18}. Observational constraints
on the net (or final) SFE of GMCs, the fraction of the initial cloud gas mass
that will ever become stars before destruction, can be obtained indirectly based
on a ``snapshot'' view of a population of clouds in different evolutionary
stages. The distribution of instantaneous SFE
$\varepsilon = M_*/(M_* + M_{\rm gas})$ suggests that GMCs in the Milky Way
convert only a small fraction of gas into stars before dispersal
\cite[e.g.,][]{myers86,murray11,lee16}. The fraction of gas mass turned into stars per
freefall time (or SFE per freefall time, $\varepsilon_{\rm ff}$), is also very
small with large scatter (e.g.,
\citealt{krumholz07,lee16,vutisalchavakul16,ochsendorf17,utomo18,schruba19}; for a recent review,
see Section 3.2 of \citealt{krumholz19}). Curiously, however,
\citet{lee16} and \citet{vutisalchavakul16} found no significant correlation between
estimated $\varepsilon_{\rm ff}$ and the instantaneous $\avir$, while an
anticorrelation is predicted by analytic theory and numerical simulations of
star formation \citep[e.g.,][]{krumholz05,federrath12,padoan12}.

The question of
whether GMCs are transients or long-lived objects has historically been
controversial \citep{heyer15}, but recent investigation of cloud evolutionary
timelines based on scale-dependent CO-to-H$\alpha$ flux ratios shows that
molecular clouds are dispersed rapidly 
($\sim 5\Myr$)
after the onset of
massive star formation in a range of star-forming galaxies
\citep{kruijssen19,chevance20a,chevance20c,kimj20}. 
The youth of clusters
($\lesssim 5 \Myr$) associated with molecular clouds or \HII\ regions in external galaxies also 
suggests rapid disruption by stellar feedback \citep{grasha18,grasha19,hannon19,messa20}.

From a theoretical point of view, a number of authors have studied how the SFE
depends on cloud properties and the feedback processes that are included
\citep[see the reviews by][]{padoan14,krumholz14b,dale15,krumholz19}. Much
theoretical work agrees that a higher net SFE is required for the destruction of
massive, high surface density clouds and that the disruption is rapid, occurring
roughly over the gas freefall time
\citep[e.g.,][]{fall10,kim16,raskutti16,geen17,grudic18,kim18,li19,rahner19,fukushima20}.
Simple models of cloud dispersal based on 1D dynamical expansion (driven by
thermal pressure of photoionized gas and/or radiation pressure on dust) and
photoevaporation show that \HII\ regions produced by UV radiation are very
effective in clearing out gas in low- and moderate-$\Sigma$ clouds, requiring
only a few percent of SFE
\citep[e.g.,][]{whitworth79,franco94,williams97,matzner02,krumholz06,murray10,kim16,thompson16,rahner19,inoguchi20}.

In realistic turbulent clouds with multiple sources of radiation, the momentum
injection by feedback is less effective than in simple 1D analytic models due to
the escape of radiation and lack of spherical symmetry
\citep[e.g.,][]{dale17,raskutti17,kim18,kim19}. Still, simulations show that UV
radiation feedback can keep the SFE low and disperse the cloud within
$\lesssim 10\Myr$ in low- and moderate-$\Sigma$ clouds
\citep{raskutti16,grudic18,kim18,he19,fukushima20,gonzalezsamaniego20}, although
massive, high-$\Sigma$ clouds with large escape velocity are less prone to
disruption \citep{dale12,kim18}. While photoevaporation is the dominant
mechanism for cloud dispersal in low- and moderate-$\Sigma$ clouds, radiation
pressure on dust grains becomes more important than photoionization at
$\Sigma\sim 10^3\Sunit$ \citep{kim18}.

Despite the recent progress in the field, several questions still remain. First,
simulations of turbulent, star-forming clouds that use {\it
  magnetohydrodynamics} (rather than hydrodynamics) have been mostly limited to
low-mass clouds with periodic boundary conditions
\citep[e.g.,][]{federrath15,cunningham18}, and have not included effects of
massive star feedback that definitively quenches star formation. A few studies
included magnetic fields of various levels in simulations of GMC dispersal
following star formation \citep[e.g.,][]{grudic18,geen18,zamoraaviles19,he19}
and in simulations with a single, constant-luminosity source put in by hand
\citep{arthur11,geen16}, but a systematic, quantitative study of the effects of
magnetic field strength on GMC dispersal with self-consistent star formation has
been lacking. Second, both theory and simulations indicate that the
gravitational boundedness (quantified via $\avir$) is the primary parameter controlling the SFE
per freefall time, but the lack of correlation between observationally inferred
$\avir$ and $\varepsilon_{\rm ff}$ requires an explanation. Third, even at a
given kinetic energy, the evolution of a cloud will differ depending on the
relative amplitude of different modes in the spectrum of turbulence, and it is
interesting to understand how much this might contribute to observed variances
(in the SFR, SFE, and lifetime). Fourth, the usual virial parameter makes a
number of assumptions about cloud geometry and ignores the contributions from
magnetic fields, and it is important to assess quantitatively how reliable it is
as a magnetized, turbulent cloud evolves and is dispersed under the influence of
feedback. Fifth, many simulations use approximate numerical treatments of
radiation feedback \citep[e.g.,][]{dale12,grudic18,gonzalezsamaniego20} and/or
adopt an isothermal equation of state \citep[e.g.,][]{dale12,kim18}, rather than
implementing explicit radiative transfer with heating/cooling and
ionization/recombination, and it is not known how much these simplified
numerical treatments compromise the conclusions. To make progress on these and
other pressing questions, controlled numerical simulations with high-accuracy
and performance-optimized algorithms for solving the equations of radiation
magnetohydrodynamics (RMHD) are required.

In this work, we carry out a suite of RMHD simulations to study the dynamical
evolution and progress of star formation in a turbulent, magnetized molecular
cloud under the influence of UV radiation feedback. A key aspect of evolution is
the quenching of star formation by the dispersal of the cloud. In addition to
including magnetic fields, we improve upon our previous simulations
\citep{kim18,kim19} by coupling the UV radiative transfer with a simple
thermochemistry module that tracks the non-equilibrium abundances of molecular,
ionized, and atomic hydrogen and includes realistic heating and cooling
processes. We consider a cloud with fixed mass and size typical of moderate-mass GMCs (such as the Orion clouds) in the
Milky Way but vary the initial virial parameter, mass-to-magnetic flux ratio,
and specific realization of the turbulent velocity spectrum. Our simulation
suite has 10 sets of different model parameters, for a total of 50 runs.

We systematically explore how the morphological evolution, SFR, SFE,
photoevaporation, timescales of star formation and cloud destruction, and escape
fraction of radiation depend on the initial virial parameter and
mass-to-magnetic ratio, and quantify variations resulting from the initial
turbulence realization. We also test the reliability of the tradiational virial
parameter by comparing it with the true virial parameter that allows for the
cloud's total energy accounting for full distributions of gas, stars, and
magnetic fields. Finally, we study the relationship between   
$\varepsilon_{\rm ff}$ and $\avir$
measured from our simulations, as compared to previous
theoretical models of SFR and observations of star-forming clouds.

The plan of this paper is as follows. In Section~\ref{s:method}, we describe our 
numerical methods and model parameters. In Section~\ref{s:results}, we present the 
simulation results. We first describe the overall evolution of the fiducial
model (\ref{s:fiducial}) and other models (\ref{s:other}). In
Sections~\ref{s:sfh}--\ref{s:fesc}, we intercompare quantitative simulation outcomes
such as star formation history, SFE, photoevaporation fraction, evolutionary timescales, and radiation escape
fractions. In Section~\ref{s:virial}, we present our analysis of the virial
parameter. Section~\ref{s:epsff} compares our result on the SFE per freefall time with other theoretical and observational work. Finally, in Section~\ref{s:summary}
we summarize our results and discuss their implications.

\section{Methods}\label{s:method}

We carry out RMHD simulations of cloud evolution and destruction by UV radiation
feedback, focusing on the effects of the initial cloud virial parameter and
magnetic fields. Our simulations are performed using the grid-based MHD code
{\it Athena} \citep{stone08}, with additional physics modules for gravity (using
open boundary conditions), sink particles, heating/cooling, photochemistry, and
adaptive ray-tracing of radiation originating in clusters. The simulation setup
is largely similar to that used by \cite{kim17,kim18,kim19}, with the addition
of magnetic fields, UV background radiation (treated using a six-ray shielding
approximation), more detailed photochemical and heating/cooling processes, and
time-dependent luminosity of star clusters. In this section, we present our
basic equations, establish notation, outline our numerical methods, and
summarize simulation parameters.

\subsection{Basic Equations}\label{s:eq}


The set of equations we solve are
\begin{equation}\label{e:continuity}
  \frac{\p \rho}{\p t} + \nabla \cdot \left(\rho \bm{v} \right) = 0 \,,
\end{equation}
\begin{equation}\label{e:momentum}
  \frac{\p (\rho \bm{v})}{\p t} + \nabla \cdot \left[ \rho \bm{v}\bm{v} +
  P^*\mathbb{1} - \frac{\bm{B}\bm{B}}{4\pi} \right] = -
\rho \nabla \Phi + \bm{f}_{\rm rad} \,,
\end{equation}
\begin{equation}\label{e:energy}
  \frac{\p E}{\p t} + \nabla \cdot \left[ ( E + P^*) \bm{v} -
    \dfrac{\bm{B}(\bm{B}\cdot \bm{v})}{4\pi} \right] = \mathcal{G} - \mathcal{L}
  + \bm{v}\cdot (-\rho \nabla \Phi + \bm{f}_{\rm rad})\,,
\end{equation}
\begin{equation}\label{e:scalar}
  \frac{\p n_{\rm s}}{\p t} + \nabla \cdot ( n_{\rm s} \bm{v} ) = \nH\mathcal{C}_{\rm s}\,,
\end{equation}
\begin{equation}\label{e:induction}
  \frac{\p \bm{B}}{\p t} - \nabla \times (\bm{v}\times \bm{B}) = 0\,,
\end{equation}
\begin{equation}\label{e:poisson}
  \nabla^2 \Phi = 4\pi G (\rho + \rho_*)\,.
\end{equation}
Here, 
$\rho$ is gas density, $\bm{v}$ is velocity, $\bm{B}$ is magnetic field,
$P^* = P + B^2/(8\pi)$ is the sum of gas pressure and magnetic pressure,
$\bm{f}_{\rm rad}$ is the radiative force per unit volume, $\Phi$ is the total
gravitational potential due to gas and stars (with density $\rho_*$),
$E = P/(\gamma - 1) + \rho v^2/2 + B^2/(8\pi)$ is the total energy density with
the ratio of specific heats $\gamma = 5/3$, $\mathcal{G}$ and $\mathcal{L}$ are
the volumetric heating and cooling rates. The number density of hydrogen is
denoted as $\nH \equiv \rho/(\mu_{\rm H}m_{\rm H})$, where
$\mu_{\rm H}m_{\rm H} = 2.37 \times 10^{-24}\gram$ is the mean mass of all gas
per hydrogen nucleus.

The quantity $n_{\rm s}= x_{\rm s}\nH$ represents the number density of species
``s,'' where $x_{\rm s}$ is the fractional abundance relative to the hydrogen
nucleus. Each species we follow is passively advected with the velocity field,
with the right-hand side of Equation~\eqref{e:scalar} representing the net
creation rate $\mathcal{C}_{\rm s}$ due to various collisional reactions, cosmic
ray ionization, and photodestruction processes. We explicitly follow the
non-equilibrium abundances of hydrogen in molecular (${\rm H}_2$), atomic
neutral (${\rm H}^0$), and ionized (${\rm H}^+$) phases but assume equilibrium
abundances for carbon- and oxygen-bearing species (${\rm C^0}$, ${\rm C}^+$,
${\rm CO}$, ${\rm O^0}$) (see Section~\ref{s:chem}). Adopting the ideal gas law,
assuming $x_{\rm He} = 0.1$, and ignoring the abundance of trace species, the
gas temperature can be written as $T = P/[(1.1 + \xEL - \xHH) \kB \nH]$.


In addition to Equations~\eqref{e:continuity}--\eqref{e:poisson}, we solve
radiative transfer equations in the form
\begin{equation}\label{e:rt}
  \bm{\hat{n}} \cdot \nabla I = -\kappa I
\end{equation}
where $I$ is the intensity, $\kappa$ is crossection per unit volume in a given
radiation component, and $\bm{\hat{n}}$ is the direction of ray propagation.
The scattering and emission by dust grains are ignored. The radiation field is
decomposed into diffuse background and starlight: $I = I^{\rm bg} + I^{\rm *}$.
The diffuse background is the interstellar radiation field (ISRF) originating from outside the simulation volume. The starlight is the radiation produced by star particles formed in the simulation. 
As detailed in Section~\ref{s:rad},
we use the six-ray approximation to solve for the diffuse background (meaning that only $\bm{\hat n}$ aligned with the cardinal axes of the grid are considered) and use the adaptive ray-tracing technique to solve for the starlight.

\subsection{Star Formation and Radiation Feedback}\label{s:rad}


Star formation is modeled via the sink particle method of \citet{gong13}
(slightly updated as described in \citealt{kimcg20}). A sink particle is created
if a gas cell (1) exceeds the Larson-Penston density criterion for
self-gravitating collapse
($\rho_{\rm crit} = 8.86 c_{\rm s}^2/(\pi G \Delta x^2)$ for the local sound
speed $c_{\rm s} = \sqrt{\gamma P/\rho}$ and the grid spacing $\Delta x$), (2)
is located at a local minimum of the gravitational potential, and (3) has the
velocity field converging along the three Cartesian axes. For actively accreting
sink particles, we reset the density, momentum, and energy of $3^3$ cells
surrounding each sink particle (control volume) by extrapolating from
surrounding non-control volume cells after the MHD update. The accretion rates
of mass and momentum onto sink particles are determined based on the flux across
the surface of the control volume, subtracting out the difference between the
total mass and momentum on the grid inside the control volume at the beginning
and end of the step.


Once sink particles are formed, they are considered as discrete, point sources
of UV radiation. Since individual stars are not resolved in our simulations
(with typical mass of individual sink particles being a few hundred solar
masses), we assume that each sink particle represents a coeval stellar
population following the Kroupa initial mass function (IMF) with mass-weighted
mean age $t_{\rm age}$ \citep[e.g.,][]{kimcg17}. Using the stellar population
synthesis model STARBURST99 \citep{leitherer14}, the time-dependent radiative
output per unit mass $\Psi$ is calculated in three frequency bins: (1)
Photoelectric (PE; $110.8\nm < \lambda \le 206.6\nm$); (2) Lyman-Werner (LW;
$91.2\nm < \lambda \le 110.8\nm$); (3) Lyman Continuum (LyC; $\lambda \le 91.2\nm$).
The PE and LW constitute FUV (or non-ionizing) radiation and are absorbed mainly
by dust. Absorption of the FUV photons by small grains ejects electrons, which
are important for heating neutral gas. The LW photons include photodissociation
bands of ${\rm H}_2$ and ${\rm CO}$ molecules and also ionize ${\rm C}^0$. The
LyC (or ionizing) photons are responsible for ionizing ${\rm H}^0$
($h\nu > 13.6 \eV$) and ${\rm H}_2$ ($h\nu > 15.2\eV$). Based on a STARBURST99 simulation, we find that the mean energy of ionizing photons decreases with the cluster age and ranges between $16\eV < h\nu_{\rm LyC} < 20\eV$ for $t_{\rm age}<10\Myr$. 
For simplicity, we adopt a constant value $h\nu_{\rm LyC} = 18\eV$. In Appendix~\ref{s:psi}, we show $\Psi(t_{\rm age})$ in different frequency bins.


We calculate the cross sections for dust absorption and photoionization averaged
over the cluster's UV spectrum and find that the dependence on $t_{\rm age}$ is
weak. We thus take constant cross sections
$\kappa_{\rm PE} = \nH \sigma_{\rm d,PE}$ and
$\kappa_{\rm LW} = \nH \sigma_{\rm d,LW}$, where
$\sigma_{\rm d,PE} = 10^{-21}\cm^2\,{\rm H}^{-1}$ and
$\sigma_{\rm d,LW} = 1.5 \times 10^{-21}\cm^2\,{\rm H}^{-1}$. These values are
appropriate for \citet{weingartner01a}'s grain model with $R_V = 3.1$. For LyC
radiation, we take
$\kappa_{\rm LyC} = \nH \sigma_{\rm d,LyC}+ \nHI\sigma_{\rm pi,H^0} +
\nHH\sigma_{\rm pi,H_2}$, where $\sigma_{\rm d,LyC}= 10^{-21}\,{\rm cm}^{2}$,
$\sigma_{\rm pi,H^0} = 3 \times 10^{-18}\,{\rm cm}^{2}$, and
$\sigma_{\rm pi,H_2} = 6 \times 10^{-18}\,{\rm cm}^{2}$.

We note that the dust absorption cross section for LyC radiation ($\sigma_{\rm d,LyC}$) is uncertain;
in \HII\ regions small carbonaceous grains (and PAHs) can be destroyed by an
intense radiation field \citep[e.g.,][]{deharveng10,binder18,chastenet19}, while large
grains can be disrupted by radiative torque and disintegrate into smaller grains
\citep{hoang19} or swept out by radiation pressure
\citep{draine11b,akimkin15,akimkin17}. Our adopted value would be intermediate
between the case of the complete destruction and the case of no destruction
\citep[see also][]{glatzle19}. \citet{kim19} found that using a lower value of
$\sigma_{\rm d,LyC}$ does not affect the overall cloud evolution and star
formation, although the escape fraction of ionizing radiation can be boosted
significantly.


We employ the adaptive ray-tracing module of \citet{kim17} to model the
propagation of radiation from multiple point sources. The reader is referred to
\citet{kim17} for more detailed description of the ray tracing algorithm and
test results. Here we give a brief summary of the method. We inject photon
packets onto the grid at the position of each source and transport them along
radial rays, whose directions are determined by the HEALPix scheme
\citep{gorski05}. As they propagate out, the photon packets are split into
sub-rays to ensure that each cell is intersected by at least four rays per
source. The optical depths between a ray's consecutive intersection points (at
grid cell faces) are used to calculate the contribution from stellar sources to
the volume-averaged radiation energy density ($\mathcal{E}$) and flux density
($\bm{F}$) of each cell in each radiation band.

For PE and LW, in addition to radiation from embedded sources we also include
diffuse radiation originating outside of the cloud. We use the six-ray
approximation \citep[e.g.,][]{glover07,gong18} to calculate the shielding of the
FUV background. The volume averaged mean intensity in a given cell is computed
as
\begin{equation}
  J^{\rm bg} = \frac{I^{\rm bg,0}}{6}
  \sum_k \frac{e^{-\tau_k}(1 - e^{-\Delta \tau})}{\Delta \tau} \,,
\end{equation}
where the index $k$ runs over the six faces of the computational domain,
$\tau_k$ is the dust optical depth integrated from the outer boundary to the cell
face along the Cartesian axis, and $\Delta \tau = \nH \sigma_{\rm d} \Delta x$
is the cell optical depth (in the respective energy band, PE or LW). We impose
boundary conditions
$I^{\rm bg,0}_{\rm PE} = J^{\rm (ISRF)}_{\rm PE} = 1.8 \times 10^{-4}
\cm^{-2}\second^{-1}\,{\rm sr}^{-1}$ and
$I^{\rm bg,0}_{\rm LW} = J^{\rm (ISRF)}_{\rm LW} = 3.0 \times 10^{-5}\,\erg
\cm^{-2}\second^{-1}\,{\rm sr}^{-1}$, where $J^{\rm (ISRF)}$ is
\citet{draine78}'s estimate of radiation intensity in the solar neighborhood. We
assume $J^{\rm (ISRF)}_{\rm LyC}=0$. We also neglect the (small) contribution to
the flux due to the ISRF.

At any location, the angle-averaged intensity $J$ in each band is the sum of
$c{\cal E}/(4\pi)$ with ${\cal E}$ computed from the adaptive ray tracing
, plus the corresponding
$J^{\rm (bg)}$ (if any). For notational convenience, we denote the normalized
angle-averaged
intensity in the FUV bands with $\chi$:
$\chi_{\rm PE} \equiv J_{\rm PE}/J^{\rm (ISRF)}_{\rm PE}$,
$\chi_{\rm LW} \equiv J_{\rm LW}/J^{\rm (ISRF)}_{\rm PE}$, and
$\chi_{\rm FUV} = (J_{\rm PE} + J_{\rm LW})/(J^{\rm (ISRF)}_{\rm PE} + J^{\rm
  (ISRF)}_{\rm LW})$.

The radiative force on the gas in a given cell is computed using a weighted sum
of fluxes calculated from the adaptive ray tracing,
\begin{equation}
  \bm{f}_{\rm rad} = \dfrac{1}{c} \left( \kappa_{\rm LyC}\bm{F}_{\rm LyC} +
    \kappa_{\rm LW}\bm{F}_{\rm LW} +  1.25 \kappa_{\rm
      PE}\bm{F}_{\rm PE} \right) \,.\label{e:frad}
\end{equation}
In Equation~\eqref{e:frad}, the factor 1.25 for the PE band is an approximate
treatment for the additional force that would be exerted by absorption of
optical photons (not followed) by dust grains, given the ratio of optical to FUV
in the composite spectrum of young clusters (Kim, J.-G. et al. 2021 in preparation).

The LyC energy density is used to calculate the photoionization rate per H atom
\begin{equation}
  \zeta_{\rm pi} = \sigma_{\rm pi} c\mathcal{E}_{\rm LyC}/(h\nu_{\rm
    LyC})\,,
\end{equation}
where $c$ is the speed of light and $\sigma_{\rm pi}$ is the photoionization
cross section.


To model the dissociation of ${\rm H}_2$ by LW band photons, it is important to
account for the effects of both dust shielding and self-shielding by absorption
lines. We employ the self-shielding function of \citet{draine96} assuming a
constant Doppler broadening parameter $b = 3\kms$. This value is the typical
one-dimensional velocity dispersion of our simulated clouds, but we note that
the shielding factor is insensitive to the choice of $b$ at
$N_{\rm H_2} \gtrsim 10^{17} \cm^{-2}$ because absorption occurs mainly in
Lorentzian damping wings \citep{sternberg14}. The evaluation of shielding factor
requires that the adaptive ray-tracing follow the column density of molecular
hydrogen from each point source to every cell in the computational domain (and
similarly for the six-ray approximation). The ${\rm H}_2$ photodissociation rate
is calculated as
\begin{equation}
  \zeta_{\rm pd,H_2} = \zeta_{\rm pd,H_2}^{\rm (ISRF)} \chi_{\rm LW} f_{\rm
    shld,H_2,eff}\,,\label{e:H2pd}
\end{equation}
where $\zeta_{\rm pd,H_2}^{\rm (ISRF)}= 5.7\times 10^{-11}\second^{-1}$ is the
dissociation rate for unshielded gas exposed to the ISRF \citep{heays17}. The
effective self-shielding factor $f_{\rm shld,H_2,eff}$ in
Equation~\eqref{e:H2pd} is the shielding factor averaged over individual point
sources and background radiation incident from six boundary faces of the
computational domain, weighted by dust-attenuated (continuum) radiation energy
density in the LW frequency bin.

Because the threshold wavelengths for ${\rm C}^0$ ionization and ${\rm CO}$
dissociation are about the same as that of the ${\rm H}_2$ photodissociation
band ($\sim 110 \nm$) \citep[e.g,][]{heays17}, we use the LW radiation
$\chi_{\rm LW}$ to calculate the ionization rate of ${\rm C}^0$ and the
equilibrium abundance of {\rm CO}. We calculate the ${\rm C}^0$-ionizing
radiation field accounting for the dust-shielding, self-shielding, and cross-shielding by ${\rm H}_2$, following the process described in \citet{gong17}. For the {\rm CO} dissociating radiation field, we consider only the dust shielding and ignore the
{\rm CO} self-shielding and shielding by ${\rm H_2}$. The FUV energy density
($\chi_{\rm FUV}$) is used for evaluating the photoelectric heating rate.

\subsection{Thermochemistry}\label{s:chem}

The chemical reaction rates and heating/cooling rates have complex dependence on
the local gas density, temperature, radiation field, species abundance, dust
abundance, and cosmic ray ionization rate. Here we only briefly summarize the
physical processes that we include. A full description of our heating/cooling
module and test results will be presented in a forthcoming paper (J.-G. Kim et
al. 2021, in preparation).

We solve the non-equilibrium evolution for molecular/atomic/ionized hydrogen,
while adopting the equilibrium abundances for carbon- and oxygen-bearing
species. We adopt 
gas-phase abundances of C ($x_{\rm C,tot} = 1.6\times 10^{-4}$)
and O ($x_{\rm O,tot} = 3.2\times 10^{-4}$)
at solar metallicity and
rate coefficients 
used by the chemistry network in
\citet{gong17}. For molecular hydrogen, we include the formation on grain
surfaces and destruction by cosmic ray ionization, photodissociation, and
photoionization.\footnote{We do not include the destruction by collisional
  dissociation that can be important in high-velocity shocks
  \citep[e.g.,][]{hollenbach79}.} For ionized hydrogen, we consider the
formation by photoionization, cosmic ray ionization, and collisional ionization;
and the destruction by radiative and grain-assisted recombination.

The equilibrium abundance of ${\rm C}^+$ is determined by balancing radiative,
dielectronic, and grain-assisted recombination with photoionization and
collisional ionization of ${\rm C}^0$. The abundance of ${\rm CO}$ is
determined by making use of
the critical density for the ${\rm C^0}$-to-${\rm CO}$
transition that depends on (shielded) $\chi_{\rm LW}$ found by \citet{gong17}.
The abundance of ${\rm C}^0$ is determined from the closure $x_{\rm C^0} = x_{\rm C,tot} - x_{\rm C^+} - x_{\rm CO}$. Similarly, the abundance of neutral hydrogen is determined from the closure
$\xHI = 1 - 2\xHH - \xHII$. The abundance of free electrons is set to
$\xEL = \xHII + \xCII$. For the primary cosmic ray ionization rate, we adopt the
canonical value $\xi_{\rm cr} = 2\times 10^{-16}\second^{-1}$ found in diffuse
molecular clouds \citep[e.g.,][]{indriolo07,neufeld17}.

The volumetric heating rate $\mathcal{G}$ is taken as the sum of photoelectric
heating, cosmic ray heating, ${\rm H}_2$ heating, and photoionization heating.
The photoelectic emission from dust by FUV radiation is the dominant heat source
in diffuse atomic gas. The heating rate is proportional to $\chi_{\rm FUV}$ and
is calculated using the functional form provided by \citet{weingartner01b},
allowing for a heating efficiency that depends on gas temperature and charging
of grains. For the heating by cosmic ray ionization (which is important to
FUV-shielded gas), we adopt fitting formulae suggested by \citet{draine11} for
atomic regions and \citet{krumholz14} for molecular regions (see Eqs.(30)--(32)
in \citet{gong17}). We also include the heating due to ${\rm H}_2$ formation on
dust grains, photodissociation, and UV pumping, following prescriptions given in
\citet{hollenbach79}. We assume that the photoionization of ${\rm H}^0$
(${\rm H_2}$) deposits an excess energy of $4.4\eV$ ($2.8\eV$) per event.

We include cooling by collisionally excited atomic fine-structure levels in
${\rm C^+}$, ${\rm O}^0$, ${\rm C}^0$; Ly$\alpha$ emission, and recombination of
electrons on small grains \citep{weingartner01b}. We also include cooling by
rotational transitions of ${\rm CO}$ with the large velocity gradient
approximation, which becomes the dominant coolant in dense molecular gas. For
photoionized gas, we use a temperature- and  density-dependent fitting formula that
accounts for the cooling by free-free emission, recombination radiation, and
cooling by collisionally excited emission lines of heavy elements (Kim, J.-G. et al. 2021, in preparation).

\subsection{Numerical Integration}\label{s:integration}

We advance the equations of ideal MHD in time using the Roe Riemann solver,
piecewise-linear reconstruction, and a predictor-corrector type time-integrator
\citep{stone09} combined with the constrained transport method
\citep{gardiner08} that enforces the divergence-free constraint on the magnetic
field. We apply diode-like boundary conditions to boundary faces of both the
computational domain and the control volumes of sink particles. The Poisson
equation is solved via the fast Fourier transform method with open (vacuum)
boundary conditions \citep{skinner15}, allowing for the contribution of star
particles by using the triangular-shaped cloud method to map point masses to
density ($\rho_*$) on the computational grid \citep{gong13}.

After the updates of the MHD equations, gravity, and sink particles, we perform
adaptive ray tracing and six-ray transfer calculations. We then perform the
operator-split, explicit update of source terms due to chemical reactions,
heating/cooling, and the radiative force. We take substeps in updating the
chemical abundances and thermal energy, with the time-step size of each substep
chosen as the minimum of 10\% of the cooling time
($0.1P/[(\gamma-1)|\mathcal{G} - \mathcal{L}|]$) and 10\% of the chemical time
($0.1/|\mathcal{C}_{\rm s}|$). In contrast to the previous studies in which the
radiative transfer is subcycled alternatingly with the abundance update
\citep{kim17}, we perform ray tracing once per MHD update. This approach cannot
accurately follow the early evolution of R-type ionization fronts if the front
propagation speed is much greater than the maximum signal speed in the
computational domain. However, we found that it has little impact on modeling
the dynamical expansion of \HII\ regions (Kim, J.-G. et al. 2021 in preparation;
see also Figure~6 in \citealt{kim17}).

\subsection{Model Initialization and Parameters}\label{s:model}

\begin{deluxetable}{ccccccc}
\tablewidth{0pt}\tablecaption{Model parameters}
\tablehead{
\colhead{Model} &
\colhead{$\alpha_{\rm vir,0}$} &
\colhead{$\mu_{\Phi,0}$} &
\colhead{$\sigma_{\rm 1d,0}$} &
\colhead{$B_{z,0}$} &
\colhead{$\mathcal{M}_{0}$} &
\colhead{$\mathcal{M}_{A,0}$} \\
\colhead{} &
\colhead{} &
\colhead{} &
\colhead{[${\rm km}\,{\rm s}^{-1}$]} &
\colhead{[$\mu{\rm G}$]} &
\colhead{} &
\colhead{} \\
\colhead{(1)} &
\colhead{(2)} &
\colhead{(3)} &
\colhead{(4)} &
\colhead{(5)} &
\colhead{(6)} &
\colhead{(7)}}
\startdata
$\alpha$-series \\
\tableline
{\tt A1B2} & $1.0$ & $2.0$ & $2.1$ & $13.5$ & $13.6$ & $1.3$ \\
{\tt A2B2} & $2.0$ & $2.0$ & $2.9$ & $13.5$ & $19.2$ & $1.9$ \\
{\tt A3B2} & $3.0$ & $2.0$ & $3.6$ & $13.5$ & $23.5$ & $2.3$ \\
{\tt A4B2} & $4.0$ & $2.0$ & $4.1$ & $13.5$ & $27.2$ & $2.7$ \\
{\tt A5B2} & $5.0$ & $2.0$ & $4.6$ & $13.5$ & $30.4$ & $3.0$ \\
\tableline
$\beta$-series \\
\tableline
{\tt A2B05} & $2.0$ & $0.5$ & $2.9$ & $54.0$ & $19.2$ & $0.5$ \\
{\tt A2B1} & $2.0$ & $1.0$ & $2.9$ & $27.0$ & $19.2$ & $0.9$ \\
{\tt A2B2} & $2.0$ & $2.0$ & $2.9$ & $13.5$ & $19.2$ & $1.9$ \\
{\tt A2B4} & $2.0$ & $4.0$ & $2.9$ & $6.7$ & $19.2$ & $3.8$ \\
{\tt A2B8} & $2.0$ & $8.0$ & $2.9$ & $3.4$ & $19.2$ & $7.6$ \\
{\tt A2Binf} & $2.0$ & $\infty$ & $2.9$ & $0.0$ & $19.2$ & $\infty$ \\
\enddata
\tablecomments{Each model is run with 5 different random seeds for turbulence.
All models have the same initial mass $\Mcl = 10^5 \Msun$ and radius $\Rcl = 20
\pc$. Columns are as follows: (1) model name indicating the initial virial parameter ({\tt A})
and magnetic flux-to-mass ratio ({\tt B}). (2) initial virial parameter
$\avircl = 5 \sigmacl^2\Rcl/(G\Mcl)$. 
(4) initial magnetic mass-to-flux ratio $\muBcl = 2\pi\sqrt{G}
\Sigmacl/|B_{z,0}|$. (3) initial velocity dispersion. (5) initial magnetic field strength.
(6) 3D sonic Mach number of initial turbulence.
(7) 3D Alfv\'{e}n Mach number of initial turbulence.}\label{t:param}
\end{deluxetable}

\begin{figure}[t!]
  \begin{center}
    \includegraphics[width=\linewidth]{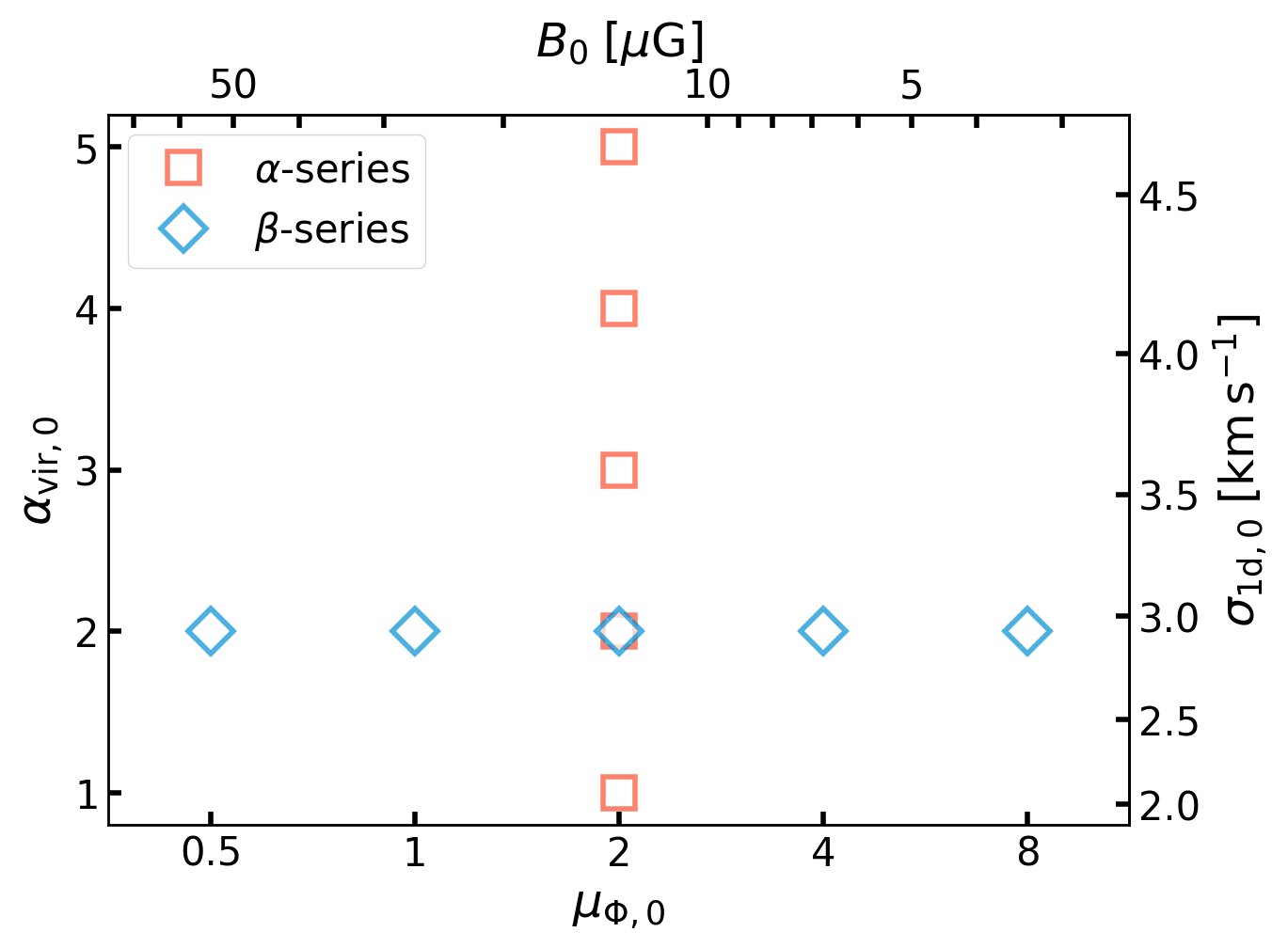}
    \caption{Parameters of simulated clouds in $\alpha$- and $\beta$-series
      models (see also Table~\ref{t:param}). All clouds have the same mass
      $\Mcl=10^5 \Msun$ and radius $\Rcl=20\pc$. The bottom $x$-axis is the
      mass-to-magnetic flux ratio $\muBcl = 2\pi\sqrt{G}\Sigmacl/B_{z,0}$ and
      the left $y$-axis is the kinetic virial parameter
      $\avircl = 5\Rcl \sigmacl^2/(G\Mcl)$. The top $x$-axis and right $y$-axis
      show the corresponding magnetic field strength $B_{z,0}$ and
      one-dimensional turbulent velocity dispersion $\sigmacl$, respectively.
      For each pair of ($\avircl$, $\muBcl$), we run five simulations with
      different random seeds for the initial turbulent velocity field.}\label{f:param}
  \end{center}
\end{figure}

We initialize the simulation by placing a uniform density\footnote{A recent numerical study by \citet{chen20} that is similar to our own shows that the overall cloud evolution and star formation properties are not sensitive to the choice of initial gas distribution.} cloud with mass
$\Mcl = 10^5 \Msun$ and radius $\Rcl = 20 \pc$ at the center of the
computational domain, which has a side length $L_{\rm box} = 4\Rcl = 80\pc$. The
number of grid cells is set to $256^3$ corresponding to the uniform grid spacing
of $\Delta x = 0.31\pc$. The cloud has the initial gas surface density
$\Sigmacl \equiv \Mcl/(\pi \Rcl^2) = 79.6 \Sunit$, hydrogen number density
$\nHcl=3\Mcl/(4\pi \Rcl^3 \mu_{\rm H}m_{\rm H}) = 86.2 \pcc$, and freefall
time 
\begin{equation}\label{e:tff0}
  \tffcl= \left(\frac{3\pi}{32 G\rho_0}\right)^{1/2} = 4.68 \Myr \,.
\end{equation}
The gas temperature is set to $24 \Kel$ and the ${\rm H_2}$ abundance to the
equilibrium value $x_{\rm H_2} = 0.43$, appropriate for UV-shielded gas with
$\xi_{\rm cr} = 2.0 \times 10^{-16}\second^{-1}$ (see Equations (15)--(18) in
\citealt{gong18}). The ambient medium is a warm neutral medium with
$\nH = 10^{-2} \nHcl$ and $T = 8200 \Kel$ and has a total mass of $0.16\Mcl$. We
note that we have tested the effects of the initial temperature choice, and
found that they are unimportant as thermal and ionization equilibrium are
rapidly achieved.

We assign the gas that is initially molecular a passive scalar field value
$s_{\rm cl}=1$ and the ambient atomic gas with $s_{\rm cl}=0$. We use
$s_{\rm cl}$ to differentiate the ``cloud material'' from the ambient medium,
but it should be noted that the use of a passive scalar cannot 
perfectly distinguish the original cloud material
due to numerical diffusion.
For analysis, we select gas that is neutral and initially molecular by using a
filter function
\begin{equation}\label{e:Theta}
    \Theta =
    \begin{cases}
      1, & \text{if $x_{\rm H^+} < 0.5$ and $s_{\rm cl} > 10^{-2}$}\\
      0, & \text{otherwise.}
    \end{cases}
\end{equation}

We initiate all clouds (except the non-magnetized model) with a uniform magnetic
field aligned along the $z$-axis. We note that a uniform magnetic field exerts
no Lorentz force and does not play a role in supporting the cloud at $t=0$ (initial clouds are not equilibria in any case), but
the nonuniform magnetic fields created by the turbulence provide support against
gravity from the early stages of evolution. The initial magnetic field strength
is measured by the dimensionless mass-to-magnetic flux ratio
\begin{equation}\label{eq:muPhidef}
\muBcl = \frac{2\pi\sqrt{G} \Sigmacl}{B_{z,0}}    
\end{equation}
which is defined such that global gravitational collapse would be suppressed in
subcritical clouds with $\muBcl \lesssim 1$, i.e., where the mass-to-flux ratio
exceeds the critical value\footnote{The numerical factor $1/(2\pi)\simeq 0.16$ is exact for an infinite cold sheet \citep{nakano78}, but a spherical cold cloud has a similar value of 0.17 \citep{tomisaka88}.} $(2 \pi \sqrt{G})^{-1}$. The parameter $\muBcl^2$ is equal to 
$10/9$ of the ratio
between the initial gravitational energy and magnetic energy of the cloud.

All simulations are initialized with a Gaussian-random turbulent velocity field
with a power spectrum $|\delta \bm{v}_k|^2 \propto k^{-4}$ for
$2 \le 2\pi/L_{\rm box} \le 64$. The corresponding structure function has
velocity dispersion increasing with scale as $\sigma(\ell)\propto \ell^{1/2}$,
consistent with observations of GMCs \citep{HeyerBrunt04}. We use the kinetic
virial parameter
\begin{equation}\label{eq:avirdef}
  \avircl = \frac{5\sigmacl^2 \Rcl}{G\Mcl}
\end{equation} 
for an isolated, uniform, spherical cloud \citep{bertoldi92} to quantify the
level of initial turbulent support, where $\sigmacl$ is the mass-weighted
one-dimensional velocity dispersion averaged over the whole cloud. The parameter
$\avircl$ is equal to twice the initial kinetic energy divided by the initial
gravitational energy of the cloud. We note that even with a given value of
$\sigmacl$ and given power spectrum, the velocity amplitudes of individual modes
differ (and therefore the cloud evolution differs) depending on the random seeds
selected.

For $\Mcl = 10^5\Msun$, $\Rcl = 20\pc$ and isothermal sound speed
$c_{\rm s,iso} = 0.26\kms$, the initial Alfv\'{e}n speed in the cloud can be
written in terms of $\muBcl$ as
$v_{\rm A}= B_{z,0}/\sqrt{4\pi \rho_0} = 5.3 \muBcl^{-1} \kms$. The 3D sonic and
Alfv\'{e}n Mach numbers of the initial turbulence are then
$\mathcal{M}_{\rm 0} = \sigma_{\rm 3d,0}/c_{\rm s,iso} = 13.6 \avircl^{1/2}$,
$\mathcal{M}_{\rm A,0} = \sigma_{\rm 3d,0}/v_{\rm A} = 0.67
\avircl^{1/2}\muBcl$, respectively. The plasma beta parameter in the cloud is
$\beta_0 = P/P_{\rm mag}= 2 c_{\rm s,iso}^2/v_{\rm A}^2 = 4.9\times 10^{-3}
\muBcl^2$.

To investigate the effects of the initial virial parameter and the magnetic
field strength, we consider $\alpha$- and $\beta$-series models, in which either
the initial kinetic or magnetic energy is held constant, while the other
changes. This is equivalent to holding either $\avircl$ (or $\sigmacl$) fixed or
holding $\muBcl$ (or $B_{z,0}$ or $\beta$) fixed, as depicted in
Figure~\ref{f:param}. In the $\alpha$-series, we vary $\avircl$ from 1 to 5,
while the initial magnetic field is set to $B_{z,0} = 13.5 \mu {\rm G}$,
corresponding to $\beta_0 = 0.02$ and $\muBcl = 2$. In the $\beta$-series, we
vary $\muB$ from 0.5 to 8 (by factors of two), while holding $\avircl = 2$
(corresponding to $\sigmacl = 2.93 \kms$). We also consider the non-magnetized
case with $\muBcl=\infty$. All model parameters are listed in
Table~\ref{t:param}. Our chosen parameter range for $\avircl$ encompasses the
observational estimates of average $\avir$ for molecular clouds in the Milky Way
and nearby galaxies \citep[e.g.,][]{mivilledeschenes17,sun18,sun20}. Except for
the subcritical case $\muBcl=0.5$ (which we include for theoretical
completeness), the range of magnetic field strength is also consistent with the
range of observed values in inter-core regions of nearby molecular clouds (a few
$\mu {\rm G}$--$30 \mu {\rm G}$) inferred from OH Zeeman observations
\citep{thompson19}.

For each model we run five simulations with different random seeds for the
initial turbulent velocity field. This allows us to study the variations of star
formation history and cloud destruction outcomes that result from different
turbulence realizations.

The name of each simulation is designated as {\tt AaBbSs}, where {\tt a}, {\tt
  b}, and {\tt s} respectively denote $\avircl$, $\muBcl$, and the label of the
set of random seeds used for initializing turbulence ({\tt seed}). We choose the
run with $\avircl=2$, $\muBcl=2$, and ${\tt seed}=4$ as the fiducial case, for
which we run additional simulations at higher ($N_{\rm cell} = 512^3$) and lower
($N_{\rm cell} = 128^3$) resolutions to check the numerical convergence.
We present the results of our  convergence test in Section~\ref{s:other} and in Table~\ref{t:result}.

\section{Evolution and Parameter Dependence}\label{s:results}

\subsection{Evolution of the Fiducial Model}\label{s:fiducial}

\begin{figure*}[t!]
  \includegraphics[width=\linewidth]{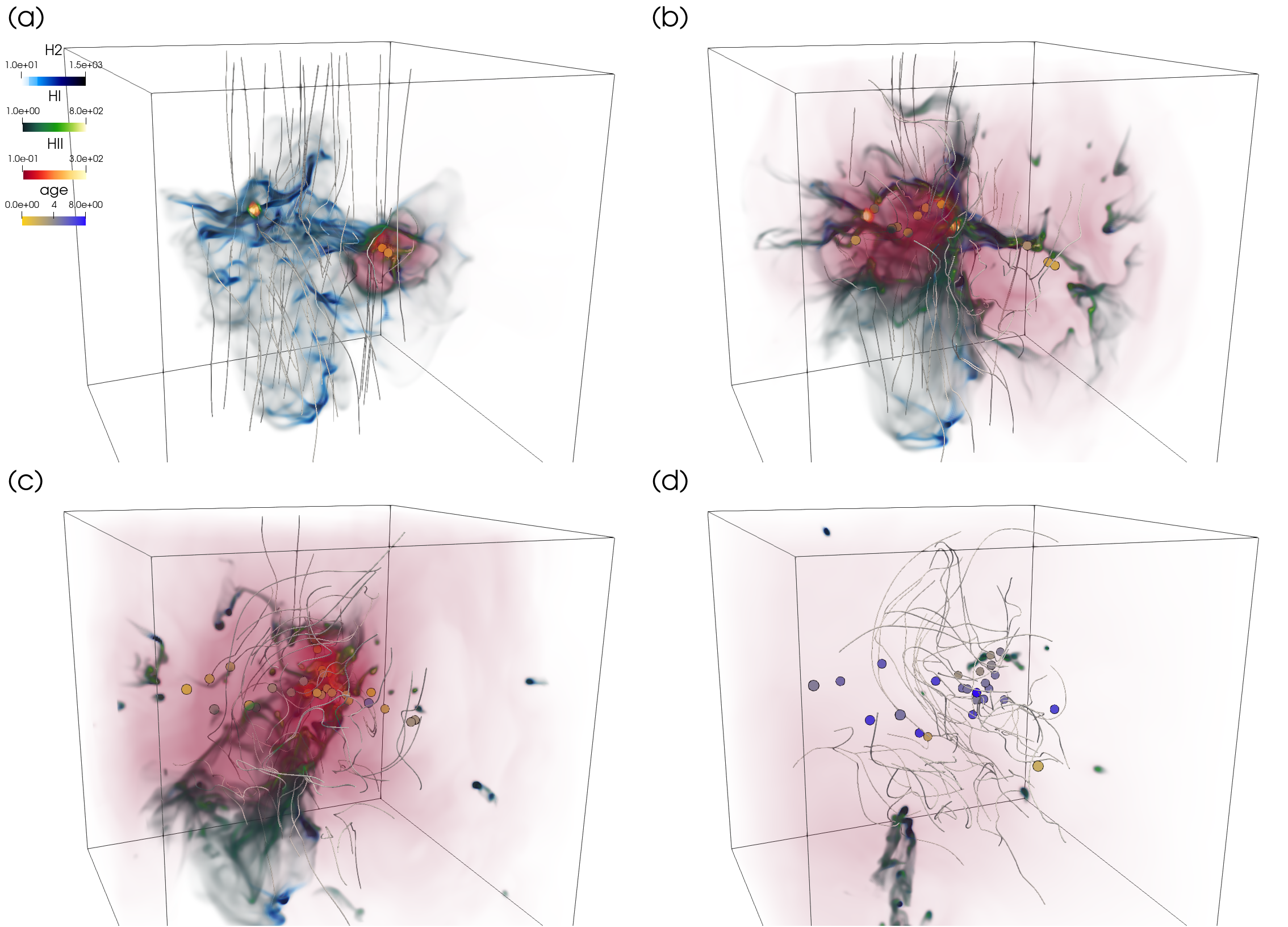}
  \caption{Volume rendering of molecular (blue-white), atomic (green-white), and
    ionized (red-orange) gas for the fiducial model {\tt A2B2S4} ($\avircl=2$,
    $\muBcl=2$, ${\tt seed}=4$) at resolution of $256^3$. Panels show snapshots
    at (a) $t= t_{*,0} + 1 \Myr$, (b) $t_{*,0}+3\Myr$, (c) $t_{*,0}+5\Myr$, and
    (d) $t_{*,0}+8\Myr$, where $t_{*,0} = 3.3\Myr$ is the time of first star
    formation. Magnetic fields are shown as lines (colored by magnitude) and
    star particles are shown as spheres (colored by $t_{\rm age}$). [An animation of this figure for the high-resolution model ({\tt A2B2S4\_N512}) is available online.]
  }\label{f:volren}
\end{figure*}

We first sketch out the evolution of the fiducial model {\tt A2B2S4} with
$(\avircl, \muBcl)=(2,2)$. Figure~\ref{f:volren} shows the volume rendering of
molecular (blue-white), atomic (green-white), and ionized (red-orange) gas of the fiducial model at
times $\tprime \equiv t - t_{*,0} = 1 \Myr$, $3\Myr$, $5\Myr$, and $8\Myr$,
where $t_{*,0}= 3.3\Myr$ denotes the time at which the first star formation
occurs. Magnetic field lines are visualized by lines (colored by magnitude) and
star particles by spheres (colored by $t_{\rm age}$). Projections of density for
snapshots at $\tprime = -0.2, 2, 5, 8\Myr$ of this model are also seen in the
comparison figures of Appendix~\ref{s:snapshot}.


The overall evolutionary sequence is similar to that of the hydrodynamic
simulations presented in \citet{kim18,kim19}. The compression of gas by
supersonic turbulence creates a network of dense filaments that have some
preference for alignment perpendicular to the direction of the large-scale
magnetic field. It also gives rise to a density probability density function
(PDF) that is approximately log-normal in shape and develops a high-density tail
as time progresses. At $t = 3.3 \Myr \equiv t_{*,0}$, the densest part of a
filament collapses and forms the first sink particle.

Ionizing radiation from the first sink particle creates a compact and confined
\HII\ region, and the pressure of photoionized gas and (subdominant) radiation
pressure force drive its dynamical expansion, suppressing further gas accretion.
Once the \HII\ region breaks out of the local star-forming clump, ionized gas
quickly fills the bulk of the computational volume, and more and more UV photons
escape the computational domain through optically-thin sight lines.
Pre-existing dense structures are carved into pillars and cometary globules and rocket away from ionizing sources due to anisotropic ablation, as seen in other simulations of expanding \HII\ regions in turbulent clouds \citep[e.g.,][]{mellema06,gritschneder10,arthur11,walch12,ali18}.
Until
$\tprime \approx 5\Myr$, stellar mass continues to increase, multiple
subclusters form, and \HII\ regions merge with each other.

A substantial fraction of molecular gas at the periphery of the \HII\ region
turns into the atomic phase by dissociating radiation and then into ionized
phase by ionizing radiation. 
Only 5\% of the initial cloud mass remains
molecular in the swept-up gas at $\tprime = 8\Myr$. 

\begin{figure*}[t!]
  \epsscale{1.0}\plotone{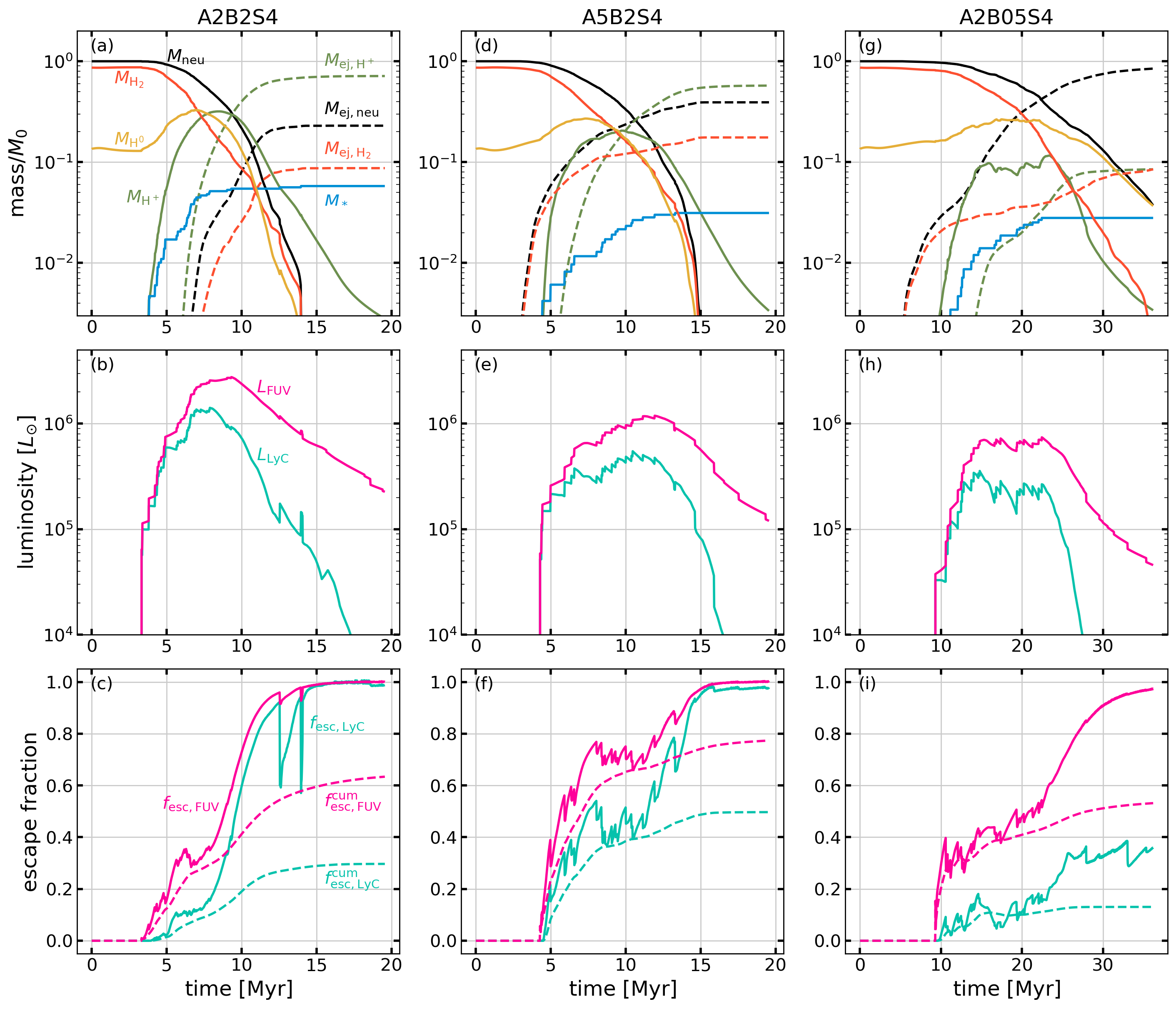}
  \caption{Time evolution of global quantities in three different models: (left)
    the fiducial model {\tt A2B2S4} with $(\avircl, \muBcl) = (2,2)$; (middle) 
    an initially
    unbound cloud {\tt A5B2S4} with $(\avircl, \muBcl) = (5,2)$; (right)
    a magnetically subcritical cloud {\tt A2B05S4} with
    $(\avircl, \muBcl) = (2,0.5)$. The three clouds have identical initial
    turbulent velocity field realizations (${\tt seed=4}$), but with different
    normalization for $\avircl=2, 5$. ({\it Top}) Stellar mass ($M_*$), cloud
    gas mass in molecular ($M_{\rm H_2}$), atomic ($M_{\rm H^0}$), and ionized
    ($M_{\rm H^+}$) phases. The dashed lines show the cumulative gas mass that
    has left the simulation domain as neutrals ($M_{\rm ej,neu}=M_{\rm ej,H_2}+M_{\rm ej,H^0}$),
    ions ($M_{\rm ej,H^+}$), and molecules ($M_{\rm ej,H_2}$). ({\it Middle}) Luminosities of ionizing
    (LyC; cyan) and non-ionizing (FUV; magenta) photons. ({\it Bottom})
    Instantaneous (solid) and cumulative (dashed) escape fractions of ionizing
    and non-ionizing photons.}\label{f:hst}
\end{figure*}

\begin{figure*}[t!]
  \begin{center}
    \includegraphics[width=\linewidth]{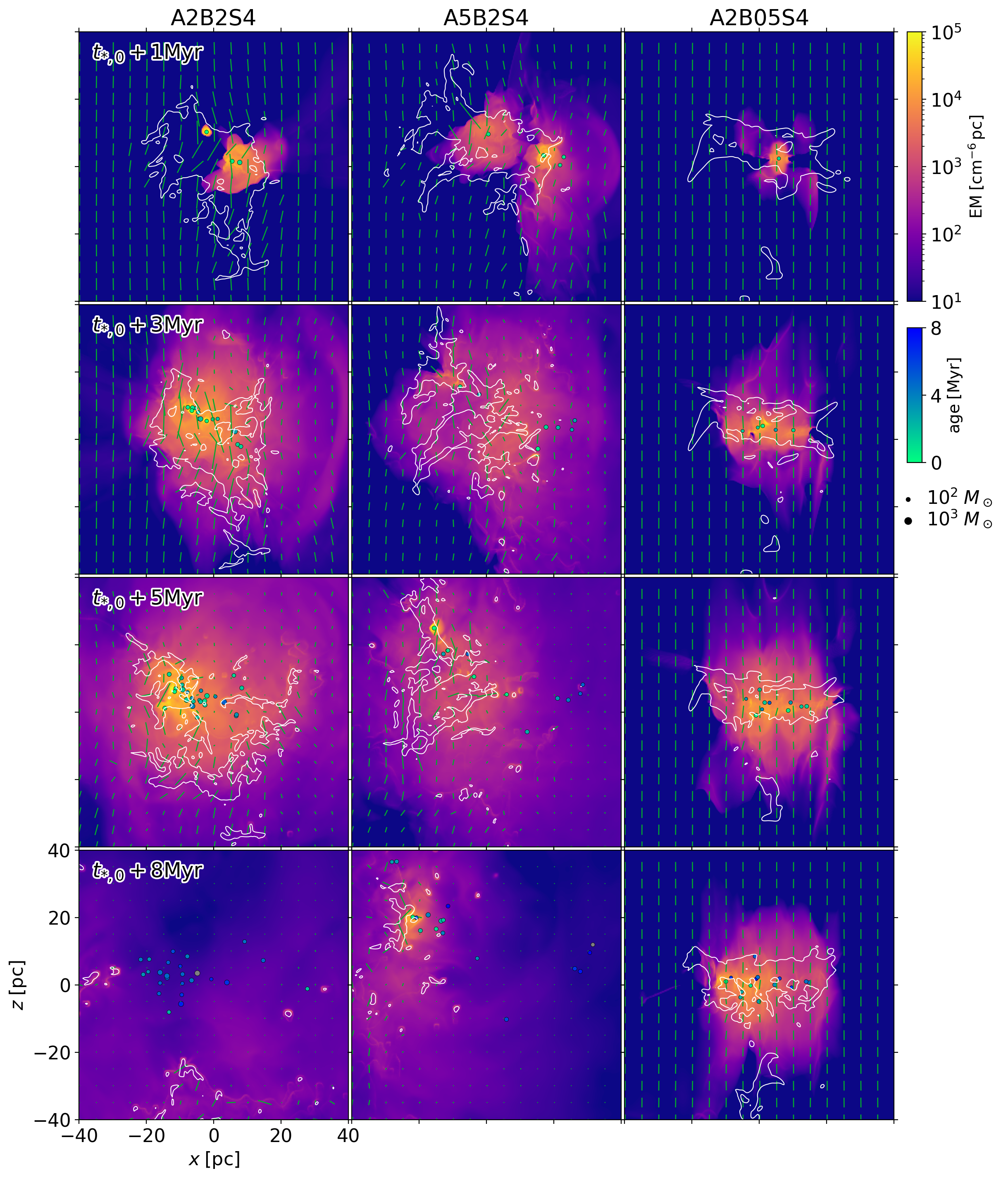}
    \caption{Emission measure along the $y$-axis of three different models ({\tt
        A2B2S4}, {\tt A5B2S4}, and {\tt A2B05S4}) at times (from top to bottom)
      $\tprime = 1$, $3$, $5$, $8 \Myr$ after the onset of star formation. 
      The projected positions of star particles are shown as circles.
      Green line segments indicate the polarization vectors of dust thermal emission
      rotated by $90^{\circ}$, indicating the direction of magnetic field. The
      length of the segment is proportional to the magnitude and is normalized
      by the initial magnetic field $B_{z,0}$. White contours show the loci
      where $A_V = 1$, corresponding to column density of neutral (atomic +
      molecular) hydrogen gas $1.87 \times 10^{21}\cm^{-2}$ ($21.4 \Sunit$).}
      \label{f:EM}
  \end{center}
\end{figure*}
\begin{figure*}[t!]
  \begin{center}
    \includegraphics[width=\linewidth]{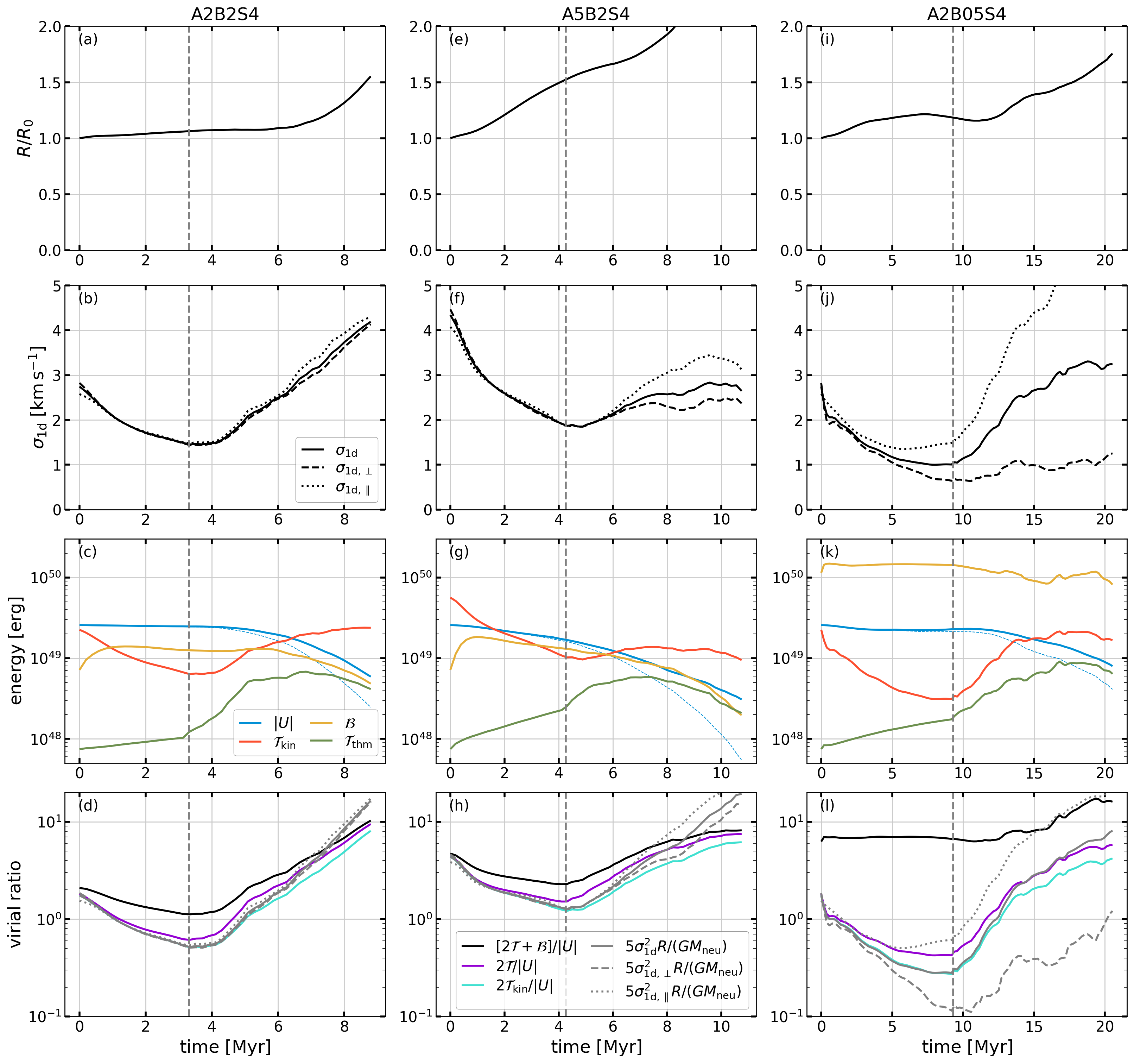}
    \caption{Time evolution of cloud size, velocity dispersion, energy, and
      virial ratios for three different models {\tt A2B2S4}, {\tt A5B2S4}, and
      {\tt A2B05S4} until $t_{*,90\%}$, when 90\% of the final stellar mass has
      been formed. The grey dashed vertical lines indicate the time of first
      star formation. (Top row) The effective radius of neutral cloud
      $R = 2^{1/3}R_{50\%}$, where $R_{50\%}$ is the half-mass radius. (Second
      row) Velocity dispersion measured perpendicular
      ($\sigma_{{\rm 1d},\perp}$) and parallel ($\sigma_{{\rm 1d},\parallel}$)
      to the background magnetic field, and averaged over all directions
      ($\sigma_{\rm 1d}$). (Third row) Gravitational ($|\mathcal{U}|$, blue),
      kinetic ($\mathcal{T}_{\rm kin}$, red), thermal ($\mathcal{T}_{\rm thm}$,
      green), and magnetic ($\mathcal{B}$, yellow) energies (see Equations
      \eqref{e:Ekin}--\eqref{e:Egrav}). A simple estimate of gravitational
      energy is also shown (blue dashed). (Fourth row) The kinetic virial
      parameter $\avir=5\sigma_{\rm 1d}^2 R/(GM_{\rm neu})$ for different
      viewing angles (gray lines) and various energy ratios.}\label{f:hst2}
    \end{center}
\end{figure*}

Figure~\ref{f:hst}(a) shows the temporal evolution of the gas and stellar mass
in the fiducial model.\footnote{In this figure and Figures~\ref{f:EM} and \ref{f:hst2}, we also show results from models with high turbulence and with high magnetic field, to be discussed in Section~\ref{s:other}.} Only gas that was originally part of the cloud
  is shown, selected, e.g., as
  $M_{\rm H_2}=\int 2 \nHH \muH m_{\rm H} s_{\rm cl} dV$ where $s_{\rm cl}$ is
  the passive scalar for the initial cloud. The stellar mass growth (blue) is
roughly linear in time and 90\% of the total star formation is complete at
$\tprime = 5.7 \Myr$. The net SFE (or lifetime/integrated/final SFE) is
$\varepsilon_{*} = M_{*,{\rm final}}/\Mcl = 0.058$. Here and elsewhere, $M_*$ is measured based on the total mass of star particles. Photodissociation is
efficient at depleting molecular gas mass: at $\tprime=3.6\Myr$, the mass of
molecular hydrogen is only $0.36\Mcl$ and the mass of atomic hydrogen increases
to $0.33\Mcl$. The dashed lines show the outflow cloud gas mass that has left
the computational domain as neutrals (black) and ions (green). Most (72\%) of
the cloud mass is photoevaporated and ejected as ions, consistent with the
findings from hydrodynamic simulations of \citet{kim18}. In addition, $22\%$ of
the initial cloud is driven out as neutrals (9\% molecular). We note that even
without radiation feedback, a small fraction ($\sim 0.1 \Mcl$) of neutrals would
have been ejected; this is the portion of gas that is unbound from the initial
turbulence \citep{raskutti16}.


In Figure~\ref{f:hst}(b) we present the evolution of the cluster luminosity in
LyC and FUV (LW+PE) frequency bins for the fiducial model. The LyC (FUV)
luminosity keeps increasing until $\tprime=4.6\Myr$ ($6.0\Myr$) at which point
$Q_{\rm LyC} = L_{\rm LyC}/h\nu_{\rm LyC} = 1.9 \times 10^{50}\second^{-1}$
($L_{\rm FUV} = 2.8 \times 10^6 \Lsun$), and decreases afterwards.

Figure~\ref{f:hst}(c) shows LyC and FUV escape fractions for the fiducial model.
Instantaneous escape fractions $f_{\rm esc}(t) = L_{\rm esc}(t)/L(t)$ are
plotted as solid lines while the cumulative (or luminosity-weighted,
time-averaged) escape fractions
$f_{\rm esc}^{\rm cum}(t) = \int_{t_{0,*}}^{t} L_{\rm esc}(t) \di
t/\int_{t_{0,*}}^{t} L(t)\di t$ are shown as dashed lines. Overall, both the
instantaneous and cumulative escape fractions keep increasing with
time.\footnote{The downward spikes of $f_{\rm esc}$ at $t=12.5 \Myr$ and
  $t=14.0\Myr$ are due to the formation of sinks in the swept up gas near the
  computational boundary.} The cumulative escape fractions measured at
$\tprime = 3\Myr$ are $6.3\%$ and $24\%$ for LyC and FUV radiation, but they
become as large as $30\%$ and $63\%$ at the end of the simulation
($t=19.6\Myr$). As we shall show, except for the magnetically subcritical clouds
in which the evolutionary timescale is significantly longer than the lifetime of
radiation sources, most of LyC radiation escapes in the first $5\Myr$ after the
star formation.

Figure~\ref{f:EM} depicts maps of emission measure
(${\rm EM} = \int \nEL^2 \di \ell$) from ionized gas projected along the
$y$-axis, with snapshots from the fiducial model at $\tprime = 1$, $3$, $5$,
$8 \Myr$ shown in the left column of panels. Also overlaid (white contours) are
the loci where $A_V=1$ ($N_{\rm H} = 1.87\times 10^{21}\cm^{-2}$) in the neutral
gas, outlining the overall projection of the cloud. Although at
$\tprime = 1\Myr$ the \HII\ region remains relatively confined, it has already
broken out of the cloud. Over the next several Myr the region of high EM rapidly
expands to cover the area of the neutral cloud and beyond. The coincidence
between high EM and $A_V>1$ suggests that bright H$\alpha$ and CO gas would
overlap for several Myr until the cloud is substantially dispersed, as seen in
observations \citep[e.g.,][]{Schinnerer19,chevance20a}.

Figure~\ref{f:EM} includes overlaid line segments indicating the magnetic field
orientation in the $x$-$z$ plane obtained from synthetic maps of Stokes $Q$ and
$U$ parameters of polarized dust emission, assuming a spatially constant dust
temperature, opacity, and intrinsic polarization fraction
\citep[e.g.,][]{kimcg19}. The length of a segment indicates the magnetic field
strength normalized by the initial field strength $B_{0,z}$.

At early times a fraction of the initial kinetic energy is converted into
turbulent magnetic fields. The magnetic intensity in neutral gas increases
(sublinearly) with density and becomes as high as $\sim 10^2\,\mu{\rm G}$ in
densest ($\nH \sim 10^4\pcc$) regions. While the large-scale magnetic field
orientation remains unchanged before significant gas dispersal, the small-scale
magnetic field fluctuates and its orientation exhibits deviation from the
large-scale orientation. Regions of higher and lower fractional polarization
also reflect the facts the parallel component of the magnetic field strength is
enhanced across shock fronts, but it is weakened significantly across ionization
fronts after the onset of star formation
\citep[e.g.,][]{redman98,draine11,kim14}.

The evolution of a cloud's size and velocity dispersion reflect its dynamical
evolution, both before and after the onset of radiation feedback. There are many
ways to compute cloud size\footnote{We have also considered the rms distance
  from the center of gas mass, as well as the effective radius
  $[3 \int \Theta dV/(4\pi)]^{1/3}$, and found similar results.}, but here we
define the effective cloud radius as $R = 2^{1/3}R_{50\%}$, where $R_{50\%}$ is
the half-mass radius of the neutral gas and the factor $2^{1/3}$ applies to a
homogeneous sphere ($R = \Rcl$ at $t=0$). The mass-weighted, line-of-sight
velocity dispersions are measured along the Cartesian axes
$\sigma_{\rm 1d,i}^2 = \int \rho (v_i - \bar{v}_i)^2 \Theta dV / \int \rho \Theta dV$,
where $\bar{v}_i$ is the mean velocity (see Equation~\eqref{e:Theta} for
definition of $\Theta$, which selects only neutral ``cloud'' gas). Because the
gas motion is anisotropic in the presence of strong magnetic field, we consider
the velocity dispersions perpendicular and parallel to the $z$-axis and their
average separately:
\begin{align}
  \sigma_{\rm 1d,\parallel} & = \sigma_{{\rm 1d},z}\,, \label{e:sigma1da} \\
  \sigma_{\rm 1d,\perp} & = \frac{1}{\sqrt{2}}\left(
                          \sigma_{{\rm 1d},x}^2 + \sigma_{{\rm 1d},y}^2
                          \right)^{1/2}\,,  \label{e:sigma1db} \\
  \sigma_{\rm 1d} & = \frac{1}{\sqrt{3}}\left(
                    \sigma_{{\rm 1d},x}^2 + \sigma_{{\rm 1d},y}^2 + \sigma_{{\rm 1d},z}^2
                    \right)^{1/2}\,. \label{e:sigma1dc}
\end{align}

The time evolution of $R$ and of $\sigma_{\rm 1d}$, $\sigma_{\rm 1d,\parallel}$,
and $\sigma_{\rm 1d,\perp}$ for the fiducial model are shown in
Figure~\ref{f:hst2}(a) and (b). We show the time evolution until $t_{*,90\%}$,
the time at which 90\% of the final stellar mass has been assembled. The time of
the first star formation is indicated by grey dashed vertical lines. Although
gas in the cloud is compressed internally by turbulence and gravity, the overall
cloud size changes very little until $\tprime \sim 5\Myr$, when feedback begins
to dominate the evolution. The decrease of the velocity dispersion before
$t=t_{*,0}$ reflects the rapid decay of turbulence within a flow crossing time
\citep[e.g.,][]{ostriker01}, but the level of the velocity dispersion begins to
increase after $\tprime \sim 0.5\Myr$ due to expanding motions induced by the
radiation feedback. The velocity dispersion becomes weakly anisotropic after
$t > 6\Myr$.

Evolution of individual energies and their ratios are shown for the fiducial
model in Figure~\ref{f:hst2}(c) and (d); these will be discussed in detail in
Section~\ref{s:virial}. Here, we simply note that as expected, the cloud's
gravitational energy ($|\mathcal{U}|$) is near constant at first and then
declines (mirroring evolution of the cloud radius $R$), while kinetic energy
($\mathcal{T}_{\rm kin}$) initially drops and then increases (tracking the
velocity dispersion). Magnetic energy ($\mathcal{B}$) at first increases slightly as
magnetic turbulence is driven, and then declines as the cloud is dispersed.

\paragraph{PDFs of Density, Radiation, and Pressure}

\begin{figure*}[t!]
  \begin{center}
    \includegraphics[width=\linewidth]{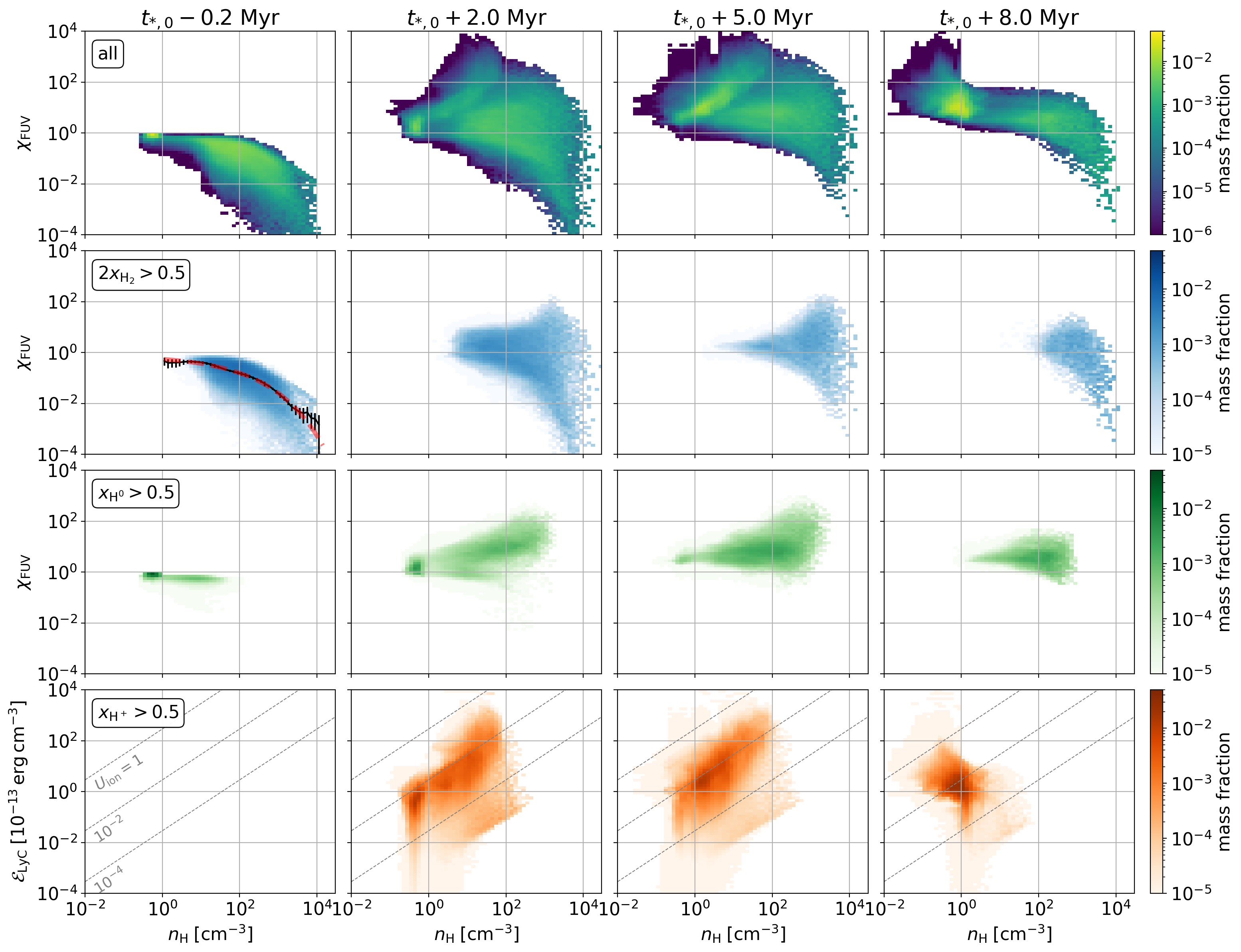}
    \caption{Mass-weighted PDFs of radiation intensity and gas density for the
      fiducial model ({\tt A2B2S4}). at times, from left to right,
      $t_{*,0}-0.2\Myr$, $t_{*,0}+2\Myr$, $t_{*,0}+5\Myr$, $t_{*,0}+8\Myr$,
      where $t_{*,0}$ is the time of first star formation. From top to bottom,
      rows show the PDF of all gas, mostly-molecular gas ($2\xHH > 0.5$),
      mostly-atomic gas ($\xHI > 0.5$), and mostly-ionized gas ($\xHII > 0.5$).
      The top three panels show the intensity in the FUV band, while the bottom
      row shows the intensity in the LyC band. In the left panel of the second
      row, the black line with error bars shows the mass-weighted average FUV
      intensity in molecular gas before $t_{*,0}$. This density-dependent
      attenuation of the ISRF is well fit by Equation~\eqref{e:tau_eff}, shown
      as the red dashed line. The gray dashed diagonal lines in the bottom row
      are the loci of constant ionization parameters (top to bottom)
      $U = \mathcal{E}_{\rm LyC}/(\nH h\nu_{\rm LyC})=1$, $10^{-2}$,
      $10^{-4}$.}\label{f:pdf-rad}
  \end{center}
\end{figure*}

\begin{figure*}[t!]
  \begin{center}
    \includegraphics[width=\linewidth]{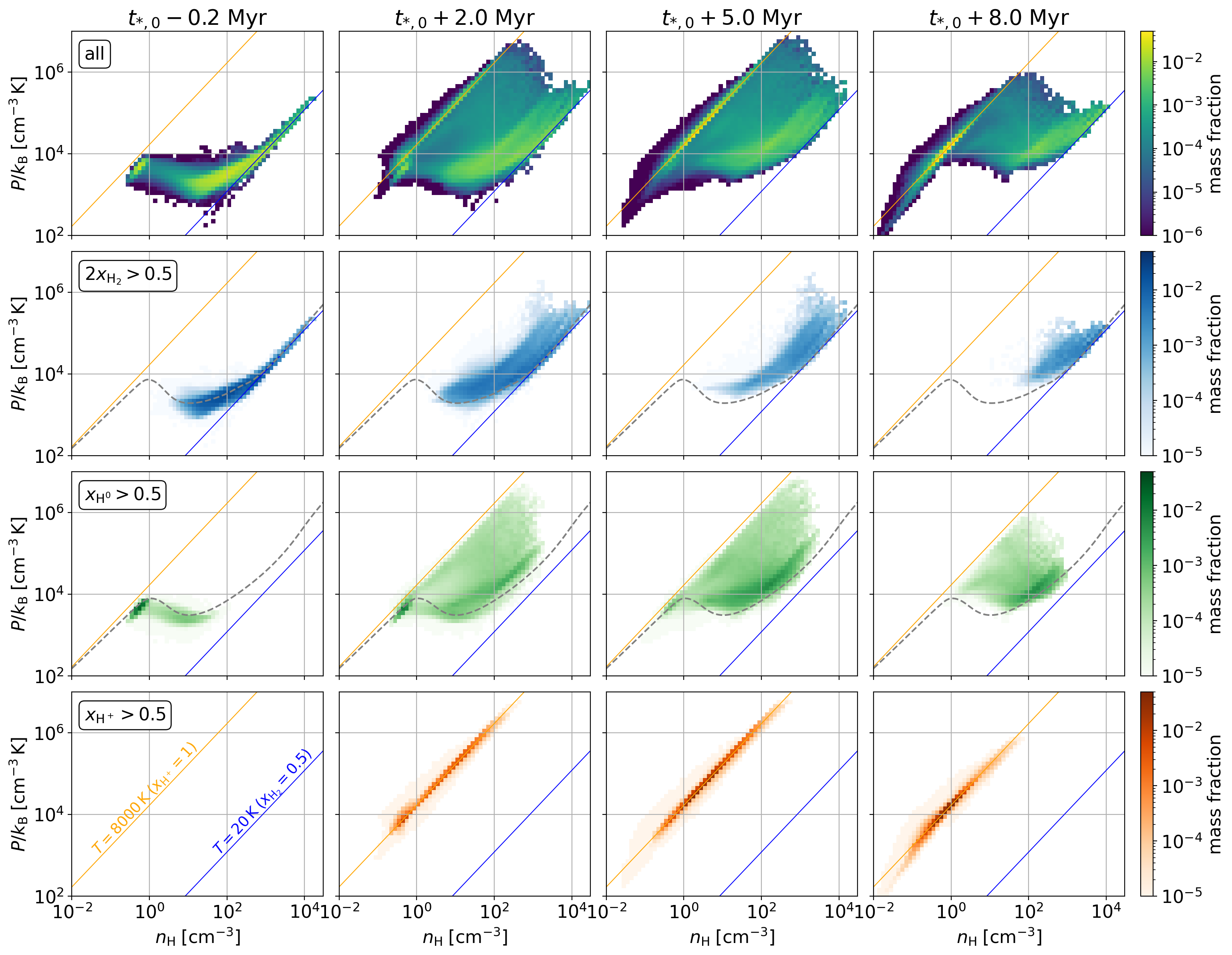}
    \caption{Mass-weighted PDF of thermal pressure and density of the fiducial
      model ({\tt A2B2S4}). Temporal snapshots and gas phases are as in
      Figure~\ref{f:pdf-rad}. The blue and orange diagonal lines show constant
      temperature $20\Kel$ and $8000\Kel$ for fully molecular ($\xHH=0.5$) and
      ionized ($\xHII=1$) gas. In the second and third rows, the grey dashed
      lines plot the equilibrium pressure expected for gas exposed to the
      background UV radiation with density-dependent shielding (see text for
      details). }\label{f:pdf-pok}
  \end{center}
\end{figure*}


The gas in our simulations is found in a range of physical conditions.
Figure~\ref{f:pdf-rad} shows the mass-weighted PDF in the density ($\nH$) and
radiation intensity ($\chi_{\rm FUV}$ or $\mathcal{E}_{\rm LyC}$) plane at times
$\tprime = -0.2$, $2$, $5$, $8\Myr$. The top row shows the PDF of the entire
gas, while the other rows select gas in molecular ($2\xHH > 0.5$; blue), neutral
atomic ($\xHI > 0.5$; green), and ionized ($\xHII > 0.5$; orange) phases. In the
top three rows of Figure~\ref{f:pdf-rad}, we show the distribution in the
$\nH$--$\chi_{\rm FUV}$ plane; in the bottom row we plot the distribution in the
$\nH$--$\mathcal{E}_{\rm LyC}$ plane. Similarly, Figure~\ref{f:pdf-pok} shows
mass-weighted PDFs in the density and pressure ($P/\kB$) plane.

Before the onset of massive star formation ($t_{*,0}-0.2\Myr$ panels), gas is
subject to the background FUV radiation only. The ambient gas is optically thin,
and is therefore exposed to the ISRF with $0.8 \lesssim \chi_{\rm FUV} \le 1$;
it has low density and is in the warm ($T \sim 8000 \Kel$) atomic phase. The
width of the density PDF of the ambient warm gas is narrow as turbulence is
subsonic.

The cold, dense gas in the cloud is subject to a range of FUV intensity and
becomes increasingly shielded at higher density. Most of the cloud is molecular, 
as shown in Figure~\ref{f:hst}(a).
For the fiducial cloud, we compute
the mass-weighted mean FUV intensity in each density bin for snapshots before
the first star formation. In the left panel of the second row in
Figure~\ref{f:pdf-rad}, the black line shows the time-averaged mean FUV
intensity $\chi_{\rm FUV,avg}$ for snapshots prior to the onset of star formation, with the error bars indicating the standard deviation.
We find that $\chi_{\rm FUV,avg}$ has weak temporal variation and is
well described 
by a local shielding approximation. The red dashed line shows
that the average FUV radiation field in molecular gas with $\nH > 1\pcc$ is well
fit by the function\footnote{We fix the factor 0.9 that accounts for the dust shielding by the ambient medium, i.e., $0.9 = \exp \left( -10^{-2} \nHcl \sigma_{\rm d,PE} \times (2\Rcl) \right)$.}  $\chi_{\rm FUV,avg} = 0.9 e^{-\tau_{\rm FUV,eff}}$, 
where the effective optical depth is fitted as a power-law function of gas density:
\begin{equation}
  \tau_{\rm FUV,eff} = 1.8 \left( \frac{\nH}{10^2
      \pcc}\right)^{0.30}\,;\label{e:tau_eff}
\end{equation}
$1.8$ and $0.30$ are the best-fit parameters.
That is, the effective optical depth increases sublinearly with density, such
that if we define $\tau_{\rm FUV,eff} \equiv \sigma_{\rm d}\nH L_{\rm shld}$,
this corresponds to the shielding length of $L_{\rm shld}\sim 6\pc$ when
$\nH=10^2 \pcc$.  

Based on this result, we calculate the equilibrium chemical abundances and
temperature of molecular gas as a function of $\nH$ assuming constant
$\xi_{\rm cr} = 2\times 10^{-16}\second^{-1}$, and density-dependent FUV
intensity $\chi_{\rm FUV,avg}$. We also calculate the equilibrium curve for
unshielded atomic gas assuming $\xi_{\rm cr} = 2\times 10^{-16}\second^{-1}$,
$\chi_0=1$, and $\tau_{\rm FUV,eff}=0$. The gray dashed lines in center two rows of
Figure~\ref{f:pdf-pok} show the resulting equilibrium curves; before the onset of star formation most gas remains close to these equilibrium curves. The low-density
molecular gas with $10\pcc < \nH < 10^3\pcc$ exhibits a range of temperature
$20\Kel < T < 100\Kel$, while the high-density molecular gas for which the
heating is dominated by cosmic ray ionization has a constant temperature of
$\sim 20\Kel$ (blue diagonal lines in Figure~\ref{f:pdf-pok}).\footnote{In
  reality, the cosmic ray ionization rate may be significantly attenuated within
  dense molecular gas \citep[e.g.,][]{neufeld17}, leading to lower gas
  temperature. However, for practical purposes the thermal energy density is
  already extremely low compared to kinetic and magnetic energy densities.} The
thermal pressure of the atomic and diffuse molecular gas ranges from
$P/k_{\rm B} \sim 10^3$--$10^4 \pcc\Kel$, while it can be as high as
$\sim 2 \times 10^5 \pcc \Kel$ at the density of sink particle formation
($\sim 2 \times 10^4 \pcc$).

After the onset of massive star formation ($\tprime > 0$),
both neutrals and ions are subject to intense UV radiation
from newborn stars, and the normalized FUV radiation intensity can be as high as
$\chi_{\rm FUV}\sim 10^2$ in the photon dominated region (PDR). \footnote{For
  reference, this is equivalent to the intensity $10\pc$ away from a point
  source of luminosity $10^6 \Lsun$.} The temperature of 
  lightly-shielded dense molecular/atomic gas in the PDR is significantly elevated because of enhanced
heating, but most of the gas is still shielded and remains cold, at
$T \sim 20$--$100\Kel$. The photoionized gas has temperature $\sim 8000 \Kel$
(orange diagonal lines in Figure~\ref{f:pdf-pok}). The ionized gas has a range of
ionization parameter $U_{\rm ion} = \mathcal{E}_{\rm LyC}/(h\nu_{\rm LyC} \nH)$,
but most falls in the range $10^{-2.5}$--$10^{-1.5}$ (see gray dashed diagonal
lines in the bottom row of Figure~\ref{f:pdf-rad}).

\subsection{Evolution of Other Models}\label{s:other}

While the overall evolution of other models is similar to that of the fiducial
model, there are a few notable differences. Here we give an overview of other
simulations and take two extreme cases, one with the strongest initial
turbulence/highest $\avircl$ ({\tt A5B2S4}) and the other with the strongest
initial magnetic field/lowest $\muBcl$ ({\tt A2B05S04}), to compare and contrast
with the fiducial model. Results from these two extreme models are shown in the
middle and right columns of Figures~\ref{f:hst}, \ref{f:EM}, \ref{f:hst2}. In
addition, in Appendix~\ref{s:snapshot} we compare a series of snapshots of projected
column density from members of the $\alpha$-series in Figure~\ref{f:snapshot-a}
(all with ${\tt seed}=4$), members of the $\beta$-series models in
Figure~\ref{f:snapshot-b} (all with ${\tt seed}=4$), and the fiducial model ({\tt
  A2B2}) with initial different turbulence realizations (i.e., varying {\tt
  seed} but the same initial kinetic and magnetic energy) in
Figure~\ref{f:snapshot-s}.

Our key simulation results are summarized in Table~\ref{t:result}. Quantitative
comparisons of star formation history, SFE, ionized and neutral gas ejection,
timescale for star formation/cloud destruction, and escape fraction of radiation
from all models will be presented in Sections~\ref{s:sfh}--\ref{s:fesc}.
Comparisons of virial parameter evolution and efficiency per freefall time, also
connecting with observations, will be discussed in Sections~\ref{s:virial} and
\ref{s:epsff}, respectively.

The bottom three rows of Table~\ref{t:result} show results from the fiducial model (${\tt A2B2S4}$) run at higher ($N_{\rm cell} = 512^3$) and lower ($N_{\rm cell} = 128^3$) resolutions.
These tests demonstrate numerical convergence of our results, within the
range of variation that is expected from turbulence-induced stochasticity in the evolution.

\subsubsection{Dependence on $\avircl$}

Even at quite early times, the extent of the molecular cloud systematically
changes with $\avircl$. The left panels of Figure~\ref{f:snapshot-a} in
Appendix~\ref{s:snapshot} show that the cloud with the lowest initial virial
parameter ({\tt A1B2S4}, with $\avircl=1$) initially contracts in size as
turbulent pressure support is insufficient to counteract gravity. In contrast,
clouds that are initially unbound have a larger fraction of their material at
velocity greater than the escape speed of the cloud, and expand rapidly over the
first few Myr of evolution. This has the effect of both directly enhancing the
mass loss of neutrals and reducing the gas surface density during the
star-forming epoch, both of which contribute to a systematically lower net SFE
at higher $\avircl$.

In part because the turbulent shear is more effective in dispersing overdense
structures before they collapse in the higher-$\avircl$ models, the time of the
first star formation $t_{*,0}$ is nearly 1 Myr (on average) earlier in the
lowest-$\avircl$ models compared to the highest--$\avircl$ models. Comparing
Figure~\ref{f:hst}(a) and (d), even though the lifetime SFE is lower in the
$\avircl=5$ model compared to the $\avircl=2$ model, ``active'' star formation
continues over a longer period in the $\avircl=5$ model (up to $t \sim 11\Myr$,
compared to $t \sim 9\Myr$) because the cloud is not as rapidly dispersed when  the 
star formation and resulting feedback is less vigorous. Still, there is late-time residual star
formation in the $\avircl=2$ model, and the overall beginning-to-end duration of
star formation does not show strong dependence on $\avircl$ across models (see
Section~\ref{s:timescales}).

We also note, as can be seen from the comparisons in Figure~\ref{f:EM} and
Figure~\ref{f:snapshot-a}, that distribution of radiation sources is less
compact in high-$\avircl$ compared to low-$\avircl$ models.

\subsubsection{Dependence on $\muBcl$}\label{s:depmu}

Our simulation suite includes both strongly magnetized clouds (with the
$\muBcl=0.5$ case even subcritical) and weakly magnetized clouds. The latter
(models $\muBcl=4$ and $8$) have initial $B_{z,0} < 10 \mu{\rm G}$, and the
magnetic field remain energetically subdominant -- with super-Alfv\'{e}nic
turbulence -- throughout their lifetimes. Their overall evolution and
morphological characteristics are largely similar to the fiducial model. We also include an unmagnetized model.
Compared to the non-magnetized run, the first star formation event in moderate-magnetization models  is a bit
delayed ($\sim 0.5$--$1\Myr$), and SFR and net SFE are slightly lower.

The magnetically subcritical and critical clouds ($\muBcl=0.5$ and $1$) have
initial magnetic fields $54\mu{\rm G}$ and $27\mu{\rm G}$, with initially
sub-Alfv\'{e}nic turbulence. Figure~\ref{f:snapshot-b} shows that overdense
filamentary structures preferentially aligned perpendicular to $\bm{B}_0$ along the $z$-axis (to
an extreme degree in the subcritical case). The strong magnetic field constrains
motions to proceed primarily along the $z$-axis, and as a result the cloud
expansion driven by feedback at late times is highly anisotropic in these
models.

Compared to the fiducial model, in the subcritical and critical cases the strong
magnetic field both delays the onset of star formation and reduces the SFR. As a
result, it takes longer to build up enough stellar mass that the feedback is
sufficient to disperse the residual gas. In addition, 
the reduced gas compressibility and strong magnetic tension force make cloud structure less porous and \HII\ regions remain confined for a longer
period of time. This makes the escape of ionized gas and ionizing radiation more
difficult. For these reasons, only a small fraction of gas is photoevaporated by
LyC in the critical and subcritical cases. Compared to the fiducial model, in
which most of the mass loss is in ionized gas, nearly 10 times as much gas is
dispersed in the neutral phase as the ionized phase.

In fact, the evolution of magnetically subcritical clouds are unusual in several
aspects. In {\tt A2B05S4}, the first star formation occurs at $9.3\Myr$ and the
net SFE is only $2.8\%$. The stellar mass builds up on a timescale that is long
compared to the lifetime of ionizing luminosity (see Figure~\ref{f:hst}g, h).
Due to the low luminosity, gas dispersal takes place steadily over a very long
period ($\sim 20\Myr$), the mass loss by photoevaporation is much less effective
than in the fiducial model, and the LyC escape fraction does not reach a high
value even at late times (Figure~\ref{f:hst}(g)). The evolution of
$\sigma_{\rm 1d}$ (Figure~\ref{f:hst2}(j)) shows that the motion of neutral gas
becomes highly anisotropic after the onset of feedback, with motions almost
entirely parallel to the background magnetic field. The map of EM in
Figure~\ref{f:EM} shows that ionized gas outflows follow the magnetic field
lines, and that the cloud dispersal is incomplete even at $\tprime=8\Myr$ (see
also the right panels in Figure~\ref{f:hst}). The final frame of
Figure~\ref{f:snapshot-b} for the $\muBcl=0.5$ model shows an essentially
columnar outflow. Lack of evidence for these unusual morphological features in
observations suggests that at GMC scales, real clouds are generally magnetically
critical or supercritical. Some lower mass dark clouds (without massive star formation) may, however, be magnetically subcritical (see Section~\ref{s:summary}).
In addition, smaller-scale bipolar \HII\ regions have been identified by \citet{deharveng15,eswaraiah17}, suggesting localized strong magnetic fields, perhaps enhanced by the collapse that created the exciting clusters.

\subsubsection{Dependence on Turbulence Realization}

For each model in the $\alpha$- and $\beta$-series, we run five simulations with
different random realizations for the initial turbulence. The initial
large-scale velocity field and its orientation relative to the field line affect
the specifics of when and where dense structures form, which in turn affect the
subsequent cluster formation and cloud evolution. As a result, the simulation
outcome exhibits a moderate degree of variation with random seeds.
Figure~\ref{f:snapshot-s} shows snapshots of the fiducial model with different
turbulence realizations. We find that runs with ${\tt seed}=1$ and $5$ form
centrally-concentrated dense filaments, which are favorable to an early, rapid
burst of star formation (see second column in Figure~\ref{f:snapshot-s}). In
contrast, overdense structures in runs with ${\tt seed}=2$ and $3$ are spatially
separated from each other, and these models have slower SFR and lower SFE. While
some of the same trends with {\tt seed} persist in other models with different
$\avircl$ or $\muBcl$, not all quantitative outcomes systematically vary with
the random seed (see Sections~\ref{s:sfh}--\ref{s:fesc}).

For a given total kinetic energy, lower SFR and SFE in some turbulent
realizations may simply be due to more of the large-scale modes (which contain
most of the energy) favoring expansion or shear rather than compression.

\subsection{Star Formation History and Rate}\label{s:sfh}

\begin{deluxetable*}{ccccccccccccc}
\tabletypesize{\scriptsize}
\tablewidth{0pt}\tablecaption{Simulation results}
\tablehead{
\colhead{Model} &
\colhead{$t_{*,0}$} &
\colhead{$\avirttot(t_{*,0})$} &
\colhead{$\varepsilon_*$} &
\colhead{$\varepsilon_{*,{\rm 3\,Myr}}$} &
\colhead{$\varepsilon_{\rm ion}$} &
\colhead{$\varepsilon_{\rm ej,neu}$} &
\colhead{$t_{\rm SF}$} &
\colhead{$t_{\rm dest,H_2}$} &
\colhead{$t_{\rm dep,0}$} &
\colhead{$\varepsilon_{\rm ff,0}$} &
\colhead{$\langle f_{\rm esc,LyC}\rangle$} &
\colhead{$\langle f_{\rm esc,FUV}\rangle$} \\
\colhead{} &
\colhead{[Myr]} &
\colhead{} &
\colhead{[\%]} &
\colhead{[\%]} &
\colhead{[\%]} &
\colhead{[\%]} &
\colhead{[Myr]} &
\colhead{[Myr]} &
\colhead{[Myr]} &
\colhead{[\%]} &
\colhead{[\%]} &
\colhead{[\%]} \\
\colhead{(1)} &
\colhead{(2)} &
\colhead{(3)} &
\colhead{(4)} &
\colhead{(5)} &
\colhead{(6)} &
\colhead{(7)} &
\colhead{(8)} &
\colhead{(9)} &
\colhead{(10)} &
\colhead{(11)} &
\colhead{(12)} &
\colhead{(13)}} 
\startdata
$\alpha$-series\\
{\tt A1B2} & $3.5^{+0.6}_{-0.4}$ & $0.86^{+0.07}_{-0.07}$ & $9.5^{+2.9}_{-1.8}$ & $5.6^{+1.3}_{-0.6}$ & $73.2^{+1.1}_{-2.9}$ & $17.4^{+3.5}_{-2.2}$ & $5.0^{+0.3}_{-0.2}$ & $7.4^{+0.7}_{-0.8}$ & $58^{+6}_{-13}$ & $8.0^{+2.4}_{-0.8}$ & $21.0^{+4.6}_{-7.6}$ & $52.7^{+4.3}_{-10.5}$ \\
{\tt A2B2} & $3.8^{+0.5}_{-0.5}$ & $1.10^{+0.02}_{-0.10}$ & $5.8^{+2.1}_{-1.4}$ & $2.6^{+2.4}_{-1.1}$ & $71.7^{+2.2}_{-0.7}$ & $21.9^{+2.8}_{-2.6}$ & $5.4^{+0.3}_{-1.1}$ & $7.6^{+0.2}_{-0.8}$ & $99^{+52}_{-39}$ & $4.7^{+3.0}_{-1.6}$ & $26.7^{+3.0}_{-0.5}$ & $62.4^{+1.0}_{-5.1}$ \\
{\tt A3B2} & $3.7^{+1.3}_{-0.4}$ & $1.43^{+0.02}_{-0.14}$ & $4.1^{+2.2}_{-1.5}$ & $1.9^{+2.0}_{-0.7}$ & $68.3^{+3.4}_{-7.7}$ & $28.7^{+8.2}_{-6.3}$ & $5.5^{+1.1}_{-1.1}$ & $7.5^{+0.5}_{-0.1}$ & $149^{+78}_{-52}$ & $3.2^{+1.7}_{-1.1}$ & $27.6^{+12.9}_{-3.4}$ & $65.9^{+3.0}_{-4.8}$ \\
{\tt A4B2} & $4.1^{+1.1}_{-0.9}$ & $1.77^{+0.13}_{-0.16}$ & $3.6^{+0.6}_{-1.9}$ & $1.7^{+0.4}_{-0.7}$ & $63.1^{+4.5}_{-15.9}$ & $33.2^{+17.9}_{-4.8}$ & $6.3^{+0.2}_{-1.2}$ & $7.4^{+0.5}_{-0.6}$ & $163^{+190}_{-11}$ & $2.9^{+0.2}_{-1.5}$ & $34.2^{+8.3}_{-9.4}$ & $67.8^{+4.6}_{-0.1}$ \\
{\tt A5B2} & $4.3^{+1.3}_{-0.9}$ & $2.12^{+0.17}_{-0.09}$ & $2.1^{+1.1}_{-1.2}$ & $1.2^{+0.2}_{-0.6}$ & $48.3^{+9.4}_{-19.5}$ & $49.6^{+20.4}_{-10.4}$ & $5.3^{+1.5}_{-0.9}$ & $8.3^{+0.5}_{-1.3}$ & $255^{+313}_{-31}$ & $1.8^{+0.3}_{-1.0}$ & $31.2^{+18.5}_{-17.1}$ & $68.8^{+8.6}_{-3.2}$ \\
\tableline
$\beta$-series\\
{\tt A2B05} & $8.6^{+4.7}_{-2.1}$ & $6.66^{+1.76}_{-0.74}$ & $2.4^{+1.2}_{-0.5}$ & $0.5^{+0.3}_{-0.1}$ & $12.4^{+5.0}_{-7.5}$ & $83.5^{+2.2}_{-4.7}$ & $12.0^{+3.5}_{-3.1}$ & $17.3^{+4.8}_{-1.4}$ & $464^{+401}_{-163}$ & $1.0^{+0.5}_{-0.5}$ & $13.1^{+0.3}_{-4.8}$ & $56.5^{+1.7}_{-6.8}$ \\
{\tt A2B1} & $4.4^{+2.4}_{-0.9}$ & $2.18^{+0.34}_{-0.20}$ & $4.3^{+1.8}_{-1.8}$ & $1.5^{+0.5}_{-0.9}$ & $35.3^{+17.7}_{-8.7}$ & $54.2^{+5.5}_{-13.9}$ & $6.9^{+1.3}_{-1.1}$ & $9.4^{+0.7}_{-1.0}$ & $182^{+103}_{-59}$ & $2.6^{+1.2}_{-0.9}$ & $21.1^{+2.8}_{-9.5}$ & $58.3^{+2.5}_{-1.7}$ \\
{\tt A2B2} & $3.8^{+0.5}_{-0.5}$ & $1.10^{+0.02}_{-0.10}$ & $5.8^{+2.1}_{-1.4}$ & $2.6^{+2.4}_{-1.1}$ & $71.7^{+2.2}_{-0.7}$ & $21.9^{+2.8}_{-2.6}$ & $5.4^{+0.3}_{-1.1}$ & $7.6^{+0.2}_{-0.8}$ & $99^{+52}_{-39}$ & $4.7^{+3.0}_{-1.6}$ & $26.7^{+3.0}_{-0.5}$ & $62.4^{+1.0}_{-5.1}$ \\
{\tt A2B4} & $3.1^{+0.1}_{-0.4}$ & $0.93^{+0.05}_{-0.16}$ & $6.9^{+1.6}_{-1.9}$ & $3.4^{+1.0}_{-0.6}$ & $79.0^{+4.8}_{-1.9}$ & $13.9^{+2.8}_{-6.1}$ & $4.9^{+2.2}_{-0.7}$ & $6.9^{+2.1}_{-0.4}$ & $84^{+20}_{-22}$ & $5.5^{+2.0}_{-1.1}$ & $25.8^{+10.9}_{-2.5}$ & $61.3^{+3.5}_{-5.3}$ \\
{\tt A2B8} & $2.6^{+0.2}_{-0.1}$ & $0.92^{+0.12}_{-0.15}$ & $7.4^{+2.3}_{-0.1}$ & $3.7^{+0.7}_{-0.1}$ & $82.3^{+2.4}_{-1.4}$ & $9.2^{+1.5}_{-1.2}$ & $5.4^{+1.0}_{-1.1}$ & $7.4^{+0.7}_{-0.4}$ & $75^{+16}_{-13}$ & $6.3^{+1.3}_{-1.1}$ & $33.0^{+4.5}_{-9.3}$ & $64.2^{+2.6}_{-7.8}$ \\
{\tt A2Binf} & $2.1^{+0.2}_{-0.5}$ & $1.00^{+0.27}_{-0.07}$ & $8.2^{+1.3}_{-1.1}$ & $5.1^{+1.7}_{-1.1}$ & $82.6^{+2.7}_{-0.6}$ & $8.1^{+1.7}_{-1.6}$ & $4.1^{+0.8}_{-0.2}$ & $6.4^{+0.4}_{-0.5}$ & $61^{+8}_{-14}$ & $7.7^{+2.2}_{-0.9}$ & $30.7^{+5.5}_{-2.5}$ & $63.3^{+3.4}_{-1.9}$ \\
\tableline
{\tt A2B2S4\_N128} & $3.1$ & $1.23$ & $6.1$ & $3.1$ & $68.1$ & $25.9$ & $6.5$ & $7.9$ & $109$ & $4.3$ & $31.4$ & $63.1$ \\
{\tt A2B2S4\_N256} & $3.3$ & $1.12$ & $5.8$ & $2.5$ & $71.7$ & $23.0$ & $5.7$ & $7.6$ & $99$ & $4.7$ & $29.7$ & $63.4$ \\
{\tt A2B2S4\_N512} & $3.2$ & $1.11$ & $5.0$ & $1.9$ & $71.6$ & $23.9$ & $5.1$ & $7.9$ & $126$ & $3.7$ & $28.7$ & $62.1$ \\
\enddata
\tablecomments{For each model in the $\alpha$- and $\beta$-series there are 5 realizations of turbulence. The reported values are medians, and the superscript (subscript)
refers to the difference to 
the maximum (minimum) value. 
Results from tests  at lower and higher (only $\tt seed$=4) resolution are given in bottom two rows.
Columns are as follows (1): model
name indicating the initial virial parameter ({\tt A}) and the magnetic flux-to-mass ratio
({\tt B}).
(2): time of first star formation. 
(3): virial parameter $\avirttot = (2\mathcal{T} + \mathcal{B})/|\mathcal{U}|$ at $t_{*,0}$. 
(4): final SFE $\varepsilon_*=M_*/M_0$.
(5): SFE at $\tprime=3\Myr$. 
(6): photoevaporation efficiency.
(7): ejected neutral efficiency.
(8): star formation duration needed to assemble 90\% of the final stellar mass.
(9): cloud destruction time needed to reach $M_{\rm H_2} = 0.05\Mcl$.
(10): gas depletion time $t_{\rm dep,0} = \Mcl/\langle \dot{M}_*\rangle$.
(11): SFE per freefall time $\varepsilon_{\rm ff,0} = t_{\rm ff,0}/t_{\rm dep,0}$
(12)--(13): cumulative (or luminosity-weighted mean) escape fractions of
ionizing (LyC) and non-ionizing (FUV) radiation.}\label{t:result}
\end{deluxetable*}

\begin{figure*}[t!]
  \begin{center}
    \includegraphics[width=\linewidth]{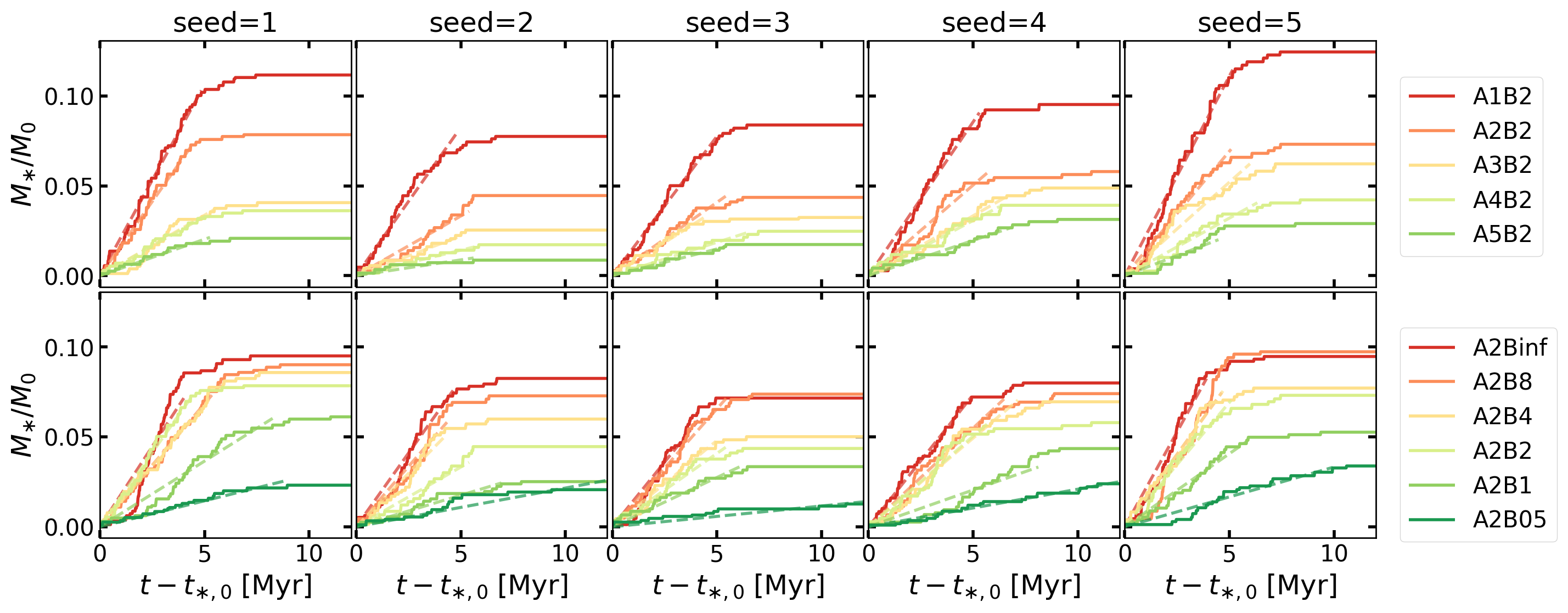}
    \caption{Star formation history of all models with different seeds. The
      upper/lower panels show $\alpha$/$\beta$-series models. Time is measured
      from the moment of first star formation $t_{*,0}$. The dashed lines
      represent a linear least square fit for the time range
      $t_{*,0} < t < t_{*,90\%}$, where $t_{*,90\%}$ is the time at which 90\%
      of the final stellar mass has been assembled. We use the slope of the line
      as the time averaged SFR $\langle \dot{M}_* \rangle$.}\label{f:sfhst}
  \end{center}
\end{figure*}

We compare star formation history of all models in Figure~\ref{f:sfhst}, which
shows the stellar mass $M_*$ as a function of $\tprime = t - t_{*,0}$ in the
$\alpha$-series (top) and $\beta$-series (bottom) models. Different initial
turbulence realizations are labeled by {\tt seed}$=1$--$5$ (left to right).

The stellar mass growth occurs over several Myrs in a ``stair-stepping'' fashion
due to our finite mass resolution (typical initial sink mass is about
$10^{-3}\Mcl$). It is roughly linear during the main phase of star formation and
levels off once the radiative feedback takes over. Some models exhibit a few
episodic star formation events that take place in swept-up gas at the periphery
of \HII\ regions at late times.

We quantify the time-averaged SFR, $\langle \dot{M}_* \rangle$, during the main
period of star formation by performing a least square fit to a function
$M_*(t) = \langle \dot{M}_* \rangle t$ for the time interval
$(t_{*,0},t_{*,90\%})$. The results are shown as dashed lines in
Figure~\ref{f:sfhst}, suggesting that linear mass growth is a good approximation
for most of our models. Except for the runs with ${\tt seed}=1$ in the
$\beta$-series,
the mean SFR monotonically increases with decreasing $\avircl$
and with increasing $\muBcl$. That is, higher kinetic and magnetic energy reduce
the SFR, for a given cloud mass and size.

\subsection{Star Formation Efficiency}

\begin{figure*}[t!]
  \begin{center}
  \includegraphics[width=0.7\linewidth]{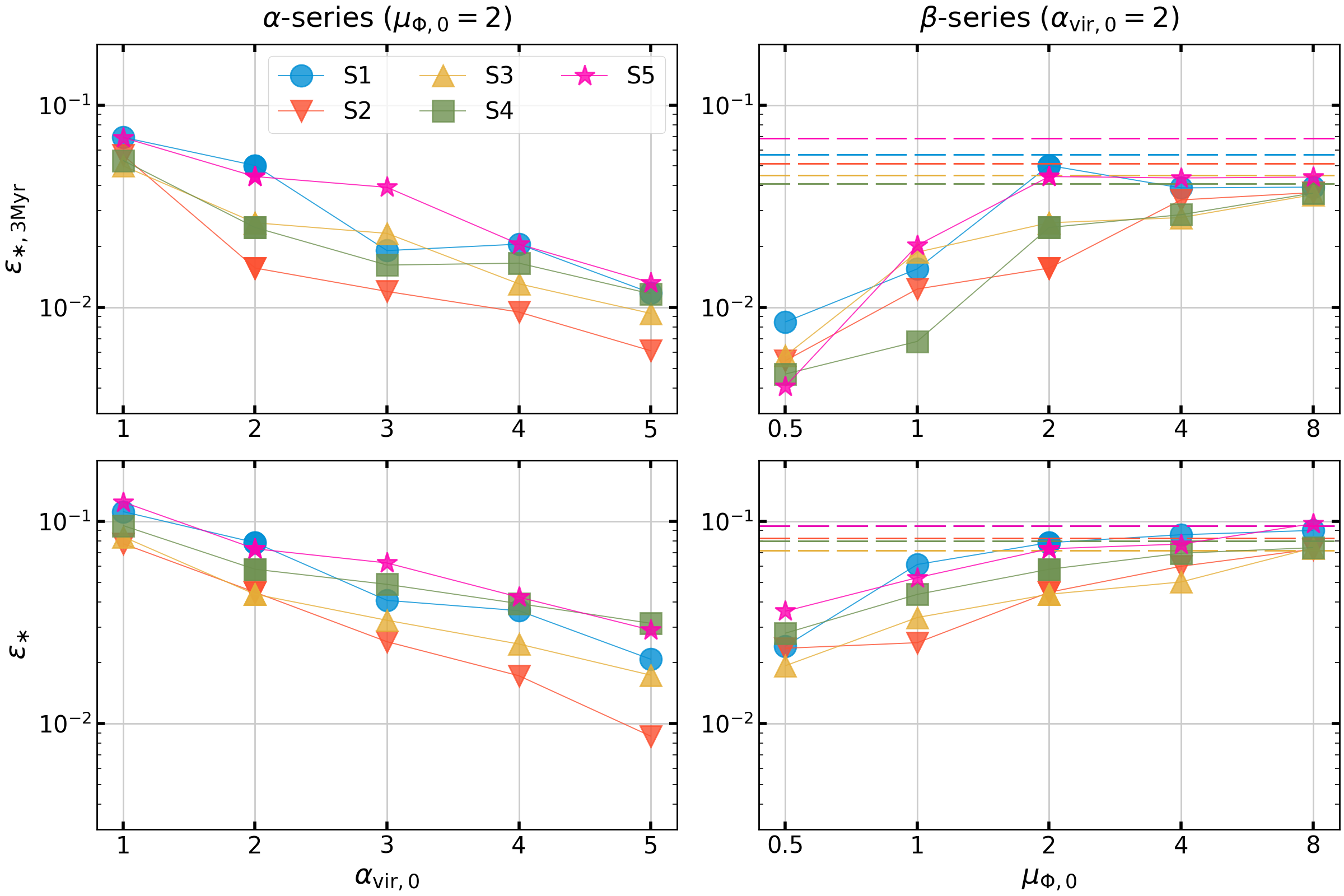}
  \caption{SFE at time $3\Myr$ after the first star
    formation ($\varepsilon_{*,3\,{\rm Myr}}$; top) and final star formation
    efficiency ($\varepsilon_{*}$; bottom), in models with varying initial
    kinetic or magnetic energy or turbulence realization. The $\alpha$-series is
    shown to the left and $\beta$-series to the right, with simulation runs
    having different turbulent realizations as marked by {\tt seed} with
    different colors/symbols. Horizontal dashed lines in the right panels
    indicate the values for the unmagnetized models ({\tt A2Binf}). Higher
    initial turbulence level (larger $\avircl$) or magnetization (smaller
    $\muBcl$) reduces the SFE.}\label{f:SFE}
  \end{center}
\end{figure*}

Figure~\ref{f:sfhst} shows that clouds in our simulations convert only a small
fraction of the initial gas mass into stars over their lifetimes. In
Figure~\ref{f:SFE}, we show the SFE at $\tprime=3\Myr$ ($\varepsilon_{*,3\Myr}$)
and the lifetime SFE ($\varepsilon_{*}$) in the $\alpha$- (left) and
$\beta$-series (right) models. For each model, we show runs with different
turbulent seeds with different symbols and colors. For the $\beta$-series
models, the results of non-magnetized runs are shown as horizontal dashed lines.
The time $\tprime=3\Myr$ is chosen to provide an estimate of the SFE at the time
of the first supernova. In Columns (4) and (5) of Table~\ref{t:result} we list
result for medians (over {\tt seed}) of $\varepsilon_{*}$ and
$\varepsilon_{*,3\Myr}$, and give differences to the minimum/maximum values with
superscripts/subscripts. For example, the fiducial model ({\tt A2B2}) has net
SFE of $\varepsilon_{*}=5.8^{+2.1}_{-1.4}\%$ for the 5 different runs.

For fixed turbulent realization, Figure~\ref{f:sfhst} shows that final stellar
mass decreases with increasing $\avircl$ and decreasing $\muBcl$. The median (over seed) net
SFE in the $\alpha$-series ranges between $\varepsilon_{*}=2.1\%$ and $9.5\%$, increasing for
lower kinetic energy (lower $\avircl$). The median net SFE in the $\beta$-series
models ranges between $\varepsilon_{*}=2.4\%$ and $8.2\%$, increasing for lower magnetization
(higher $\muBcl$). In all models, different turbulence realizations can produce
variations at a level of a few percent in the SFE. Although the absolute
variations in SFE are small, this amounts to up to factor of $2$--$3$ range in
$\varepsilon_{*,3\Myr}$ and $\varepsilon_*$ for different turbulence
realizations. Most of our simulated clouds have formed about
$\sim 50^{+20}_{-20}\%$ of the final stellar mass at $\tprime = 3\Myr$. The
exceptions are the very strongly-magnetized models {\tt A2B05} and {\tt A2B1},
in which $\varepsilon_{*,3\Myr}/\varepsilon_{*} \sim 0.25$ as they have
significantly longer star formation duration (see Section~\ref{s:timescales}).

The simulations of \citet{kim18} were purely hydrodynamic, including a model
analogous to ${\tt seed}=1$ and $\muBcl=\infty$, which resulted in
$\varepsilon_{*}=13\%$, somewhat higher than $9.6\%$ of our model
${\tt A2BinfS1}$. This difference is likely to be caused by the inclusion of
more realistic thermochemical processes in our new simulations; the temperature
of molecular gas in the PDR is raised significantly due to the FUV heating,
which makes the collapse of dense structures more difficult at late times
\citep[e.g.,][]{inoguchi20}. Considering the stochastic nature of the system,
these differences can also induce other changes; as we have shown, different
realizations of turbulence lead to variations in SFE of a few percent.

\subsection{Photoevaporation and Ejection Efficiencies}\label{s:evej}

\begin{figure*}[t!]
  \begin{center}
  \includegraphics[width=0.7\linewidth]{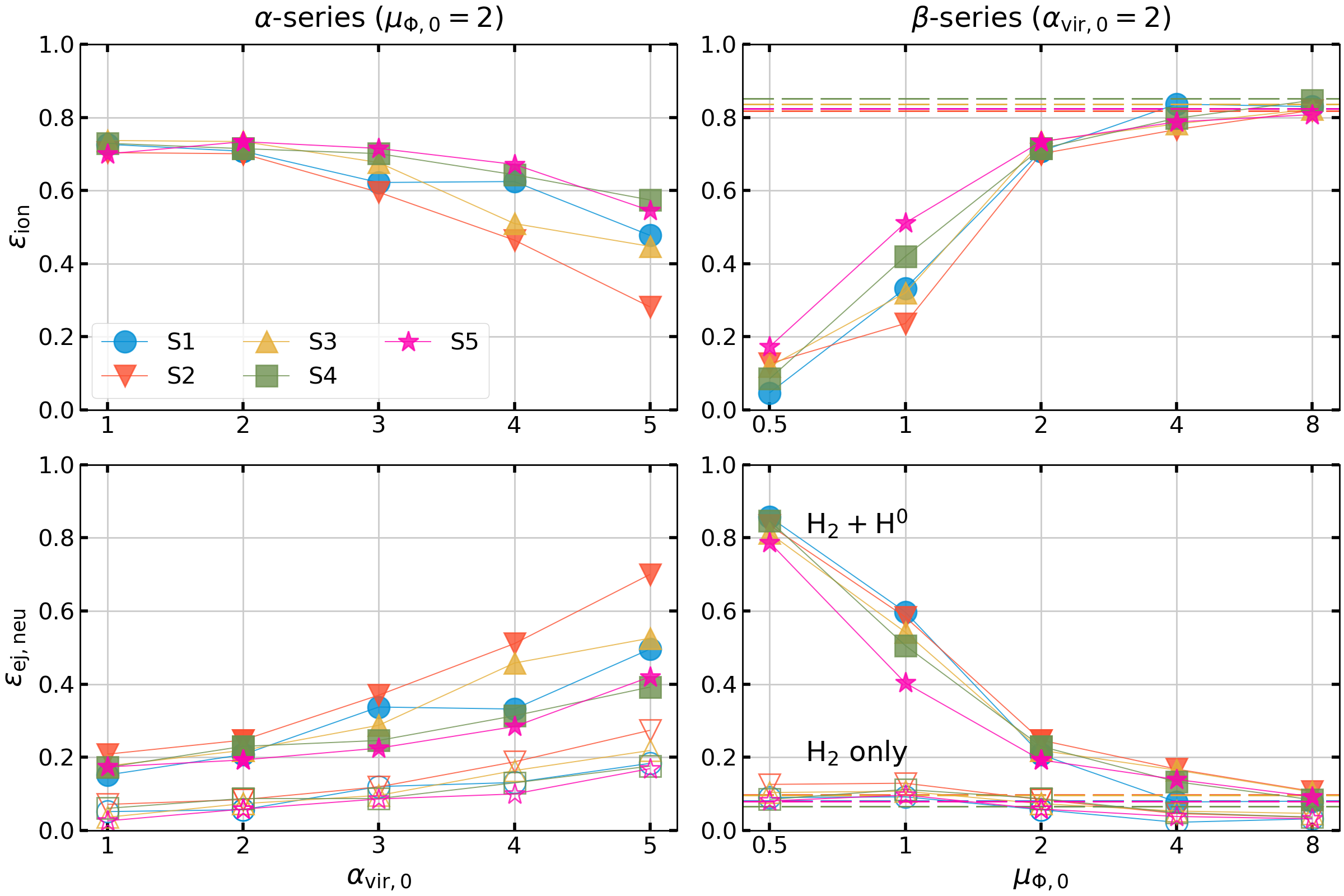}
  \caption{(Top) Fraction of initial cloud mass converted into ionized gas
    (photoevaporation efficiency; $\varepsilon_{\rm ion}$) and (bottom) ejected
    from the computational domain as neutrals (ejected neutral efficiency;
    $\varepsilon_{\rm ej,neu}$). The open symbols show the ejection efficiency
    of molecular gas. The left/right column shows results from the
    $\alpha$/$\beta$-series simulations, with different turbulent {\tt seed} and
    unmagnetized models indicated as in Figure~\ref{f:SFE}.}\label{f:evej}
  \end{center}
\end{figure*}

Clouds in our simulations are efficiently dispersed by UV radiation feedback,
which converts the cold molecular gas that would otherwise form stars into
atomic/ionized gas and drives outflows out of the computational domain. We
define the photoevaporation efficiency ($\varepsilon_{\rm ion}$) as the
fraction of the initial cloud mass turned into ionized gas, and the neutral
ejection efficiency ($\varepsilon_{\rm ej,neu}$) as the fraction ejected from
the simulation box in the molecular and atomic phases. Both
$\varepsilon_{\rm ion}$ and $\varepsilon_{\rm ej,neu}$ are measured at the end
of the simulation (see also \citealt{kim18}).

Figure~\ref{f:evej} shows $\varepsilon_{\rm ion}$ (top) and
$\varepsilon_{\rm ej,neu}$ (bottom) of the $\alpha$- (left) and $\beta$- (right)
series models. The median/minimum/maximum values are listed in
Table~\ref{t:result}. Except for strongly magnetized ($\muBcl \le 1$) clouds,
the mass loss is dominated by photoevaporation, consistent with the results of
hydrodynamic simulations by \citet{kim18}. In the $\alpha$-series models, as
$\avircl$ increases from 1 to 5, the median value of $\varepsilon_{\rm ion}$
decreases from 73\% to 48\% while $\varepsilon_{\rm ej,neu}$ increases from 17\%
to 50\%. This result can be understood as initially unbound clouds ejecting a
larger amount of neutral gas by initial turbulence and having lower ionizing
photon rate. In the $\beta$-series models, the median value of
$\varepsilon_{\rm ion}$ is higher than $70\%$ for clouds with lower
magnetization ($\muBcl \ge 2$). In the magnetically subcritical (critical) case,
however, only 13\% (35\%) of gas is photoevaporated by LyC photons and most of
outflow mass is in the atomic phase. Although magnetically subcritical clouds
have extended lifetime, the photoevaporation efficiency is low because of the low
ionizing photon rate and magnetically confined \HII\ region geometry. We also
remark that outflows in magnetically subcritical clouds are unrealistically
anisotropic; more than $\sim 80\%$ of the mass loss occurs along the $\hat z$  direction
of the background magnetic field (through the vertical  boundaries of the
computational domain).

\subsection{Timescales}\label{s:timescales}

\begin{figure*}[t!]
  \epsscale{0.9}\plotone{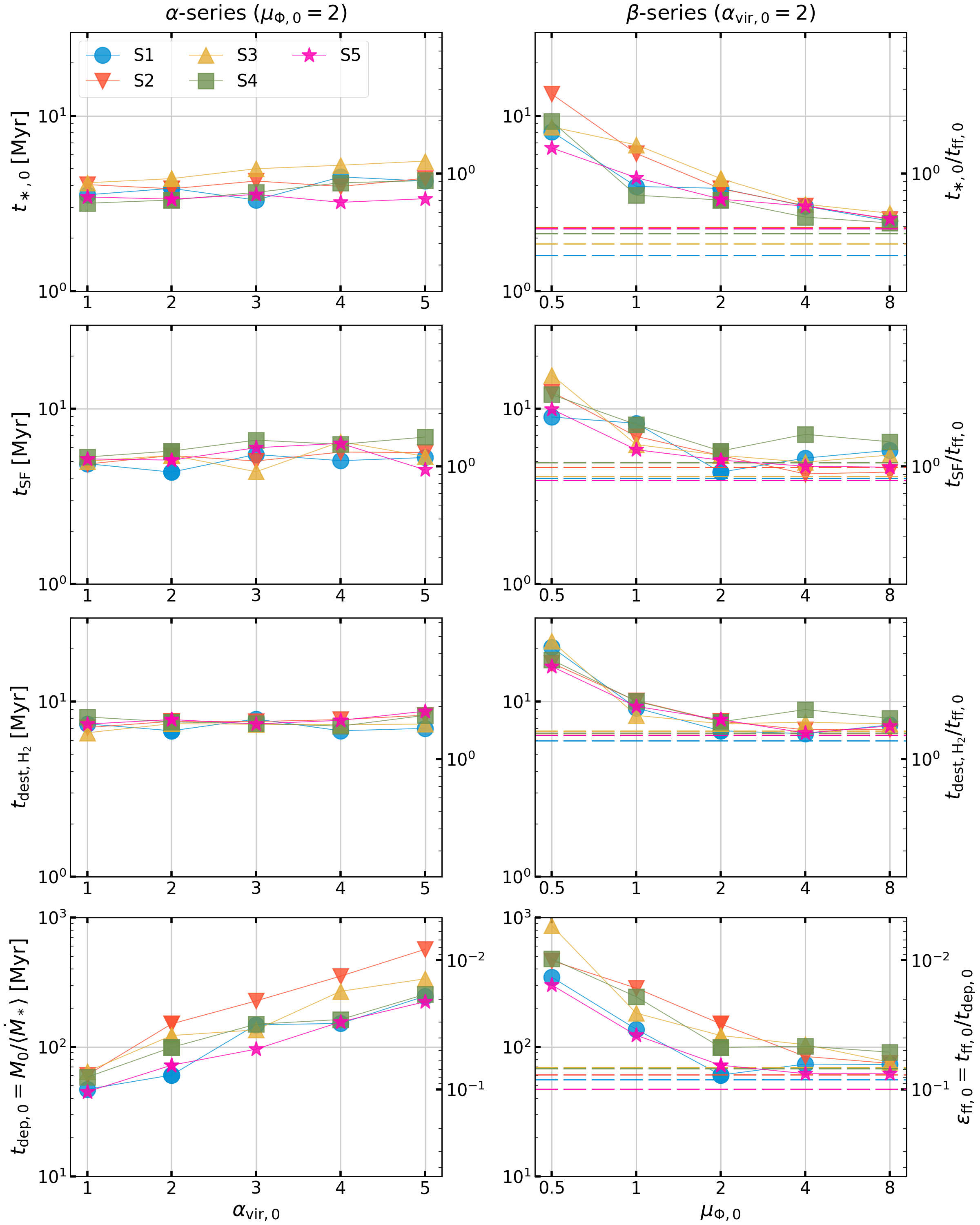}
  \caption{Evolutionary timescales with varying initial kinetic or magnetic
    energy or turbulence realization. From top to bottom, each row shows
    time of first star formation event $t_{*,0}$, 
    duration of star formation $t_{\rm SF} = t_{*,90\%} - t_{*,0}$, 
    timescale for cloud destruction $t_{\rm dest, H_2} = t_{\rm H_2,5\%} - t_{*,0}$, and 
    gas depletion time $t_{\rm dep,0} = \Mcl/\langle \dot{M}_* \rangle$. Left axes give times in
    Myr. Right axes (top three rows) show times in units of the initial
    freefall time $t_{\rm ff,0} = 4.68\Myr$, while the bottom row shows the
    inverse, equivalent to the SFE per freefall time. The
    left/right column shows the $\alpha$/$\beta$-series simulations, with
    different turbulent {\tt seed} and unmagnetized models indicated as in
    Figure~\ref{f:SFE}.}\label{f:t_all}
\end{figure*}

Our simulations allow us to characterize several evolutionary timescales of GMCs
dispersed by radiative feedback: the time of first star formation, timescales
for star formation and cloud destruction, and the gas depletion time. Similar to
the definition used by \citet{kim18}, we define the duration of star formation
as $t_{\rm SF} = t_{\rm *,90\%} - t_{\rm *,0}$. The cloud destruction time is
defined as the time for the molecular gas mass\footnote{We find that the
  destruction timescale based on the total neutral gas
  ($M_{\rm H^0} + M_{\rm H_2}$) gas mass is about $10\%$ longer than the
  destruction timescale based on $M_{\rm H_2}$, except for the model {\tt A2B05}
  for which $t_{\rm dest,neu}$ is about $30\%$ longer than $t_{\rm dest,H_2}$.}
to be reduced to $5\%$ of the initial cloud mass ($0.05\Mcl$),
$t_{\rm dest,H_2} = t_{\rm H_2,5\%} - t_{*,0}$. The gas depletion time is
calculated using the initial cloud mass and the time-averaged SFR as
$t_{\rm dep,0} = \Mcl/\langle \dot{M}_* \rangle$. This should be distinguished
from the observed gas depletion time, based on instantaneous gas mass and SFR.
We note that the ``instantaneous'' depletion time $M_{\rm gas}(t)/\dot{M}_*(t)$
will be close to $t_{\rm dep,0}$ if the SFR is relatively constant and only a
small fraction of the mass has been dispersed by feedback; from Figure~\ref{f:hst}
and Figure~\ref{f:sfhst}, these conditions are generally satisfied for our
simulations in the middle of the active star-forming phase. An inherent issue in
observations, however, is how the SFR is measured; we return to this in
Section~\ref{s:obscomp}. Finally, for the purposes of comparing to observations, we note that for constant SFR the mean age of an observed cluster in an actively star-forming cloud would be $t_{\rm SF}/3$. This assumes an equal number of stars in each age bin $\{t_{\rm SF}/N, 2t_{\rm SF}/N, \cdots, t_{\rm SF} \}$ within a given cloud and equal representation of clouds at each age, so that the luminosity-weighted average age of clusters would be $t_{\rm SF}(N+2)/(3N)$.

Figure~\ref{f:t_all} shows, from top to bottom, $t_{*,0}$, $t_{\rm SF}$,
$t_{\rm dest,H_2}$, $t_{\rm dep,0}$ of the $\alpha$- (left) and $\beta$-series
(right) models. Results for all turbulence realizations (labeled by {\tt seed})
are shown separately; median, minimum, and maximum values for each model are
summarized in Table~\ref{t:result}. In the $\alpha$-series models, the median
value of $t_{*,0}$ ranges from $3.5\Myr$ to $4.3\Myr$, equivalent to
$0.8-0.9 \tffcl$, increasing only weakly with $\avircl$. The duration of star
formation is also roughly constant (slightly increasing with $\avircl$) at
$t_{\rm SF}\sim 5$--$6 \Myr$, which is $\sim 1.1$--$1.3 \tffcl$. Cloud destruction occurs
within $\sim 1.6$--$1.8\tffcl$, a few Myrs after $t_{*,90\%}$, consistent with
the results of hydrodynamic simulations in \citet{kim18}.

In the $\beta$-series models, $t_{*,0}$ systematically increases with decreasing
$\muBcl$ while $t_{\rm SF}$ and $t_{\rm dest,H_2}$, have mild variations with
$\muBcl$ for $\muBcl \gtrsim 2$, but all timescales become significantly longer
in the magnetically critical and especially subcritical cases. For example, the
model {\tt A2B05S4} has $t_{*,0} = 9.3\Myr$, $t_{\rm SF} = 12.0\Myr$, and
$t_{\rm dest,H_2} = 17.3\Myr$ (see also Figure~\ref{f:hst}(g)).

\begin{figure*}[t!]
 \epsscale{0.9}\plotone{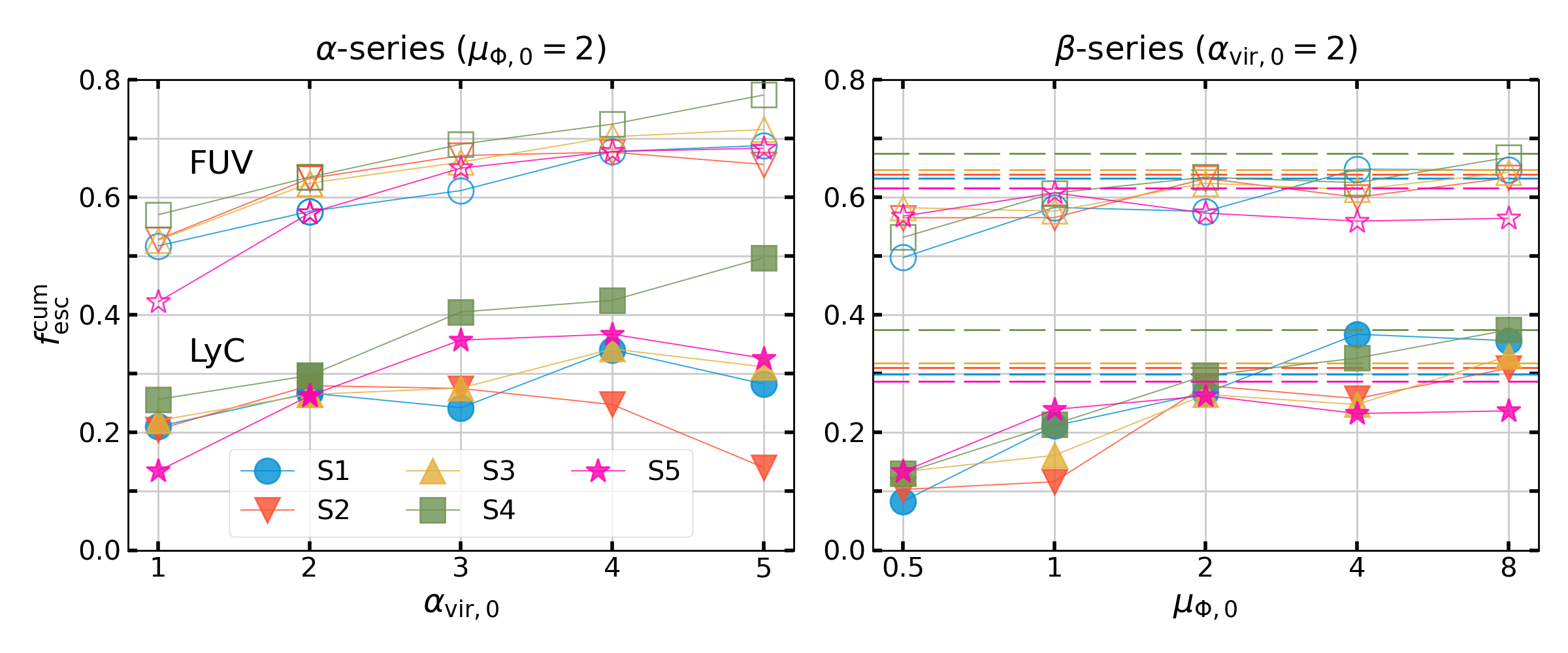}
 \caption{Cumulative escape fractions of LyC and FUV radiation for our
   $\alpha$-series (left) and $\beta$-series models (right). Different turbulent
   {\tt seed} and unmagnetized models are indicated as in
   Figure~\ref{f:SFE}.}\label{f:fesc}
\end{figure*}

The depletion time $t_{\rm dep,0}$ systematically increases with increasing
$\avircl$, with median values ranging over $58$--$255 \Myr$. The median
depletion time also increases from $61$--$464 \Myr$ from unmagnetized to
subcritically-magnetized models. The ratio $t_{\rm ff,0}/t_{\rm dep,0}$ is
equivalent to the SFE per initial freefall time of the cloud (we shall introduce the notation $\eff$ for this ratio in Equation~\eqref{e:epsff_model}).
For our models, the median value of $t_{\rm ff,0}/t_{\rm dep,0}$ increases from 1.8\% to
8.0\% as $\avircl$ decreases, and from 1.0\% to 7.7\% as $\muBcl$ increases. We
further discuss how our results on efficiency per freefall time relate to
theoretical models and observations in Section~\ref{s:epsff}.

\subsection{Escape Fraction of Radiation}\label{s:fesc}

Figure~\ref{f:fesc} shows the cumulative escape fractions of LyC and FUV
radiation measured at the end of simulations (see also Table~\ref{t:result}).
Because neutral hydrogen acts as an additional source of opacity for ionizing
radiation and most of LyC photons are emitted before the cloud destruction, the
escape fraction of LyC radiation is smaller than that of FUV radiation.

In the $\alpha$-series, median values of $f_{\rm esc,LyC}^{\rm cum}$ are in the
range $21$--$34\%$, generally increasing with $\avircl$, although there is a
significant variation with random seeds at large $\avircl$. In the
$\beta$-series, $f_{\rm esc,LyC}^{\rm cum}$ increases with decreasing magnetic
field strength, with medians in the range $13$--$33\%$. For point radiation
sources, the escape fraction is determined by the solid-angle PDF of the optical
depth as seen from the source and is higher if the width of the PDF is higher
\citep{safarzadeh16,kim19}. Therefore, the trend in the $\beta$-series is a
result of the reduced gas compressibility and the reduced width of the column density PDF in strongly magnetized clouds \citep[e.g.,][]{ostriker01}.

In all of our models, $f_{\rm esc,LyC}^{\rm cum}$ measured at $\tprime = 3\Myr$
is only $\sim 2$--$10\%$. In contrast, the hydrodynamic simulations of
\citet{kim19} found that $f_{\rm esc,LyC}^{\rm cum}=30\%$ at $\tprime = 3\Myr$
for the same cloud with $\Mcl=10^5 \Msun$, $\Rcl=20 \pc$, ${\rm seed}=1$,
$\muBcl=\infty$. This discrepancy is likely due to different treatments of the
ionizing photon production rate per stellar mass in the two simulations. In the
present simulations, $\Psi_{\rm LyC}$ decreases with time and
$\Psi_{\rm LyC}(t_{\rm age} = 3\Myr) \approx 0.5\Psi_{\rm LyC}(0)$; the
cumulative value is heavily weighted by values at early times when
$f_{\rm esc,LyC} \ll 1$. In contrast, the conversion factor adopted by
\citet{kim19} was constant with time, but depended on the total cluster mass to
account for the effects of poor sampling of the IMF at the high mass end. This
effectively made the LyC photon production rate per unit mass lower at the
earliest times (when $M_{*,\rm total} \lesssim 10^3$) compared to later time.
Thus, the luminosity-weighted escape fraction was more heavily weighted to later
times when photons escape more easily.

\section{Virial Ratios}\label{s:virial}

The virial theorem
\begin{equation}
  \frac{1}{2}\ddot{I} + \frac{1}{2}\frac{\di}{\di t} \int \rho r^2 \bm{v} \cdot d \bm{S} = 2(\mathcal{T} - \mathcal{T}_{\rm s}) + (\mathcal{B} - \mathcal{B}_{\rm s}) + \mathcal{U} \label{e:virial}
\end{equation}
describes the rate of change of a gaseous system's moment of inertia
($I = \int \rho r^2 \di V$) due to kinetic and thermal ($\mathcal{T}=\mathcal{T}_{\rm kin} + \mathcal{T}_{\rm thm}$), magnetic ($\mathcal{B}$), and gravitational ($\mathcal{U}$)
energies associated with it \citep[e.g.,][]{mckee92}. 

The terms with subscript ``s'' are integrals involving Reynolds and Maxwell stresses over a bounding surface that 
fully encloses the system of interest, given by
\begin{equation}\label{eq:Tsurf}
2\mathcal{T}_{\rm s} = \int_S \bm{r}\cdot \left( P \mathbb{1} + \rho \bm{v}\bm{v} \right) \cdot d\bm{S}    
\end{equation}
and 
\begin{equation}\label{eq:Bsurf}
\mathcal{B}_{\rm s} = - \frac{1}{4\pi}\int_S \bm{r}\cdot \left( \bm{B}\bm{B} - (B^2/2)\mathbb{1} \right) \cdot d\bm{S}.    
\end{equation} 
If the surface stresses are isotropic,
these may be evaluated as 
$2\mathcal{T}_{\rm s}=
3 [ P + \rho \sigma_{\rm 1d}^2 ]_{\rm s} V$ and 
$\mathcal{B}_{\rm s} 
= 3 [ B_{\rm 1d}^2/(8\pi)]_{\rm s} V$, where the subscript ``s'' indicates an average over the bounding surface, $V$ is the enclosed volume and ``1d'' denotes the component along any single direction.   Thus, $2\mathcal{T}_{\rm s}$  corresponds to what would be twice the sum of ``ambient'' thermal and kinetic energy within $V$, while 
$\mathcal{B}_{\rm s}$ corresponds to what would be the ``ambient'' magnetic energy within $V$.
Equivalently, $(2 \mathcal{T}_{\rm s} + \mathcal{B}_{\rm s})/V=3P_{\rm amb}$, where the total ambient pressure $P_{\rm amb}$ includes thermal, turbulent, and magnetic terms.

The virial theorem has been widely invoked to study 
cloud stucture, stability, and evolution
\citep[e.g.,][]{zweibel90,bertoldi92,mckee92,shu92,mckee99,ballesteros-paredes99,dib07,ballesteros-paredes09} and to obtain mass estimates for molecular clouds \citep[e.g.,][]{solomon87,bolatto08}.

Some useful insights can be obtained by considering simple limiting cases. When Equation~\eqref{e:virial} is averaged over an ensemble of turbulent clouds in different microstates, the terms on the left hand side would average out to zero. 
In weakly self-gravitating cases, $\mathcal{T} \approx \mathcal{T}_{\rm s}$ and $\mathcal{B} \approx \mathcal{B}_{\rm s}$, with one (or both) $\gg |\mathcal{U}|$. In this situation, 
molecular clouds would just represent overdense, UV-shielded parts of the turbulent ISM, with internal terms balancing the surface terms; the ambient material could be either diffuse molecular or atomic gas. Indeed, the study of \citet{schruba19} suggests that there are both atomic-dominated and molecule-dominated cases with $\mathcal{T}_{\rm s} \approx \mathcal{T} \gg |\mathcal{U}|$.

More generally, for an ensemble of clouds where the time-dependent terms are zero, we would have 
\begin{equation}
\frac{2\langle \mathcal{T} \rangle}{\langle {|\mathcal{U}|} \rangle  }=\frac{\langle \mathcal{T}\rangle}{\langle \mathcal{T} - \mathcal{T}_{\rm s}\rangle + \langle\mathcal{B} -\mathcal{B}_{\rm s}\rangle/2 }.\label{e:virial_bal0}
\end{equation}
If the magnetic terms are individually small, or if the surface and volume magnetic terms are comparable (which would be true for a uniform magnetic field), this ratio would be greater than unity.  Alternatively, if surface terms $\mathcal{T}_{\rm s}$ and $\mathcal{B}_{\rm s}$ are small compared to corresponding volume terms, this ratio would be less than unity.  Thus, while ``virialized'' is often taken as synonymous with having $2\mathcal{T}/|\mathcal{U}| \approx 1$, this is not true in general \citep[and does not appear to be satisfied for observed molecular clouds in many environments -- see][]{schruba19}. Rather, for an ensemble of clouds in statistical equilibrium (some forming and others dispersing) we would expect 
\begin{equation}
1 \approx \frac{2\langle \mathcal{T} -\mathcal{T}_{\rm s}\rangle   + \langle\mathcal{B} - \mathcal{B}_{\rm s}\rangle }{\langle |\mathcal{U}| \rangle} \label{e:virial_bal1}
\end{equation}
or
\begin{equation}
\frac{2\langle \mathcal{T}\rangle}{\langle |\mathcal{U}| \rangle} \left( 1 + \frac{\langle \mathcal{B} \rangle}{2 \langle \mathcal{T}\rangle} \right) \approx 1 + \frac{2\langle \mathcal{T}_s \rangle + \langle \mathcal{B}_s \rangle}{\langle |\mathcal{U}| \rangle}\,. \label{e:virial_bal2}
\end{equation}
The analogous expressions without angle brackets would also be true for an individual cloud if the ``acceleration'' terms on the left-hand side of Equation~\eqref{e:virial} are small.

The ratios $2\langle \mathcal{T}\rangle/\langle |\mathcal{U}| \rangle$ and $(2\langle \mathcal{T}_s \rangle + \langle \mathcal{B}_s \rangle)/\langle |\mathcal{U}| \rangle$ in Equation~\eqref{e:virial_bal2} can be estimated observationally based on the measurements of cloud-scale internal pressure and large-scale pressure that is in balance with vertical weight of the ISM.
First, the ratio $2\langle \mathcal{T}\rangle/\langle |\mathcal{U}| \rangle$ is proportional to the ratio between the velocity dispersion squared and 
the gravitational potential 
of molecular gas, both averaged over cloud populations 
weighted by the CO intensity. It is also proportional to the ratio between the internal (turbulent plus thermal) pressure and self-gravitational weight of molecular gas \citep[e.g.,][]{schruba19}.\footnote{Specifically, for spherical, uniform-density clouds of surface density $\Sigma_{\rm mol}=(2/3)\rho_{\rm mol}D$ filling a beam of diameter $D$ (so that the effective volume $V=\pi D^3/6$),
$2 \langle \mathcal{T}\rangle = 3V\langle P_{\rm turb} + P_{\rm th} \rangle $ and 
$\langle \mathcal{U}\rangle =  (6 V/5)\langle \mathcal{W}^{\rm self}_{\rm cloud} \rangle$, where 
$\langle P_{\rm turb} + P_{\rm th} \rangle = \langle 3\Sigma_{\rm mol}(\sigma_{\rm 1d}^2 + c_{\rm s,iso}^2) \rangle / (2D)$ and $\langle \mathcal{W}^{\rm self}_{\rm cloud} \rangle = (3\pi/8) G \langle \Sigma_{\rm mol}^2 \rangle$ is the self-gravitational contribution to the internal cloud weight.  We note that if the gas and stellar distributions outside of a beam are effectively uniform, they do not contribute to the gravitational energy $\langle \mathcal{U}\rangle$, but they do contribute additional terms beyond $\mathcal{W}^{\rm self}_{\rm cloud}$ to the total weight experienced by a cloud \citep{sun20a}.
In the realistic case that clouds do {\it not} fill the beam, but instead have typical diameter $d$, the true value of $|\mathcal{U}|$ will be a factor $D/d$ larger than the beam-averaged estimate, while $\mathcal {T}$ is unaffected.}
Second, to the extent that surface stresses are isotropic when ensemble averaged, the surface term $\langle 2\mathcal{T}_{\rm s} + \mathcal{B}_{\rm s}\rangle$ is equal to $3 V \langle P_{\rm amb}\rangle$  (see Equations~\eqref{eq:Tsurf}--\eqref{eq:Bsurf} and following).
Numerical simulations show that the total ambient pressure at the midplane is generally in good agreement with vertical dynamical equilibrium predictions for the large-scale ISM, so standard expressions based on the gas surface density and the stellar density may be used to estimate $P_{\rm amb}$ \citep[e.g.,][]{KKO2013,KO2015}.
Given the knowledge of these quantities, one may infer the fractional contribution from magnetic support on cloud sales, $\mathcal{B}/(2\mathcal{T})$ or $\mathcal{B}/|\mathcal{U}|$.

While it is observationally very difficult, if not impossible, to determine the
individual terms in Equation~\eqref{e:virial} with precision
\citep[e.g,][]{singh19}, observers have traditionally employed the simple
kinetic virial parameter $\avir = 5 \sigma_{\rm 1d}^2 R/(G M)$ based on
estimates of size, velocity dispersion, and mass to assess whether a cloud is
gravitationally bound ($\avir < 2$) or in virial equilibrium
($\avir \approx 1$). Strictly speaking, these diagnostics using $\avir$ apply to
a steady-state homogeneous, spherical, non-magnetized cloud with
no external forces acting on it. 
In particular, simple approaches treat clouds
as isolated, while in reality GMCs are often in close proximity to each other
and/or surrounded by cold \ion{H}{1} envelopes with significant mass. 
The former would reduce the magnitude of the gravitational energy because the effective
zero of the gravitational potential lies near clouds rather than at infinite
distance, while the latter would make the effective potential deeper \citep[e.g.,][]{ballesteros-paredes06}. By analyzing the dynamical state of structures found in kpc-scale simulations of the multiphase, star-forming ISM with self-consistent star formation and supernova feedback, \citet{mao20} found that $\avir$ is often inadequate as a measure of the gravitational boundedness of a structure, because tidal gravity, magnetic terms, and other effects can be large.

While in the present simulations there are no tidal forces, our model clouds otherwise have quite realistic physics and very high resolution. It is therefore interesting to
test how well the simple estimate $\avir$ represents the true boundedness of our
simulated clouds. To do this, we directly measure the virial parameter
accounting for kinetic, thermal, and magnetic energies and compare with $\avir$
computed via the traditional estimate. 

We calculate the
energies of the neutral gas in the cloud as
\begin{equation}
  \mathcal{T}_{\rm kin} = \frac{1}{2}\int \rho v^2 \Theta \di V
  \,,\label{e:Ekin}
\end{equation}
\begin{equation}
  \mathcal{T}_{\rm thm} = \frac{3}{2}\int P \Theta \di V\,, \label{e:Ethm}
\end{equation}
\begin{equation}
  \mathcal{B} = \frac{1}{8\pi}\int B^2 \Theta
  \di V\,,\label{e:Emag}
\end{equation}
\begin{equation}
  \mathcal{U} = -\int \rho \bm{r} \cdot \nabla \Phi \Theta \di
  V\,. \label{e:Egrav}
\end{equation}
\citep[e.g.,][]{mckee92}. The gravitational potential $\Phi$ in Equation~\eqref{e:Egrav} includes contributions from not only the neutral cloud gas but also ionized gas and stars in the computational domain. 
Because $\Phi$ includes ``external'' terms, we cannot re-express $\mathcal{U}$ as $-(1/2)\int \rho \Phi dV$; the latter expression would only apply if we were solely considering the self-gravitational energy of the gas cloud.

In the analysis of our simulations, we shall ignore the surface terms 
$\mathcal{T}_{\rm s}$ and $\mathcal{B}_{\rm s}$.
Given the low density and high temperature of the ambient medium in our simulations, in fact $\mathcal{T}_{\rm kin,s}$ is much smaller than $\mathcal{T}_{\rm kin}$ and $\mathcal{T}_{\rm thm,s}$ is comparable to $\mathcal{T}_{\rm thm}$. 

In principle, the magnetic energy associated with the cloud $\mathcal{B} - \mathcal{B}_{\rm s}$ can be calculated as the difference between the total magnetic energy of the cloud and magnetic energy of the ambient field in the absence of the cloud \citep{mckee92}.
However, our simulations impose an initially uniform magnetic field rather than e.g., some kind of hourglass geometry (which we avoided since it would be arbitrary). As such, 
the ambient field strength can be larger than realistic values.  While this does not affect dynamics within the cloud itself, it would imply an unrealistically large measured $\mathcal{B}_{\rm s}$ compared to $\mathcal{B}$.
If we had begun with a larger-scale simulation that followed cloud formation, the ambient magnetic field would be lower than the cloud value, and 
$\mathcal{B}_{\rm s}$ would be small compared to $\mathcal{B}$.  In particular, if magnetic flux is conserved in the formation of a cloud of size $R_{\rm cl}$ from a region of crossection $\sim R_d^2$ in the diffuse ISM, the volume magnetic term would exceed the surface term by a factor of order $R_{\rm d}/R_{\rm cl}\gg 1$.  
Therefore, our $\mathcal{B}$ can be considered as an approximation of (and an upper limit to) the true magnetic term.

With the above definitions, the more exact equivalent of the traditional kinetic virial parameter defined (for the initial cloud) in Equation~\eqref{eq:avirdef} is
\begin{equation}\label{e:alphavt}
  \avirt \equiv \frac{2\mathcal{T}_{\rm kin}}{|{\cal U}|};
\end{equation}
we use the tilde to emphasize the distinction.

The third row in Figure~\ref{f:hst2} plots the evolution of kinetic, thermal,
magnetic, and gravitational energies of the cloud in models {\tt A2B2S4}, {\tt
  A5B2S4}, and {\tt A2B05S4}. In the fiducial model, the magnitude of the
gravitational energy $\mathcal{U}$ remains very close to the initial value $3G\Mcl^2/(5\Rcl)$
at early times. At late times ($t \gtrsim 5\Myr$), it begins to decrease as gas
is photoevaporated and dispersed. A naive estimate $3GM_{\rm neu}^2/(5 R)$ for
the gravitational energy using the half-mass radius
($R=2^{1/3} R_{50\%}$) is also shown using a blue dashed curve; the true
gravitational energy is larger than this at late times. This is because the simple estimate ignores  substructure in the mass distribution, which is characterized
by a few widely-separated clusters of gas clumps (which themselves contain
closely packed dense cores), leading to the overestimation of the effective size
of the system at late times as the cloud disperses.

The evolution of the kinetic energy $\mathcal{T}_{\rm kin}$ follows that of the velocity dispersion
(second row in Figure~\ref{f:hst2}), although the increase after $t \sim 5\Myr$
is less significant due to mass loss. The magnetic energy is $2 G M_0^2/(3R_0 \muBcl^2)$ initially, then increases slightly due to turbulence-induced magnetic fields, and saturates at a level that is about half
the initial gravitational energy, until it decreases due to gas dispersal at late times. The thermal energy is unimportant at early times, but later becomes comparable to $\mathcal{B}$ and $|\mathcal{U}|$ due to the FUV heating.\footnote{Also, at late times the mean temperature is enhanced due to warm, partially-ionized (but $\xHII < 0.5$) gas on the cloud surface.}

The other two models show similar evolutionary trends for gravitational and kinetic terms. It is worth noting
that $|\mathcal{U}|$ in ${\tt A5B2S4}$ decreases starting from $t=0$ due to the
initial expansion of the unbound cloud.  The magnetic energy in the subcritical model ${\tt A2B05S4}$ is larger than the other terms throughout the simulation.

\begin{figure*}[t!]
  \epsscale{1.0}\plotone{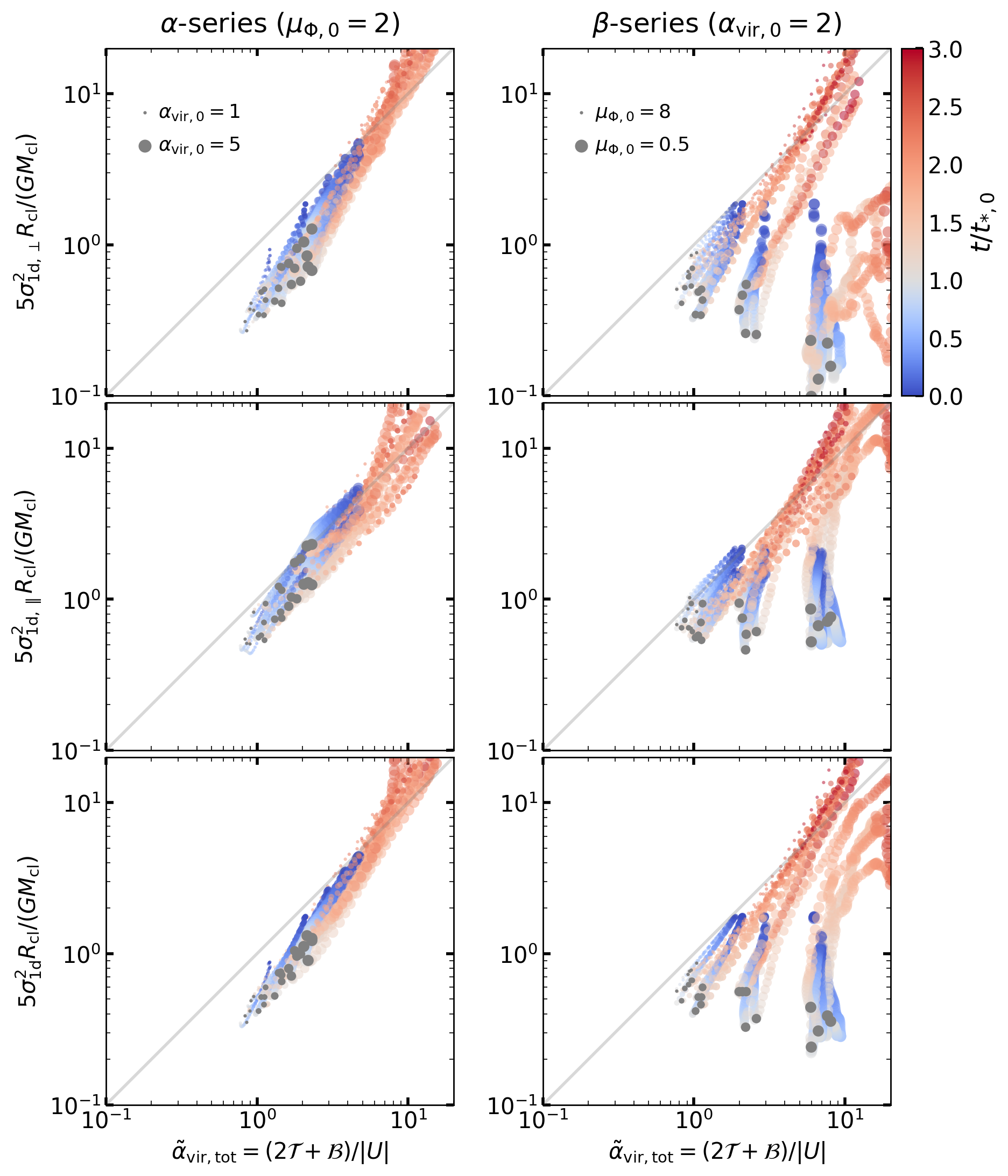}
  \caption{Comparison of the virial ratio
    $\avirttot=(2\mathcal{T} + \mathcal{B})/|\mathcal{U}|$ and the traditional
    kinetic virial parameter defined as
    $\avir= 5\sigma_{\rm 1d}^2 R/(5 M_{\rm neu})$ in the $\alpha$-series (left)
    and $\beta$-series (right) models. Each circle represents one snapshot and
    the color indicates the time of the snapshot normalized by the time of the
    first star formation. The grey circles mark values at $t=t_{*,0}$. The size
    of circles increases with increasing $\avircl$ and $B_{z,0}$ (decreasing
    $\muBcl$). Rows show the traditional $\avir$ calculated with velocity
    dispersion perpendicular (top) and parallel (middle) to the 
    background magnetic field
    and averaged along all three Cartesian axes (bottom).}\label{f:avir}
\end{figure*}

In the fourth row of Figure~\ref{f:hst2}, we show as gray lines the ``observed'' virial
parameter $\avir$ computed from the instantaneous $M_{\rm neu}$ and
$R = 2^{1/3}R_{50\%}$, and three different measures of the velocity dispersion:
$\sigma_{\rm 1d,\perp}$, $\sigma_{\rm 1d,\parallel}$, and $\sigma_{\rm 1d}$. We
compare to the ratio $\avirt=2\mathcal{T}_{\rm kin}/|\mathcal{U}|$ (cyan). The
kinetic virial parameter $\avir=5\sigma_{\rm 1d}^2R/(G M_{\rm neu})$ (solid gray)
agrees very well with $\avirt=2\mathcal{T}_{\rm kin}/|\mathcal{U}|$ at early
times, but becomes larger than $\avirt$ when cloud disruption is underway, as
the true gravitational binding energy is larger than the simple estimate
$3GM_{\rm neu}^2/(5R)$.

Because clouds are supported jointly by thermal energy, turbulence, and magnetic
fields, it is natural to consider virial parameters including additional terms.
The virial parameter including both nonthermal and thermal gas motions is
$2\mathcal{T}/|\mathcal{U}|$. The virial parameter accounting for the total
energy is
\begin{equation}\label{eq:alphavtot}
  \avirttot = \frac{2\mathcal{T} + \mathcal{B}}{|\mathcal{U}|}\,.
\end{equation}
The time evolution of $2\mathcal{T}/|\mathcal{U}|$ (purple) and
$(2\mathcal{T} + \mathcal{B})/|\mathcal{U}|$ (black) are plotted in the bottom
row of Figure~\ref{f:hst2}. The thermal terms do not make a large difference,
but the magnetic term can lead to 
a factor of two increase in $\avirttot$ compared to $\avirt$, or even more for the most strongly magnetized model.

The scatter plots in Figure~\ref{f:avir} show the relationship between the
traditional kinetic virial parameter estimate ($\avir$) and total virial ratio
($\avirttot$) for all of the simulation snapshots taken before $t_{*,90\%}$ in
the $\alpha$- (left) and $\beta$-series (right) models. From top to bottom, the
ordinate shows the kinetic virial parameter computed using
$\sigma_{\rm 1d,\perp}$, $\sigma_{\rm 1d,\parallel}$, and $\sigma_{\rm 1d}$. The
symbols are colored by time normalized to $t_{*,0}$. Grey dots mark the values
at $t=t_{*,0}$.

\begin{figure*}[t!]
  \epsscale{1.0}\plotone{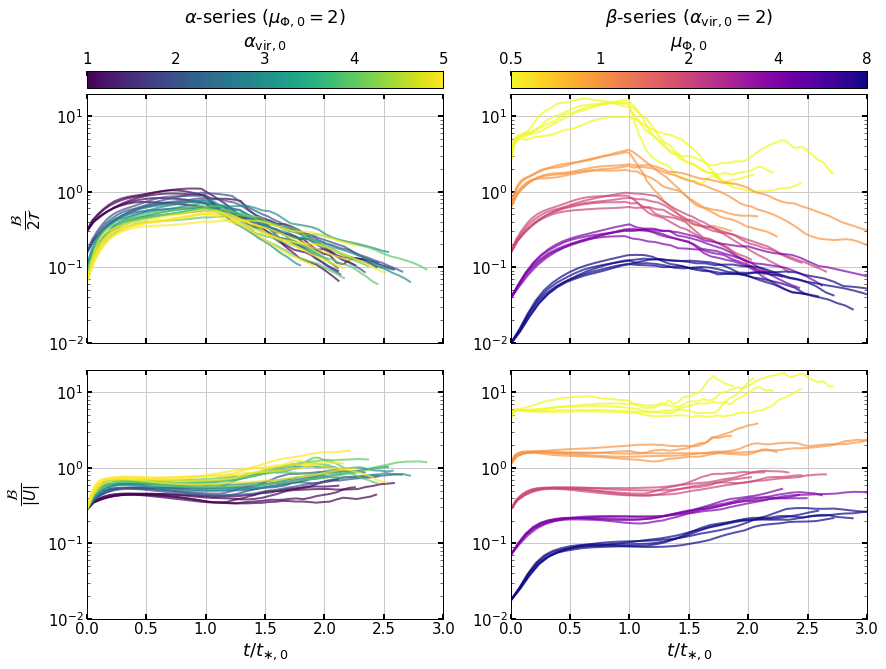}
  \caption{Time evolution of the ratio of the magnetic energy ($\mathcal{B}$) to twice the kinetic energy ($2\mathcal{T}$; top) and to the gravitational ($|\mathcal{U}|$) energy until $t_{*,90\%}$ in all of the $\alpha$- (left) and $\beta$-series (right) models. Models with lighter colors have stronger initial turbulence ($\alpha$-series) or magnetic field strength ($\beta$-series). The time is normalized by the time of first star formation $t_{*,0}$.}\label{f:Bratio}
\end{figure*}

Except for the strongly magnetized models, $\avirttot$ initially decreases with time due to the decay of turbulence, and then turns around after the onset of radiative feedback. 
Remarkably, the total virial parameter at $t=t_{*,0}$ in the $\alpha$-series falls in a fairly narrow range $\avirttot \sim 0.9$--$2.1$, suggesting that small-scale collapse starts when the cloud reaches a near-virial state.  Except for the subcritical clouds {\tt A2B05}, $\avirttot \sim 0.9$--$2.2$ at $t=t_{*,0}$ in the $\beta$-series as well. In contrast, the magnetically subcritical clouds ({\tt A2B05}) remain globally unbound throughout their evolution.

The results of the $\alpha$-series models suggest that, for moderate magnetic
field strength ($\muBcl=2$), $\avir$ can 
slightly underestimate or more significantly overestimate the true gravitational boundedness depending on the relative orientation between the mean
magnetic field and the observer's line of sight. 
The bottom panels of
Figure~\ref{f:avir} suggest that, even with the correct estimate of kinetic
energy, the kinetic virial parameter would systematically overrate 
the
boundedness due to the neglect of magnetic support. We find that the median value of the ratio $\avirttot/\avir$ for the $\alpha$-series ranges between $1.48$--$1.65$.

The results of the
$\beta$-series models (the bottom right panel in Figure~\ref{f:hst2} and the top
right panel in Figure~\ref{f:avir}) suggest that the use of the traditional
kinetic virial parameter can significantly underestimate  the kinetic energy in
magnetically critical and subcritical clouds when the line of sight is
orthogonal to the field direction. 
Still, except for extreme magnetizations and
orientations, the traditional $\avir$ estimates boundedness within a factor of
$\sim 2$.  We find that the median value of the ratio $\avirttot/\avir$ for the $\beta$-series except for {\tt A2B05}  is $1.07$--$1.94$. For the magnetically subcritical clouds, the median value of $\avirttot/\avir$ is $5.61$.

To quantify the relative importance of magnetic support, in Figure~\ref{f:Bratio} we show the time evolution of the magnetic energy divided by twice the kinetic (top) or the gravitational (bottom) energy in the $\alpha$- (left) and $\beta$-series (right) models. In all models, the ratio $\mathcal{B}/(2\mathcal{T})$ increases at early times (due to kinetic turbulence decay and magnetic turbulence growth),  and decreases at late times due to feedback-driven dispersal; the ratio $\mathcal{B}/|\mathcal{U}|$ exhibits a mild temporal variation (see also third row of Figure~\ref{f:hst2}). In the $\alpha$-series models, 
$\mathcal{B}/(2\mathcal{T})$ ($\mathcal{B}/|\mathcal{U}|$) falls in a range $\sim 0.22$--$1.10$ ($0.34$--$1.09$) for $0.5 \le t/t_{*,0} \le 1.5$. The right panels based on the $\beta$-series show that the importance of magnetic support widely varies depending on the initial magnetic field strength; $\mathcal{B}/(2\mathcal{T})$ can be as large (small) as 17 (0.07) in strongly (weakly) magnetized clouds for $0.5 \le t/t_{*,0} \le 1.5$.

We emphasize that the above results are for isolated clouds; tidal forces will
reduce the gravitational energy, which would increase $\avirttot$. This would
enhance the discrepancy between the traditional virial estimate and the true
boundedness shown in the bottom panels of Figure~\ref{f:avir} and the ratio $\mathcal{B}/|\mathcal{U}|$ shown in the bottom panels of Figure~\ref{f:Bratio}.

\section{SFE per Freefall Time}\label{s:epsff}

\begin{figure}[t!]
  \begin{center}
  \epsscale{1.0}\plotone{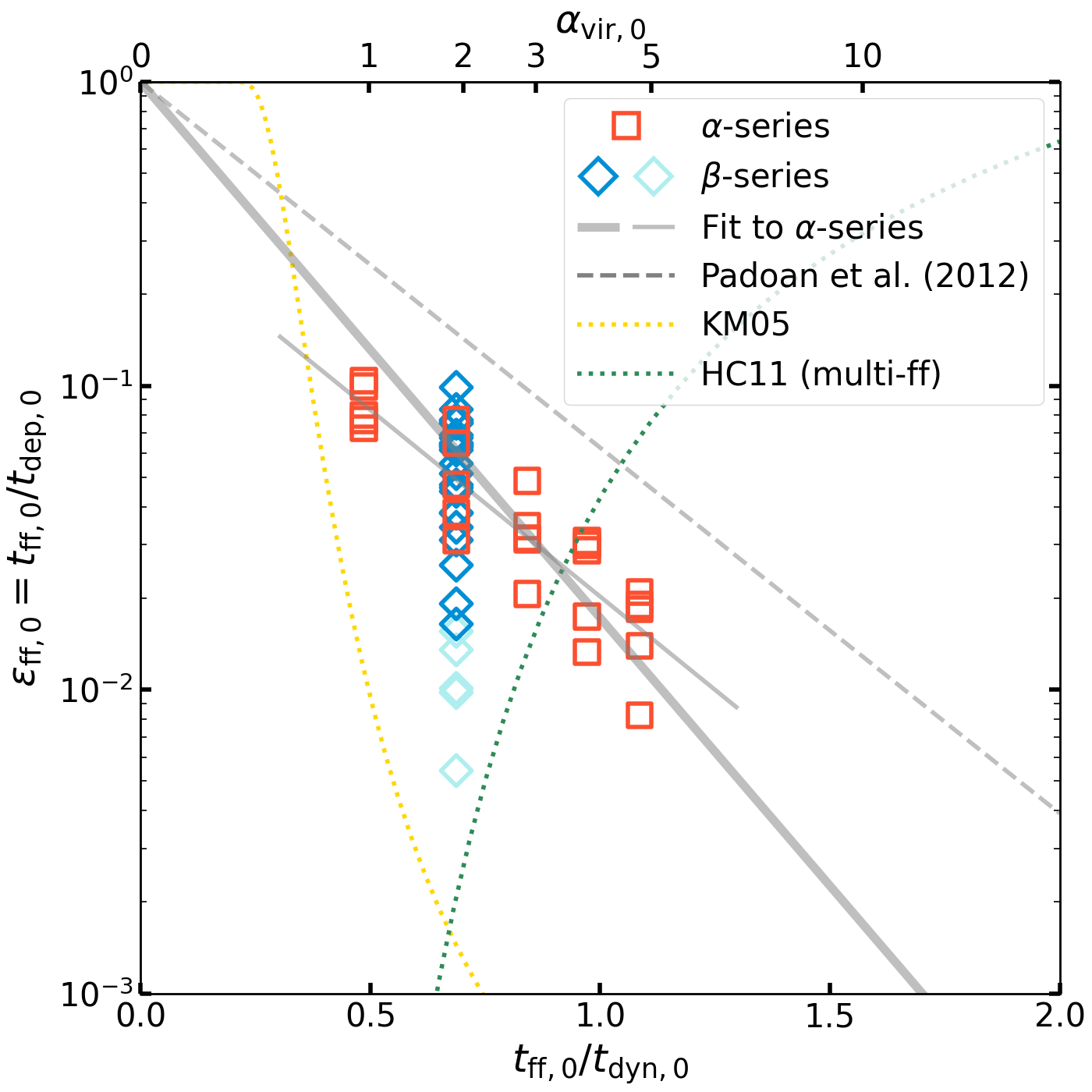}
  \caption{Time-averaged SFE per free fall time defined as
    $\eff = t_{\rm ff,0}\langle \dot{M}_*
    \rangle/\Mcl = t_{\rm ff,0}/t_{\rm dep,0}$ as a function of $\avircl$ (top $x$-axis) or the ratio
    between the initial freefall and dynamical times (bottom $x$-axis) in the $\alpha$-series (red
    squares) and the $\beta$-series (diamonds in dark blue for critical critical/supercritical and light blue for subcritical) models. The
    numerical result 
    $\exp(-1.6\sqrt{3} t_{\rm ff,0}/t_{\rm dyn,0})$ found by \citet{padoan12} is
    shown as a gray dashed line. The thick gray solid line shows the fit to the
    $\alpha$-series simulations $\eff=\exp(-4.06 t_{\rm ff,0}/t_{\rm dyn,0})$. 
    The thin gray solid line segment shows the results of a two-parameter fit $\eff=0.34\exp(- 2.83 t_{\rm ff,0}/t_{\rm dyn,0})$. 
    The dotted lines show the predictions from the \citeauthor{krumholz05} and \citeauthor{hennebelle11} analytic models (see text for  details).}\label{f:epsff}
  \end{center}
\end{figure}

The freefall time $t_{\rm ff} = \sqrt{3\pi/(32G\rho)}$ is a natural
characteristic timescale in a self-gravitating gaseous system, representing the
shortest possible timescale for gravitational collapse at a given mean density.
In common practice, the SFR is expressed as
\begin{equation}
  \dot{M}_* \equiv \varepsilon_{\rm ff} \frac{M_{\rm gas}}{t_{\rm ff}},
\end{equation}
which defines the SFE per freefall time $\varepsilon_{\rm ff}$; the freefall
time is evaluated at the mean density of the system \citep[e.g.,][]{krumholz05}.
Since the SFR is also defined in terms of the depletion time,
$\dot{M}_*\equiv M_{\rm gas}/t_{\rm dep}$, these are related by
$\varepsilon_{\rm ff} = t_{\rm ff}/t_{\rm dep}$. A cloud collapsing under its
own gravity without any hindrance would form stars at maximal efficiency
($\varepsilon_{\rm ff} \approx 1$, $t_{\rm dep}\approx t_{\rm ff}$). A wealth of
observational studies however suggest that both galactic and cloud-scale star
formation proceeds much slower than the freefall rate
($\varepsilon_{\rm ff} \ll 1$ or $t_{\rm dep} \gg t_{\rm ff}$)
\citep[e.g.,][]{zuckerman74,williams97,krumholz07,
  kennicutt12,evans14,lee16,barnes17,leroy17,ochsendorf17,utomo18}.

In our simulations, we measure the SFE per freefall time using the mass and
freefall time of the cloud at $t=0$ and the time-averaged SFR
$\langle \dot{M}_* \rangle$ as
\begin{equation}
\eff = \tffcl \frac{\langle \dot{M}_*
    \rangle}{\Mcl} = \frac{\tffcl}{t_{\rm dep,0}}\,. \label{e:epsff_model}
\end{equation}
This is similar to the approach adopted by \citet{padoan12,raskutti16}. Since
$t_{\rm ff,0}$ is the same ($4.68\Myr$) for all of the present models, $t_{\rm dep,0}$ and
$\varepsilon_{\rm ff,0}$ are inversely proportional to each other.

The right $y$-axes in the bottom row of Figure~\ref{f:t_all} indicates that
$\eff$ ranges between $0.008$ and $0.1$ in the $\alpha$-series and between
$0.006$ and $0.1$ in the $\beta$-series, with scatter of $0.3$--$0.5\,{\rm dex}$
for each model. Note that the ratio between $\varepsilon_{\rm *}$ and
$\eff$ corresponds to the duration of star
formation in units of $\tffcl$ (second row in Figure~\ref{f:t_all}), which is in
the range $\sim 1$--$2$ (except for the subcritical models).

Our simulations show that $\eff$ decreases when the virial parameter increases,
which is qualitatively consistent with some previous work. In the rest of this
section, we study more quantitatively the relationship between the SFE per
freefall time and the virial parameter in comparison to other simulations,
theoretical models, and observations.

\subsection{Comparison with Simulations and Theory}

In Figure~\ref{f:epsff}, we plot $\eff$ of all models against the ratio
$t_{\rm ff,0}/t_{\rm dyn,0} = \sqrt{\pi^2/40}\alpha_{\rm vir,0}^{1/2}$, where
$t_{\rm dyn,0} \equiv \Rcl/\sigma_{\rm 1d,0}$ is the dynamical time (or
turbulent crossing time) at $t=0$. 
The $\alpha$-series results locally follow a trend $\eff = 0.34 \exp(- 2.83 t_{\rm ff,0}/t_{\rm dyn,0})$; this local fit is shown as a thin grey solid line in Figure~\ref{f:epsff}.

The exponential decline of $\eff$ with increasing $t_{\rm ff,0}/t_{\rm dyn,0}$ is qualitatively consistent with the findings of \citet{padoan12}. Motivated by results from their adaptive-mesh refinement simulations
of star formation with driven turbulence in a periodic box (see below), they proposed  a functional form  
$\eff=D \exp \left(-C t_{\rm ff,0}/t_{\rm dyn,0} \right)$, where $D$ is the core-to-star conversion efficiency on small scales (from protostellar outflow interactions), and $C$ is set by cloud-scale processes.  We 
fit the $\alpha$-series models to this form, adopting a fixed value of $D=1$ because all locally collapsing gas is converted into stars in our simulations.  We find a best-fit
value $C=4.06$; this is shown as a thick grey solid line in Figure~\ref{f:epsff}.\footnote{We also found that a two-parameter fit $\exp \left( -c(t_{\rm ff,0}/t_{\rm dyn,0})^d \right)$ with $c=3.90$ and $d=1.07$ well describes the $\alpha$-series models (not shown).}

The \citet{padoan12} set of simulations covered a range of 3D
sonic Mach number ($\mathcal{M}=10,\,20$), Alfv\'{e}n Mach number ($1.25 \le \mathcal{M}_A \le 33$), and 
timescale ratio $0.23 \le t_{\rm ff}/t_{\rm dyn} \le 1.78$ (or $0.2 \le \avir \le 13$), and 
they measured $\varepsilon_{\rm ff}$ from the slope of a linear fit between 
$M_*(t)/M_0$ and $t/t_{\rm ff,0}$ (the same as our procedure). They found that $\varepsilon_{\rm ff}$ has
weak dependence on $\mathcal{M}$ and $\mathcal{M}_{\rm A}$, and 
fitting an exponential yields $C=1.6\sqrt{3}=2.77$ for the slope (with $D=1$ fixed)
\footnote{
The factor of $\sqrt{3}$ comes in because
  \citet{padoan12} defined the dynamical timescale as
  $t_{\rm dyn}^{\prime} = L/(2\sigma_{\rm 3D})$, where $L$ is the box size;
  taking $L = 2\Rcl$ and $\sigma_{\rm 3D}=\sqrt{3} \sigma_{\rm 1d}$ gives $t_{\rm dyn}^{\prime} = t_{\rm dyn,0}/\sqrt{3}$.},
which is shown as a gray dashed line in Figure~\ref{f:epsff}. 
Our best-fit
value $C=4.06$ is steeper than $2.77$ found by \citet{padoan12}, 
but we consider it quite interesting that
a similar exponential dependence of
$\eff$ on $\avircl^{1/2}$ holds true in global cloud as well as periodic-box simulations, and
for realistic feedback as well as driving in $k$-space.

To explain the low observed SFR, a class of theoretical models has been
proposed to describe how (magnetized) turbulence regulates SFR (e.g.,
\citealt{krumholz05,padoan11,hennebelle11}; see also the reviews by
\citealt{federrath12} and \citealt{padoan14}) In brief, these models assume that in the (magnetized) supersonically turbulent clouds where star formation takes place, the distribution of mass with density follows a lognormal PDF:
$p(s) = (\sqrt{2\pi}\sigma_s)^{-1}\exp\left[- (s - s_0)^2/(2 \sigma_s^2)
\right]$, where $s = \ln (\rho/\rho_0)$, $s_0 = \sigma_s^2/2$, $\rho_0$ is the
mean density, and $\sigma_s$ increases at higher Mach number $\mathcal{M}$, with
some reduction for stronger magnetic fields.

The collapse is assumed to occur in regions where the support by thermal plus
turbulent pressure cannot overcome gravity, and it is argued that this leads to
a critical density for star formation, with all gas at  $\rho > \rho_{\rm crit}$ collapsing. Various different forms have
been proposed for $\rho_{\rm crit}$ as a function of $\alpha_{\rm vir}$,
$\mathcal{M}$, and $\beta$ (see Table~1 in \citealt{federrath12}). The fraction
of mass undergoing collapse per freefall time (hence SFE per freefall time)
can then be obtained by integrating the mass PDF above the critical density. A
``multi-freefall'' factor $t_{\rm ff}(\rho)/t_{\rm ff}(\rho_0)$ is included
inside the integral in some models because the freefall time of gas varies with density.

We compare our results with the analytic models of SFR. Of the six models
presented in Table 1 of \citet{federrath12}, we consider the
\citeauthor{krumholz05} model without the multi-freefall factor (KM), and the
multi-freefall version of the \citeauthor{hennebelle11} model (multi-ff HC),
both of which are extended to include the effect of magnetic field. In both models, we have assumed fixed values of 
the turbulence forcing parameter $b=0.5$ and plasma beta $\beta = 0.02$ and $\epsilon_{\rm core}/\phi_t=1$, and author-favored values
of fudge factors $\phi_x = 1.12$ for the KM model and $y_{\rm cut} = 0.1$ for
the multi-ff HC model, respectively (see \citealt{federrath12}).
Because in the $\alpha$-series models the Mach number and $\avircl$ vary together rather than independently of each other (see Table~\ref{t:param}), for the analytic model comparisons we   assume $\mathcal{M} = 13.5\alpha_{\rm vir,0}^{1/2}$. These predictions are shown as dotted lines in Figure~\ref{f:epsff}.

The qualitative prediction from the KM model agrees with our simulation results, although the quantitative results differ.
The KM model predicts $\rho_{\rm crit} \propto \avir \mathcal{M}^2$, which for  $\mathcal{M} = 13.5\alpha_{\rm vir,0}^{1/2}$ yields  $\rho_{\rm crit} \propto \avir^2$.
Higher overdensity is required for collapse at higher $\avir$ because the importance of turbulent stress, which tends to tear apart structures, increases relative to gravity.
After integrating the lognormal, the end result is that $\varepsilon_{\rm ff}$ decreases with increasing $\avir$. 
By contrast, the HC multi-ff model predicts $\rho_{\rm crit} \propto \avir \mathcal{M}^{-2}$, which with $\mathcal{M} = 13.5\alpha_{\rm vir,0}^{1/2}$  becomes  $\rho_{\rm crit} = {\rm const}$. 
Therefore, while the dependence of $\varepsilon_{\rm ff}$ on $\avir$ is similar to the KM model for fixed $\mathcal{M}$, the behavior is quite different for $\mathcal{M} \propto \alpha_{\rm vir,0}^{1/2}$. Since the lognormal PDF width increases at higher $\alpha_{\rm vir,0}^{1/2}$, at fixed $\rho_{\rm crit}$ the fraction of gas that is above-threshold would increase, such that $\varepsilon_{\rm ff}$ increases monotonically with $\avircl$.

\subsection{Comparison with Observations}\label{s:obscomp}

\begin{figure*}[t!]
  \epsscale{1.0}\plotone{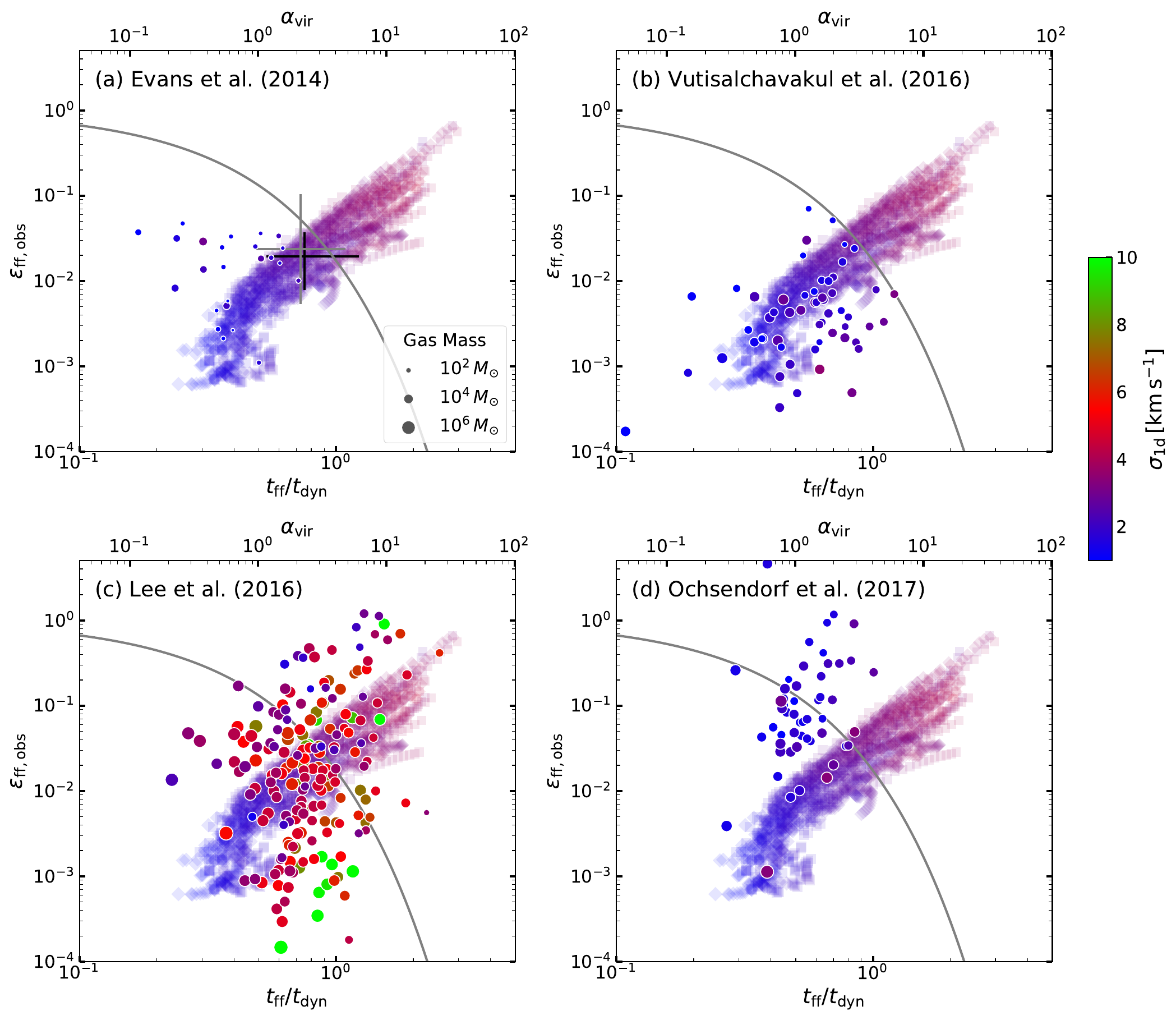}
  \caption{Relation between instantaneous
    $t_{\rm ff}/t_{\rm dyn} = (\pi^2/40)^{1/2}\avir^{1/2}$ {\it vs.}
    $\varepsilon_{\rm ff,obs}$ in simulations (squares and diamonds; $\alpha$-
    and $\beta$-series models) and observations (circles). For simulations, we
    use the instantaneous mass, size, and velocity dispersion of the gas cloud
    and cluster's bolometric UV luminosity. We show snapshots in the time
    interval $t_{*,0} < t < t_{*,90\%}$. The circles in different panels show
    data from different observational studies: (a) low-mass, nearby clouds
    compiled by \citet{evans14}; (b) molecular clouds in the Galactic Plane with
    massive SFR $M_* > 5\Msun \Myr^{-1}$
    \citep{vutisalchavakul16}. (c) GMCs with associated star-forming complexes
    in the Milky Way \citep{lee16}. (d) GMCs with associated star-forming
    complexes in the Large Magellanic Cloud \citep{ochsendorf17}. All data
    points are colored by one-dimensional velocity dispersion $\sigma_{\rm 1d}$
    and their sizes denote gas mass. 
    As in Figure~\ref{f:epsff}, the solid gray line shows the fit to the $\alpha$-series models.
    In panel (a), the
    grey bars indicate the ranges of $\eff$ and $t_{\rm ff,0}/t_{\rm dyn,0}$ 
    in the $\alpha$-series
    plotted in Figure~\ref{f:epsff}, while the black bars indicate the ranges of
    median values of $\varepsilon_{\rm ff,obs}$ and $t_{\rm ff}/t_{\rm dyn}$ for
    snapshots between $t_{*,0}$ and $t_{*,90\%}$. }\label{f:epsff_obs}
\end{figure*}

Next, we compute the time-dependent, ``observed'' SFE per freefall time
$\varepsilon_{\rm ff,obs}$ using mock observations of simulated clouds. For this
exercise, we use the instantaneous cloud mass ($M_{\rm neu}$, assumed to be
recoverable from molecular lines or dust emission), size
($R = 2^{1/3}R_{50\%}$), and velocity dispersion ($\sigma_{\rm 1d}$) to compute
the dynamical time $t_{\rm dyn} = R / \sigma_{\rm 1d}$ and the freefall time
$t_{\rm ff} = \sqrt{3\pi/(32 G \bar{\rho})}$, where
$\bar{\rho} = 3M_{\rm neu}/(4\pi R^3)$. The instantaneous ratio
$t_{\rm ff}/t_{\rm dyn} \propto \sigma_{\rm 1d}(R/M_{\rm neu})^{1/2}$.

Assuming that an observer is able to recover the cluster's bolometric UV luminosity
from massive star formation tracers such as H$\alpha$ and/or infrared fluxes,
the instantaneous SFR is calculated as
$\dot{M}_{*} = \Psi^{-1} L_{\rm UV}/t_{\rm UV}$, where
$L_{\rm UV} = L_{\rm LyC} + L_{\rm FUV}$ is the bolometric UV luminosity. For
the conversion factor between SFR and UV luminosity, we use the IMF-averaged
light-to-mass ratio $\Psi = 10^3 \Lsun \Msun^{-1}$ (see also
Figure~\ref{f:psi}), and we adopt an effective lifetime of UV luminosity
$t_{\rm UV} = 6.7 \Myr$ to be consistent with the relation
$\dot M_* = 1.5\times 10^{-4} \Msun \, \Myr^{-1} (L_{\rm UV}/L_\odot) $ widely
adopted in observations \citep{kennicutt12,Liu20}\footnote{This timescale is similar to the luminosity-weighted timescale of UV radiation
$6.1\Myr$ for a simple stellar population from STARBURST99 with our bandwidth.}.
This assumes a more exact age
for the stellar system is not known, although with spectral fitting cluster ages
can be obtained \citep[e.g.,][]{Adamo17}. The observed SFE per freefall time is
then $\varepsilon_{\rm ff,obs} =
(t_{\rm ff}/t_{\rm UV})L_{\rm UV}/(\Psi M_{\rm neu}) \propto M_* (R/M_{\rm
  neu})^{3/2}$.\footnote{Strictly speaking, $L_{\rm UV}/\Psi$ can underestimate
  the stellar mass if $t_{\rm SF} \gtrsim t_{\rm UV}$, but
  $t_{\rm SF} < t_{\rm UV}$ for most of our clouds.}

Figure~\ref{f:epsff_obs} plots as squares ($\alpha$-series) and diamonds
($\beta$-series) $\varepsilon_{\rm ff,obs}$ for all models as a function of
$t_{\rm ff}/t_{\rm dyn}$. The top axis also shows instantaneous
$\alpha_{\rm vir}=(40/\pi^2)(t_{\rm ff}/t_{\rm dyn})^2$. We consider snapshots
in the time range $t_{*,} < t < t_{*,90\%}$. The color of symbols indicates the
1D velocity dispersion. In contrast with $\eff$ in Figure~\ref{f:epsff}, all of
our models start from the lower left with $\alpha_{\rm vir} \sim 0.5$--$1.5$ and
$\varepsilon_{\rm ff,obs} \sim 10^{-3}$, and, as time progresses, they move
upward slightly and then to the upper right. This trajectory reflects the
evolution of cloud's gas mass, size, and velocity dispersion, and the cluster's
mass and luminosity. The stellar mass $M_*$ and UV luminosity start very small
and keep increasing during the main star formation phase (note that $t_{\rm SF}$
is comparable to or shorter than $t_{\rm UV}$, except for a small number of late
snapshots in strongly magnetized clouds). Meanwhile, radiation feedback drives
mass loss and radial expansion, leading to an increase in $R/M_{\rm neu}$,
together with an increase in velocity dispersion $\sigma_{\rm 1d}$. As a result,
evolution causes $\varepsilon_{\rm ff,obs} \propto M_* (R/M_{\rm neu})^{3/2}$
and $t_{\rm ff}/t_{\rm dyn} \propto \sigma_{\rm 1d}(R/M_{\rm neu})^{1/2}$ to
increase in tandem. The distribution of $t_{\rm ff}/t_{\rm dyn}$ extends to
larger values (by a factor of a few) than in Figure~\ref{f:epsff} due to cloud
evolution. Even more extreme is the stretch in the distribution of
$\varepsilon_{\rm ff,obs}$ by two 
orders of magnitude compared to $\eff$ in
Figure~\ref{f:epsff}.
We note, however, that in their later stages, the simulated
systems have very low gas mass (leading to $\varepsilon_{\rm ff,obs}\gtrsim 0.1$
and $t_{\rm ff}/t_{\rm dyn} \gtrsim 1$).  
While we include the full evolution for completeness, clouds in these stages would not necessarily be included in observed samples of star-forming GMCs.

Although there is a large variation during the evolution, we calculate the
``characteristic'' values of $t_{\rm ff}/t_{\rm dyn}$ and
$\varepsilon_{\rm ff,obs}$ as a median for snapshots equally spaced in time
between $t_{*,0}$ and $t_{*,90\%}$. We find
$0.54<{\rm med}(t_{\rm ff}/t_{\rm dyn})<1.08$ and
$0.8\%<{\rm med}(\varepsilon_{\rm ff,obs})<3.7\%$ in the $\alpha$-series models,
while $0.54<{\rm med}(t_{\rm ff}/t_{\rm dyn})<1.23$ and
$0.8\%<{\rm med}(\varepsilon_{\rm ff,obs})<3.5\%$ in the $\beta$-series models.
The black bars in Figure~\ref{f:epsff_obs}(a) indicate the ranges of the medians.
These values are similar to the ranges of $t_{\rm ff,0}/t_{\rm dyn,0}$ and
$\eff$ shown in Figure~\ref{f:epsff}, as indicated by gray bars in
Figure~\ref{f:epsff_obs}(a).

The circles in each panel show estimates of $t_{\rm ff}/t_{\rm dyn}$
and $\varepsilon_{\rm ff}$ for observed star-forming clouds, colored by
one-dimensional velocity dispersion. Panel (a) shows the low-mass, nearby cloud
sample taken from \citet{evans14}, in which the cloud mass and size are derived
from extinction maps, velocity dispersion from molecular linewidth (mostly
$^{13}{\rm CO}$), and SFRs from counting young stellar objects (YSOs). Panel (b)
shows the sample of larger molecular clouds in the Galactic Plane with ongoing
massive star formation, compiled by \citet{vutisalchavakul16} (their Fig. 6). In
their study, the cloud properties are derived from $^{13}{\rm CO}$ observations
and SFR is determined from mid-infrared (MIR) luminosity. Following
\citet{vutisalchavakul16}, we exclude clouds with $\dot{M}_* < 5 \Msun\yr^{-1}$
as MIR emission may seriously underestimate the true SFR at low SFR (even for
cases shown, the MIR-based SFR may be underestimated by a factor $2$--$3$).
Panel (c) shows the Milky Way GMCs with associated star-forming complexes in
\citet{lee16} (their Fig. 7). Their samples are built from the ${}^{12}{\rm CO}$
catalog of molecular clouds \citep{mivilledeschenes17}, cross-correlated with
WMAP free-free sources. Panel (d) shows the star forming complexes in the
Large Magellanic Cloud \citep{ochsendorf17}. The cloud properties are identified
by ${}^{12}{\rm CO}$ and SFR from H$\alpha$ + $24\mu {\rm m}$. We note that they
multiplied the H$\alpha$-based SFR by a factor of 2 to account for the fact that
the diffuse H$\alpha$ emission (outside the star-forming regions) constitutes
50\% of the total H$\alpha$ emission \citep{pellegrini12}.

The results for $\varepsilon_{\rm ff,obs}$ from the simulations are generally
similar to the observations, especially for the \citet{vutisalchavakul16}
sample. 
However, the observations show no significant correlation of SFE per
freefall time with $t_{\rm ff}/t_{\rm dyn}$. 
This discrepancy is partly due to the fact that points from the simulations extend towards large $t_{\rm ff}/t_{\rm dyn}$ and $\varepsilon_{\rm ff,obs}$ for clouds in the dispersing stages, which would be unlikely be included in the observational samples.
Furthermore, the (extended) theoretical fit of $\eff$ vs. $t_{\rm ff,0}/t_{\rm dyn,0}$  to our $\alpha$-series (shown in grey) 
does not follow the same trend as the correlation of time-dependent  ``observables'' from the simulations.
We conclude that evolutionary
effects (the correlated increase over time of stellar mass, cloud size, and
velocity dispersion along with the decrease of gas mass), which tend to make
$\varepsilon_{\rm ff,obs}$ increase with $t_{\rm ff}/t_{\rm dyn}$ in
Figure~\ref{f:epsff_obs}, completely obscure the underlying decrease of $\eff$ with
$t_{\rm ff,0}/t_{\rm dyn,0}$ seen in Figure~\ref{f:epsff}.

Also, while the observed SFE per freefall time is small on average, it exhibits
large intrinsic scatter.\footnote{From (a) to (d), the median value
  $\varepsilon_{\rm ff,obs}$ of each sample is $0.015$, $0.004$, $0.018$, and
  $0.075$, and the scatter is $0.5$, $0.6$, $0.8$, and $0.6$ dex. The
  systematically higher $\varepsilon_{\rm ff}$ in the LMC may partly be due to
  the difference in the method by which molecular clouds are matched to
  star-forming regions (identified by H$\alpha$ or free-free emission)
  \citep{krumholz19}.} As discussed by previous authors, the large scatter in
$\varepsilon_{\rm ff,obs}$ may arise for a number of reasons. First, as in our
mock observed clouds, sampling clouds at different evolutionary stages (low
stellar mass and high gas mass at an early evolutionary stage and the opposite
at a later stage), is an important source of scatter \citep[see
also][]{grudic19}. Second, if the SFR in a cloud accelerates with time it will
tend to produce a large scatter in $\varepsilon_{\rm ff,obs}$
\citep[e.g.,][]{lee16} (although in fact we do not find evidence for
accelerating SFR in our simulations). Third, poor sampling of the IMF at the
high-mass end can be a source of large scatter and bias in the light-to-mass
ratio and hence the inferred SFR, especially for SFR tracers relying on ionizing
radiation \citep[e.g.,][]{daSilva14,geen18}. Fourth, some variations can be
induced by the distribution of $\avircl$ and $\muBcl$ of initial clouds
\citep[e.g.,][]{federrath13}, which in reality would represent a varying
dynamical environment where molecular clouds form. In addition, there is scatter
introduced in the observational estimation of cloud mass and $t_{\rm ff}$,
largely because the former usually depends on conversion from CO luminosity via $X_{\rm CO}$, which is known from both observations and modeling to have
significant scatter \citep[e.g.,][]{pineda08,shetty11a,shetty11b,seifried20,gong20}. Adopting a constant value of $X_{\rm CO}$ at cloud scales adds to the error in the estimated gas mass, which enters as  $\varepsilon_{\rm ff,obs} \propto X_{\rm CO}^{-3/2}$ and $t_{\rm ff}/t_{\rm dyn} \propto X_{\rm CO}^{-1/2}$.

Our analysis shows that evolutionary effects make it difficult to test certain predictions from star formation theory/simulations with observations of individual molecular clouds. Looking forward, however, large extragalactic surveys that
sample varying large-scale ISM/environments as well as different stages of cloud evolution \citep[e.g., EDGE, PHANGS:][]{Bolatto_EDGE2017,sun20} may be useful for testing 
model predictions. In particular, the anti-correlation between 
$\avircl$ and $\eff$ seen in our simulations (Figure~\ref{f:epsff}) can be investigated by inter-comparing population-averaged cloud properties in different galactic environments;
at a given galactocentric radius, $\avircl$ can be estimated from the cloud-scale $\avir$ averaged over non-star-forming clouds,
while $\eff$ can be estimated by taking the cloud-scale $t_{\rm ff}$ averaged over the same clouds, multiplying by the annular SFR, and dividing by the total mass of (non-star-forming) molecular gas in the annulus.
By averaging over azimuthal angle at a given radius, it can be ensured that all stages of (early) evolution are represented. For better leverage, comparisons based on widely ranging environments are likely to be most useful -- e.g., atomic-dominated outer disks, molecule-dominated inner disks, and galactic centers (including starbursts). Galactic regions with high or low ratios of stellar-to-gas vertical gravity may also have systematically higher or lower mean values of $\avir$.

A recent compilation study by \citet{schruba19} including both Milky Way and extragalactic (8 galaxies) samples of GMCs indeed found that molecular clouds in atomic-dominated, low-pressure environment follow the decreasing trend of $\varepsilon_{\rm ff}$ with $\avir$, although this trend is lacking for clouds in molecule-dominated, high-pressure environments. \citet{schruba19} also emphasized that only by averaging cloud properties over the whole population can dependencies on environment become evident. 
Future studies utilizing homogeneous datasets will be desirable to draw more statistically significant conclusions.

\section{Summary and Discussion}\label{s:summary}

\subsection{Summary}

In this work, we have presented a suite of RMHD simulations modeling the
evolution of a turbulent, magnetized GMC and its dispersal by UV radiation
feedback from embedded star clusters. Our simulations model radiation from both
internal cluster particles (using adaptive ray tracing) and external sources
(using the six-ray approximation). The radiative transfer calculation is coupled
with a simple photochemistry module that follows the non-equilibrium abundances
of molecular, atomic, and ionized hydrogen as well as equilibrium abundances of
C- and O-bearing species. Dynamical evolution is driven by the high thermal
pressure of the ionized (photoevaporated) gas, and by the radiation pressure of
both LyC and FUV on the gas/dust.

Our model clouds all have a common initial mass $\Mcl = 10^5\Msun$, radius
$\Rcl=20\pc$, and freefall time $\tffcl = 4.68\Myr$. They differ in the initial
level of turbulence and in the initial uniform magnetic field that threads the
clouds. In the $\alpha$-series models, we vary the initial kinetic virial
parameter $\avircl$ between 1 and 5 and adopt a fixed initial mass-to-magnetic
flux ratio $\muBcl=2$ (relative to the critical mass-to-flux). In the
$\beta$-series models, we adopt a fixed $\avircl=2$, while varying $\muBcl$
between $0.5$ and $8$; we also consider the unmagnetized case
($\muBcl = \infty$). For each cloud model with a given ($\avircl$, $\muBcl$), we
run five simulations with different turbulence realizations. Our fiducial model
has $(\avircl,\muBcl)=(2,2)$, corresponding to an initial 1D velocity dispersion
of $2.9\kms$ and magnetic field strength of $13.5 \mu {\rm G}$.

The following summarizes key results from our work:
\begin{enumerate}[wide,labelwidth=!,labelindent=0pt]
\item {\it Overall Evolution.} Clouds in our simulations undergo the following
  evolutionary stages: (1) development of filamentary structures, (2) formation
  of sink particles and compact \HII\ regions within filaments, (3) breakout and
  merging of \HII\ regions that photoevaporate and dynamically eject the cold
  neutral gas, leading ultimately to quenching of star formation. Compared to
  the fiducial model, stronger turbulence in initially unbound ($\avircl>2$)
  clouds makes them expand in size and lose a larger fraction of neutral gas as
  outflows (Figure~\ref{f:snapshot-a} and Figure~\ref{f:hst2}(e)). In strongly
  magnetized clouds ($\muBcl \le 1$), filaments are aligned predominantly
  perpendicular to the direction of large-scale magnetic fields
  (Figure~\ref{f:snapshot-b}) and the motion of (both neutral and ionized) gas
  is highly anisotropic (Figures~\ref{f:EM} and \ref{f:hst2}(j)). Models with
  different turbulence realizatons show that the initial large-scale velocity
  field and its relative orientation to the magnetic field affect the details of
  structure formation and ensuing cluster formation, giving rise to moderate
  variations in simulation outcomes (Figures~\ref{f:sfhst}--\ref{f:fesc} and
  Table~\ref{t:result}).
  
\item {\it Star Formation Rate and Efficiency.} For all models, UV radiation
  feedback is very effective in dispersing gas and keeping the final cloud-scale
  SFE low. Both the final SFE and the SFR are reduced in models with strong
  turbulence and magnetic fields. The median value of the final SFE
  ($\varepsilon_{*}$) ranges between $2.1\%$ and $9.5\%$ in the $\alpha$-series
  and between $2.4\%$ and $8.2\%$ in the $\beta$-series models, systematically
  increasing with decreasing $\avircl$ and increasing $\muBcl$. Different
  realizations of turbulence lead to an additional $\sim 2$--$5\%$ variation 
  in the final SFE. The SFE at the nominal time of the first
  supernovae ($\varepsilon_{*,{\rm 3 Myr}}$) ranges between $0.5\%$ and $7\%$
  and is about $50^{+20}_{-20}\%$ of the final SFE (Figure~\ref{f:SFE}). The
  time-averaged SFR $\langle \dot{M}_* \rangle$ is systematically lower at
  higher $\avircl$ and lower $\muBcl$ (Figure~\ref{f:sfhst}). The median value
  of the gas depletion timescale
  $t_{\rm dep,0} = \Mcl/\langle \dot{M}_* \rangle$ ranges between
  $58$--$255\Myr$ in the $\alpha$-series models and $61$--$464\Myr$ in the
  $\beta$-series models.

\item {\it Mass Loss Efficiency.} In most of our simulated clouds, the majority
  of cloud mass is lost via photoevaporation followed by dynamical ejection
  ($\varepsilon_{\rm ion} \sim 0.6$--$0.8$; Figure~\ref{f:evej}). Initially
  unbound clouds have a higher fraction of gas ejected as neutrals due to
  expansion driven by initial turbulence. In strongly magnetized clouds, most of
  the gas is ejected in the atomic phase due to the low ionizing photon rate and
  magnetic confinement, which limits escape routes of the gas.

\item {\it Evolutionary Timescales.} The median value of the time at which the first
  star formation occurs ($t_{*,0}$) is $3.5$--$4.3\Myr$ in the $\alpha$-series
  models and $2.1$--$8.6 \Myr$ in the $\beta$-series models. Except for the
  magnetically subcritical clouds ({\tt A2B05}), clouds in our simulations form
  stars on timescales of $t_{\rm SF} \sim 5$--$7\Myr$ (or $1$--$1.5$ freefall
  times) and are destroyed within $t_{\rm dest,H_2} \sim 6$--$9\Myr$ (or
  $1.3$--$2.0$ freefall times) after the onset of radiation feedback
  (Figure~\ref{f:t_all}). Magnetically subcritical 
  clouds have longer
  evolutionary timescales because of the low SFR and inefficient mass loss
  processes. 
  Since SFRs are approximately constant, the average observed age of the stellar clusters in an actively star-forming cloud would be approximately $t_{\rm SF}/3\sim 2\Myr$.

\item {\it Escape of Radiation.} The cumulative median escape fraction of LyC (FUV)
  radiation ranges between $21$--$34\%$ ($53$--$69\%$) in the $\alpha$-series
  models and between $13$--$33\%$ ($57$--$64\%$) in the $\beta$-series. The escape fraction is low in
  magnetically subcritical clouds because of reduced gas compressibility,  which
  results in less porous structure.  Different turbulent realizations yield greater variation in $f_{\rm esc,Lyc}$ than in $f_{\rm esc,FUV}$.
  
\item {\it Virial Ratios.} 
We study the evolution of the virial ratio
  $\avirttot$ that accounts for kinetic, thermal, and magnetic energies of the
  neutral cloud, and compare it with the traditional kinetic virial parameter
  $\avir = 5 R \sigma_{\rm 1d}^2/(G M_{\rm neu})$ based on the cloud mass, size,
  and velocity dispersion (Figures~\ref{f:hst2} and \ref{f:avir}). The virial
  ratio decreases due to decaying turbulence at early times, but it increases to
  large values ($\avirttot > 10$) at late times as feedback disperses the cloud.
  In spite of the widely ranging initial conditions of the models,
  at the time
  of the first star formation the median value of $\avirttot$ falls in a fairly narrow range $0.9$--$2.2$ (except for the magnetically subcritical clouds, for which the median value is $6.7$).
  In strongly magnetized clouds, the kinetic energy may be
  significantly underestimated or overestimated depending on the relative
  orientation between the observer's line of sight and the direction of magnetic
  field. Averaging over directions (as would be possible with an ensemble
  average in observations), the traditional virial parameter generally agrees
  with $\avirttot$ within a factor of two. However, gravitational boundedness is
  systematically overestimated when $\avir$ is used, due to the neglect of magnetic energy.
  We also discuss approximate relations (Equations~\eqref{e:virial_bal0}--\eqref{e:virial_bal2}) 
  that would be satisfied by an ensemble of molecular clouds in statistical equilibrium,
  and how the importance of magnetic support can be indirectly inferred from observations.
  
\item {\it SFE per freefall time}. The median value of the SFE per freefall time
  based on the time-averaged SFR and the initial cloud mass and size
  ($\eff = \tffcl \langle \dot{M}_* \rangle / \Mcl$) ranges between $1.8$--$8.0\%$ in
  the $\alpha$-series and $1.0$--$7.7\%$ in the $\beta$-series, higher at lower
  $\avircl$ (weak turbulence) and higher $\muBcl$ (weak mean magnetic fields)
  (bottom row of Figure~\ref{f:t_all} and Figure~\ref{f:epsff}).
  However, the dependence of $\eff$ on $\muBcl$ is very slight in the magnetically supercritical regime. 
  The functional form
  $\eff = \exp( -C t_{\rm ff,0}/t_{\rm dyn,0} )$ \citep[similar to][from
  driven-turbulence simulations]{padoan12} well describes the results of
  the $\alpha$-series models, with the best-fit parameter $C=4.06$. The decreasing
  trend of $\eff$ with $t_{\rm ff,0}/t_{\rm dyn,0} \propto \avircl^{1/2} $ is
  also in qualitative agreement with some theoretical models for SFR based on a
  density threshold and log-normal density PDF, although others follow different trends 
  \citep[e.g.,][]{krumholz05,hennebelle11,federrath12}.
  Nevertheless, observational validation is challenging because the relation between instantaneous measurements
  $\varepsilon_{\rm ff,obs}(t)$ and $\avir(t)$ (as would be available in
  observations) shows a quite different (almost orthogonal) trend from both
  numerical measurements and theoretical models of $\eff$. The increase in
  $\varepsilon_{\rm ff,obs} \propto M_* (R/M_{\rm neu})^{3/2}$ with
  instantaneous
  $t_{\rm ff}/t_{\rm dyn}\propto \sigma_{\rm 1d}(R/M_{\rm neu})^{1/2}$ owes to
  rapidly evolving cloud and cluster properties (physical expansion of the cloud
  combined with conversion of neutral gas to stars and ionized gas) which affect
  both quantities in the same way. This ``evolutionary masking'' explains why
  observations of star-forming clouds fail to reproduce the decline of $\eff$
  with $\avircl$ predicted by the theory (Figure~\ref{f:epsff_obs}).
\end{enumerate}

\subsection{Discussion}

Our conclusions regarding the rapid and efficient dispersal of GMCs by ionizing
and non-ionizing UV are in agreement with our previous findings in
\citet{kim18}. There, we focused on how evolution and final outcomes depend on
the cloud surface density (for a range $\Sigmacl \approx 10$--$10^3 \Sunit$), and
found that in all cases star formation is quenched and clouds are dispersed
within a few freefall times or several Myr after star formation commences. Our
new simulations, with more realistic physics (addition of realistic
thermochemistry, magnetic fields, and time-dependent luminosity) confirm this
evolutionary scenario. The short timescales we find are also consistent with
recent empirical constraints on rapid cloud dispersal after the onset of star
formation \citep[e.g.,][]{kruijssen19,chevance20a,chevance20c,kimj20}.

Our findings of decreasing SFR and net SFE with increasing $\avircl$ are
consistent with findings from previous hydrodynamical simulations of global
clouds \citep[e.g.,][]{raskutti16,bertram15,howard16,dale17,kim18}, as well as
hydrodynamic and MHD simulations with driven turbulence in a periodic box
\citep{padoan12}. General theoretical expectations as well as some numerical simulations \citep[e.g.,][]{federrath15} suggest lower SFR and SFE in magnetized compared to unmagnetized clouds,
but we are not aware of a previous systematic numerical study of the dependence on global mass-to-magnetic flux ratio with massive star feedback.

While traditionally it was usually assumed that the virial parameter is close to
unity for GMCs \citep{solomon87}, recent extragalactic surveys of molecular gas
on scales $\sim 50$--$100\pc$ have instead found that the typical (kinetic) virial parameter
may be closer to 4 than 1 \citep{sun20}. Measurements of pressure also show that
only half can be accounted for by the self-gravity of the cloud \citep{sun20a},
which again suggests $\avir > 2$. Thus, our $\avir\sim 4-5$ models are likely
the most realistic, and it is these that have the lowest $\eff \sim 2-3\%$.
While the estimate of the mean $\varepsilon_{\rm ff}$ from the large
extragalactic PHANGS survey is closer to $\sim 1\%$ \citep{utomo18}, it must be
borne in mind that whole-galaxy averages include diffuse and quiescent molecular
gas in clouds prior to the onset of star formation. For our simulations, the
quiescent period in fact has comparable duration to the active star-forming
epoch; averaging over both the non-star-forming and star-forming epochs would
reduce the mean efficiency per unit time. The range of $\varepsilon_{*}$ in our
simulations is in good agreement with estimates of the integrated cloud-scale
SFE of $4$--$10\%$ obtained from the statistical analysis of ${\rm CO}$ and
${\rm H}\alpha$ emission in nearby star-forming galaxies \citep{chevance20a}.

As remarked in Section~\ref{s:depmu}, the subcritical clouds ($\muBcl=0.5$) have
evolution very different from other models in several respects. These include an
extremely long time until the first star formation followed by very slow star
formation and dispersal (the lifetime can exceed $30 \Myr$ for some {\rm seed}s,
and the final SFE $~\sim 2$--$3\%$), columnar outflow configuration, very low
fraction of gas that is ionized, and extremely low escape fraction of LyC
radiation, $f_{\rm esc, LyC}\sim 8-13\%$. The columnar outflow configuration of
the subcritical models is not seen in observed GMCs, although recent evidence suggests
that the dynamics of bipolar (or hourglass-shaped) \HII\ regions on smaller scales is
controlled by magnetic fields \citep[e.g.,][]{eswaraiah17}.
In addition, the low
$f_{\rm esc, LyC}$ from magnetically subcritical clouds appears insufficient to explain emission from diffuse ionized
gas, which requires low-density channels in the large-scale ISM and also
sufficient photons escaping from active star-forming clouds \citep[see][and references therein]{kim19,kado-fong20}. Another feature of the subcritical
models is that the cloud typically consists of massive filaments aligned
preferentially perpendicular to the mean magnetic field (e.g., top row of
Figure~\ref{f:snapshot-b}). This feature, while unlike observations of GMCs, is
reminiscent of the perpendicular alignment between magnetic fields and
high-column structures in several nearby lower-mass ($M \lesssim 10^4\Msun$)
dark molecular clouds \citep[e.g.,][]{soler16}. This raises the interesting
possibility that observed dark clouds, which have low overall SFE and lack of
high-mass stars \citep[e.g.,][]{evans09}, may be magnetically subcritical. With
slow star formation and low total mass, they may typically be destroyed (by
external supernova shocks) before they ever form high-mass stars.

We note that our reported values for $\varepsilon_{*}$ are higher than the
predictions for the SFE ($\sim 1\%$) based on idealized spherical shell expansion \citep{kim16,rahner19,inoguchi20} even though the physical processes (ionized gas pressure and radiation pressure) in those spherical models 
are included in our simulations. As explained previously in \citet{kim18},
several factors contribute to higher SFEs in realistic (non-spherical) models. One factor is the filamentary nature of the
gas distribution, which allows a significant fraction of radiation to escape
without contributing direct radiation pressure or photoevaporating neutral gas and helping to maintain the
thermal pressure in ionized gas. Also, given the turbulent nature of clouds and multiple
sources, thermal and radiation pressure forces may partially cancel, unlike the
situation for a spherical cloud with a central source. In clouds with realistic
structure, forces derived from feedback are reduced by an order of magnitude
compared to simple spherical estimates (e.g., Figure 13 of \citealt{kim18}).

While idealized analytic studies are valuable for identifying key physical
effects and predicting trends in parameter dependence, numerical simulations
with realistic geometry and accurate physical treatments are required in order
to obtain quantitative predictions for SFRs and SFEs.
Indeed, idealized spherical evolution driven by ionized gas pressure and radiation pressure does not even include the photoevaporation mechanism, which we find (as in \citealt{kim18}) is the main means for dispersing the cloud.  Although not always taken under consideration in simple treatments of cloud destruction  \citep[e.g.,][]{chevance20c}, we note more generally that photoevaporation and self-acceleration of ionized gas is a distinct mechanism from acceleration of neutral gas by the pressure of confined photoionized gas, and results in a much larger fraction of gas becoming ionized and ejected.

Ejected ions can explain the low-density ionized gas (``halo'') surrounding individual \HII\ regions \citep[e.g.,][]{anderson15,luisi20}, but this process is not likely to be the immediate source of diffuse ionized gas in the larger-scale ISM. The reason is that both the lifetime of ionizing sources in a given cloud and the gas recombination time scale are short compared to the timescale which ejected ionized gas would require to reach and fill the distant diffuse ISM. The ejected photoionized gas would become neutral on a recombination timescale ($t_{\rm rec} \sim (\nH/0.1 \pcc)^{-1}\Myr$) after the death of ionizing stars, or if the path to radiation sources becomes obstructed.

The present simulations have made a significant advance over previous work in
terms of physical realism. However, there is still more to be done. One
limitation of the present simulations is the isolated nature of clouds. The
isolation of clouds means that they are not affected by events in the
surrounding ISM, which in reality may be important in high density environments
such as spiral arms where GMCs are near neighbors. These events include
supernovae from neighboring, more mature star-forming regions, which may destroy
a cloud before star formation reaches the limit imposed by its own internal
feedback.

A second limitation is the idealized initial condition with uniform
density and magnetic field. Our model clouds become highly filamentary before
star formation commences, so we do not consider the uniform spherical initial
condition in itself a serious issue. However, our model clouds are at much
higher density than the mean ISM, and it is clearly desirable to follow the ISM
condensation processes that {\it form} GMCs at high resolution with
comprehensive physics, not least because this removes the need to artifically
set an initial virial parameter and mass-to-flux ratio. Zoom-in simulations
\citep[e.g.,][]{seifried17,haid19,seifried20} offer a route to a more complete
and holistic numerical study of molecular cloud lifecycle, from formation to
destruction.

A third limitation is related to our treatment of star formation. In our
simulations, we use sink particles, which are relatively massive, to represent
regions of the cloud that have collapsed gravitationally. Because our approach
does not resolve individual stars, we adopt the same time-dependent
light-to-mass ratio for all particles, which assumes a fully-sampled IMF. While
our clouds are massive enough that a fully-sampled IMF is reasonable over the
lifetime of the cloud, stochastic IMF sampling is not properly represented
during the earlier formation stages. Sampling from the IMF would lead to
additional statistical variations in the lifetime SFE \citep{grudic19b} on top
of those we have found from turbulence sampling, since clouds that happen to
form more high-mass stars early would be able to quench star formation more
rapidly, and vice versa for those with late formation of massive stars. Simple
methods for IMF sampling have been developed to address this \citep{sormani17},
but just as for turbulence sampling, IMF sampling greatly increases the total
computational expense because multiple runs are needed.

Beyond addressing the above issues, another important goal for future work is to
include additional forms of feedback, rather than only radiation. In particular,
stellar winds may be important at early stages after star clusters form
(especially in very dense clouds; e.g., Lancaster et al. 2020 submitted),
while supernovae may be important at late
stages (in low density clouds with long lifetimes; e.g., \citealt{fall10}). However, the reduction of the Courant–Friedrichs–Lewy
timestep necessary to follow hot gas implies considerably higher
computational expense. This underscores the imperative for numerical algorithms
that both represent the underlying equations accurately, and are
performance-optimized for modern computational platforms.

\acknowledgements


We thank the anonymous referee for helpful comments and suggestions that greatly improved the manuscript. We thank Neal Evans for reading our manuscript carefully and providing thoughtful comments.
J.-G.K. acknowledges the participants of Paris-Saclay University's Institut Pascal program ``The Self-Organized Star Formation Process'' for useful discussions. J.-G.K. benefited from the KITP program on ``Globular Clusters at the Nexus of Star and Galaxy Formation,'' which is supported by the National Science Foundation under Grant No. NSF PHY-1748958.
J.-G.K. acknowledges support from the Lyman Spitzer, Jr.
Postdoctoral Fellowship at Princeton University.
This work was partly supported by the National Science Foundation (AARG award
AST-1713949). Computational resources were
provided by the Princeton Institute for Computational Science and Engineering
(PICSciE) and the Office of Information Technology’s High Performance Computing
Center at Princeton University.

\software{{\tt Athena} \citep{stone08,stone09}, {\tt ParaView}
  \citep{ayachit15}, {\tt IPython} \citep{perez07}, {\tt NumPy}
  \citep{vanderWalt11}, {\tt xarray} \citep{hoyer17}, {\tt scipy}
  \citep{virtanen20}, {\tt Matplotlib} \citep{hunter07}, {\tt Astropy}
  \citep{astropy1, astropy2}, {\tt pandas} \citep{mckinney10}, {\tt yt}
  \citep{yt}.}

\appendix

\begin{figure}[t!]
  \begin{center}
    \includegraphics[width=1.0\linewidth]{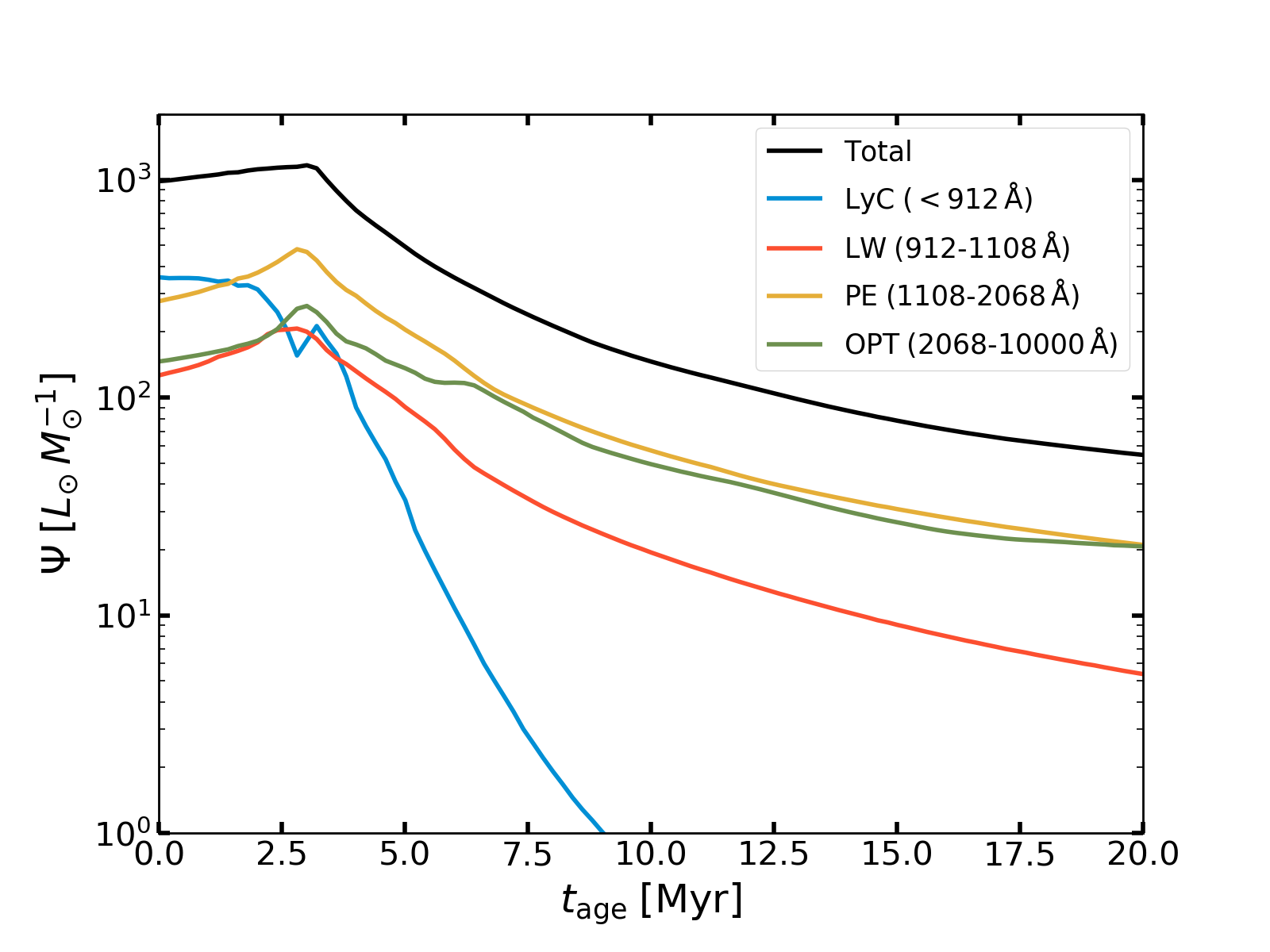}
    \caption{Luminosity per unit mass $\Psi$ as a function of the age of a star
      cluster sampling the Kroupa IMF. Different lines show
      $\Psi$ in different frequency bins: H-ionizing (LyC), photoelectric and
      ${\rm H_2}$-dissociating (LW), photoelectric (PE), and optical (OPT)
      wavelength range.}\label{f:psi}
  \end{center}
\end{figure}

\section{Radiative Output from Star Clusters}\label{s:psi}

We use the Starburst99 stellar population synthesis code \citep{leitherer14} to
calculate the radiative output per unit mass $\Psi = L_*/M_*$ from a coeval
stellar population fully sampling the Kroupa IMF in the mass range of
$0.1$--$100\Msun$. We adopt the Geneva evolutionary tracks for non-rotating,
solar metallicity stars and Pauldrach + Hillier stellar atmosphere models.
Figure~\ref{f:psi} shows the luminosity per unit mass as a function of age in
different frequency bins. The conversion factors for ionizing (LyC) and
non-ionizing (LW+PE) UV photons drop to 50\% of the initial value after $3\Myr$
and $6\Myr$, respectively. The ratios $\Psi_{\rm LW}/\Psi_{\rm PE}$ and
$\Psi_{\rm OPT}/\Psi_{\rm PE}$ range between $0.25$--$0.5$ and $0.5$--$1$ for
$t_{\rm age} < 20\Myr$.

\section{Snapshots of Representative Models}\label{s:snapshot}

To provide a sense of the differences in evolution for different models, here we
present comparison snapshots at times $t_{*,0}-0.2\Myr$, $t_{*,0}+2\Myr$,
$t_{*,0}+5\Myr$, $t_{*,0}+8\Myr$, where $t_{*,0}$ is the time of first star
formation. We show projections of gas surface density along the $y$ axis, so
that individual panels have $x$ horizontal and $z$ vertical, with the initial
magnetic field oriented along the $z$-direction.
      
\begin{figure*}[t!]
  \begin{center}
    \includegraphics[width=1.0\linewidth]{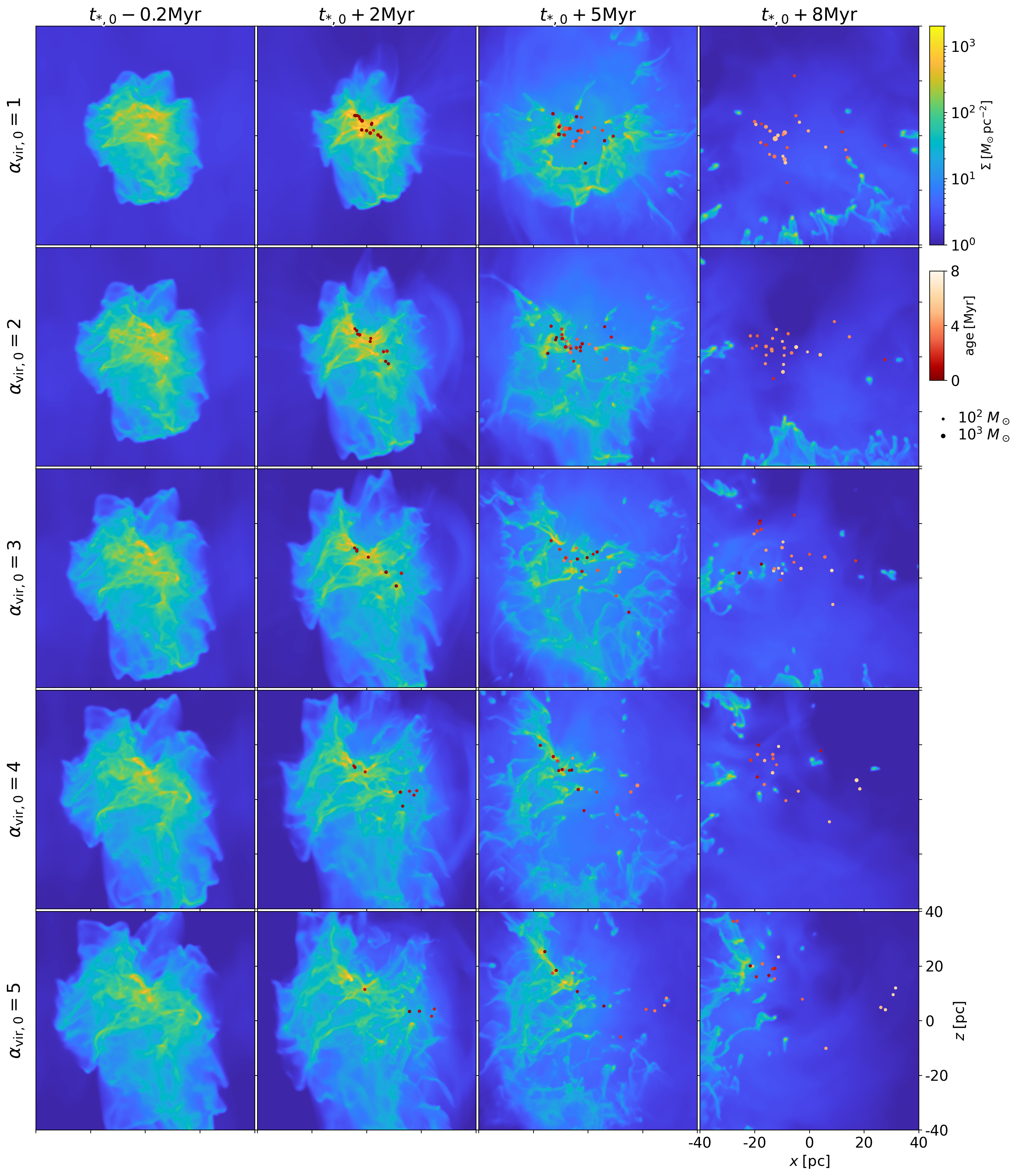}
    \caption{Snapshots of gas surface density projected along the $y$-axis from
      the $\alpha$-series models with ${\tt seed=4}$ From top to bottom, the
      initial virial parameter $\avircl$ increases from $1$ to $5$. The
      projected positions of star particles are indicated by circles. The size
      and color of circles indicate mass and age, respectively.}
      \label{f:snapshot-a}
  \end{center}
\end{figure*}

Figure~\ref{f:snapshot-a} compares members of the $\alpha$-series with different
initial kinetic energy (increasing top to bottom) but identical initial relative
amplitudes of all the different turbulent Fourier modes ({\tt seed}$=4$) and
initial mass-to-magnetic flux ratio $\muBcl=2$. The second row shows the
fiducial model. Members of the series with higher initial $\avircl$ disperse
more rapidly and have a lower SFE.
      
\begin{figure*}[t!]
  \begin{center}
    \includegraphics[width=1.0\linewidth]{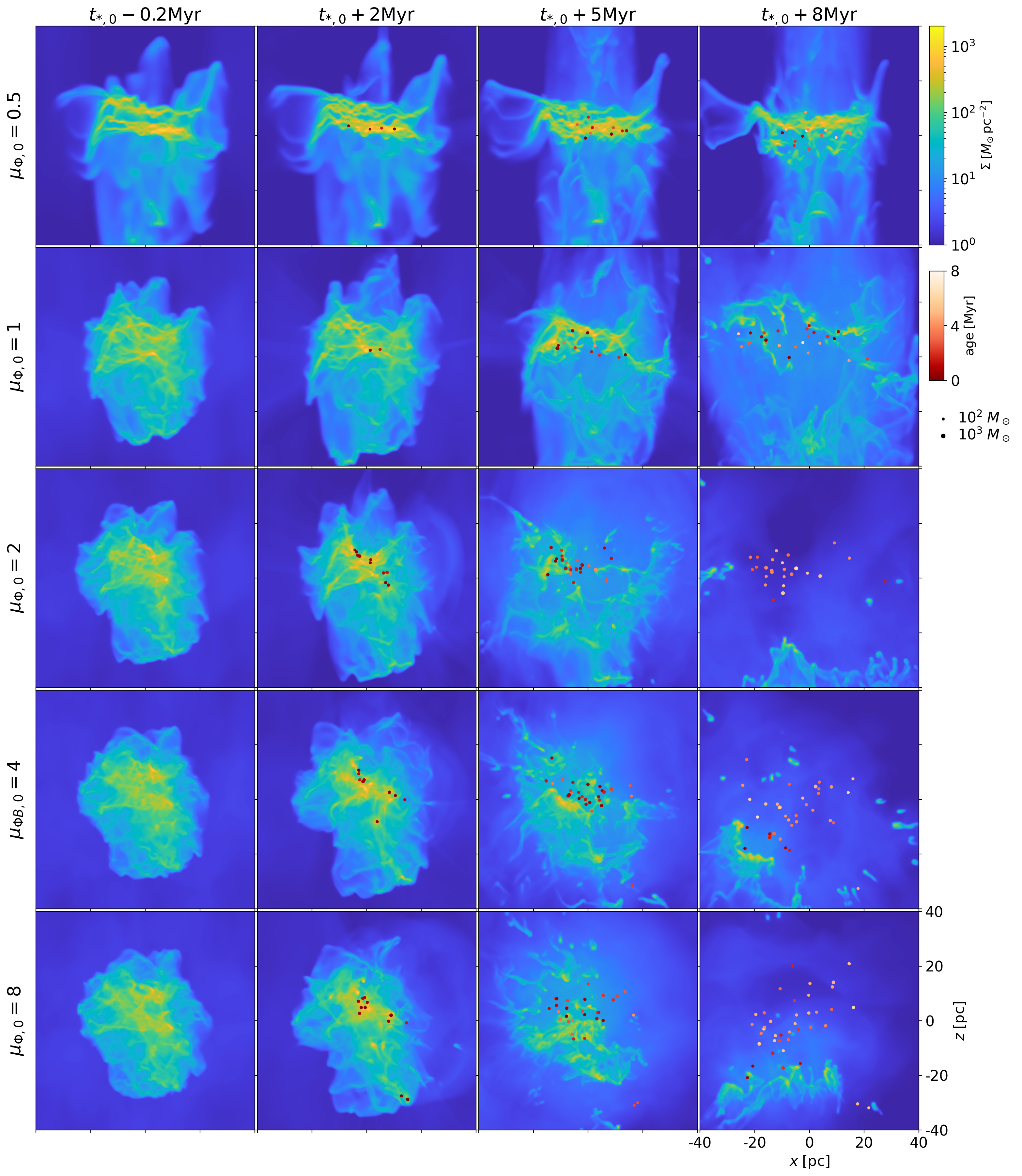}
    \caption{Same as Figure~\ref{f:snapshot-a} but for the $\beta$-series modes
      with ${\tt seed=4}$, which have identical initial turbulent velocity field
      but varying initial magnetic field strength, decreasing from top to
      bottom. The corresponding initial mass-to-magnetic flux ratio $\muBcl$
      varies from $0.5$ to $8$.}\label{f:snapshot-b}
  \end{center}
\end{figure*}

Figure~\ref{f:snapshot-b} compares members of the $\beta$-series with different
initial magnetic energy (decreasing top to bottom) and an identical initial
turbulence realization ($\avircl = 2$, {\tt seed}$=4$). The third row shows the
fiducial model. Dense filamentary structures tend to be aligned perpendicular to
the direction of the background magnetic field in strongly-magnetized models,
and outflows driven by radiation are primarily parallel to the magnetic field.
Weakly magnetized models are overall more isotropic in their evolution, evolve
more rapidly, and have higher SFE.
      
\begin{figure*}[t!]
  \begin{center}
    \includegraphics[width=1.0\linewidth]{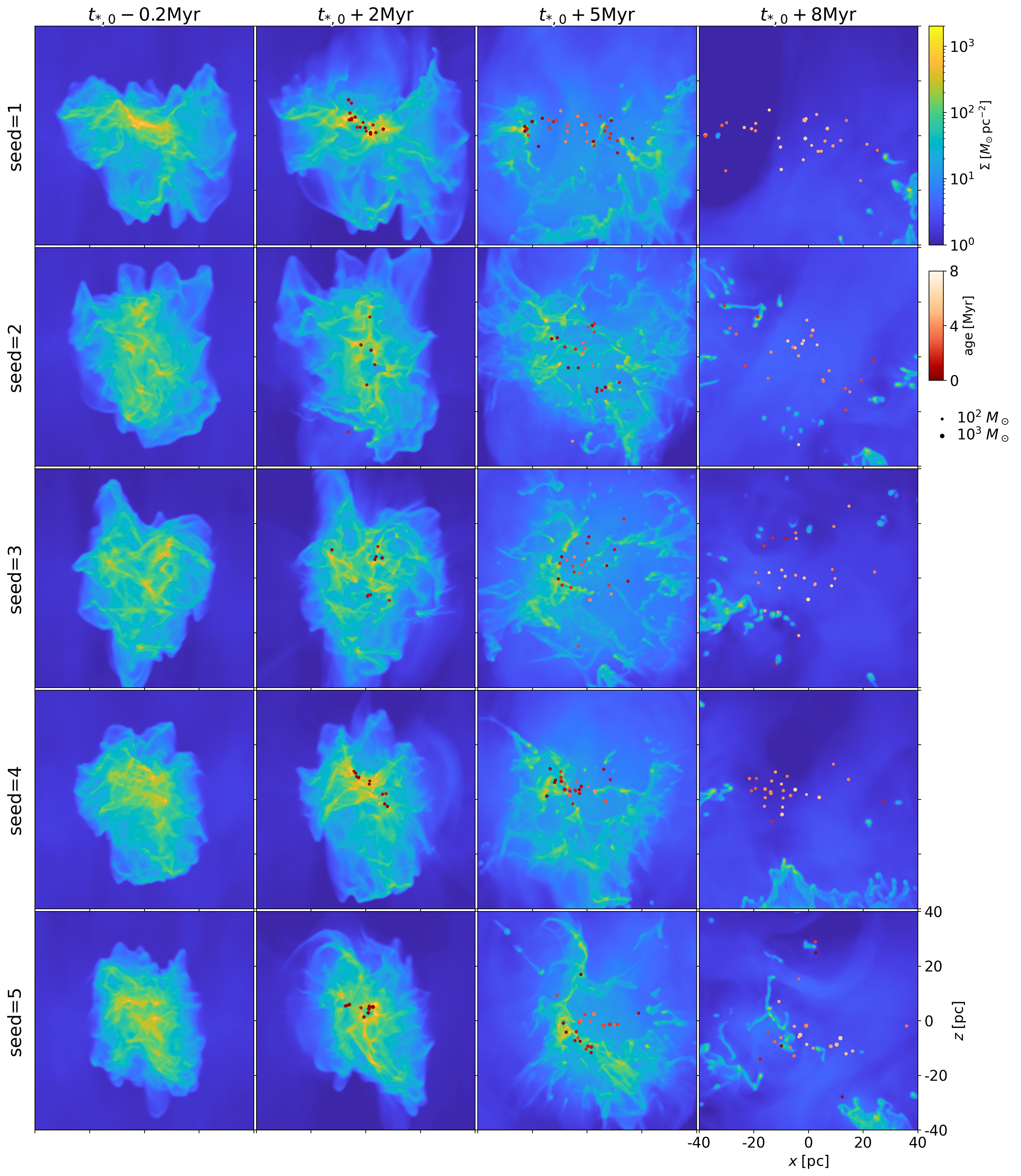}
    \caption{Same as Figure~\ref{f:snapshot-a} but for the fiducial model
      ${\tt A2B2}$ with different random seeds for the initial turbulent
      velocity field.}\label{f:snapshot-s}
  \end{center}
\end{figure*}

Figure~\ref{f:snapshot-s} compares different simulations with different initial
turbulence seeds (i.e., different initial Fourier amplitudes of the turbulence)
and the same value of $\avircl=2$, $\muBcl=2$. Differences in the detailed
turbulence realization can produce quite different morphological evolution, and
there can also be significant differences in the SFE (see Figure~\ref{f:SFE})
and other quantitative outcomes.

\clearpage

\bibliographystyle{aasjournal}
\bibliography{ref_file.bib}{}

\end{document}